\documentclass[reprint,aps]{revtex4-2}

\usepackage{amsmath,amssymb}
\usepackage{graphicx}
\usepackage{bm}
\usepackage{xcolor}
\usepackage{siunitx}
\usepackage{hyperref}
\newcommand{\bra}[1]{\langle #1|}
\newcommand{\ket}[1]{|#1\rangle}

\begin{document}
	
\preprint{APS/123-QED}

\title{The strong modulation limit of excitons and trions in moir\'e materials}
\author{Yongxin Zeng and Allan H. MacDonald} 
\affiliation{Department of Physics, University of Texas at Austin, Austin, TX 78712}
\date{\today}

\begin{abstract}
The optical properties of weakly-doped two-dimensional materials are dominated by strong exciton and trion
absorption and luminescence features.  In this article we examine the influence of moir\'e patterns in semiconductor 
heterobilayers on exciton and trion states in the limit of strong moir\'e modulation potentials,
commenting on similarities and differences compared to the case of excitons and trions in semiconductor quantum dots.
We discuss strategies for using optical properties as quantitative probes of moir\'e materials, and 
the prospects for exploiting moir\'e materials to design unique light emitters.
\end{abstract}

\maketitle


\section{Introduction}
Moir\'e superlattices form when two-dimensional (2D) crystals are stacked with a small twist angle or lattice mismatch.  When the 
host materials are semiconductors or semimetals, moir\'e superlattices are accurately described by continuum models
with the periodicity of the moir\'e pattern, giving rise to artificial two-dimensional moir\'e materials with 
controllable large lattice constants.  
Extensive theoretical and experimental work on twisted bilayer graphene \cite{bistritzer2011moire,cao2018correlated,cao2018unconventional,lu2019superconductors,yankowitz2019tuning,po2018origin,koshino2018maximally,kerelsky2019maximized,serlin2020intrinsic,kang2018symmetry,isobe2018unconventional,wu2018superconductivity,lian2019twisted,xie2020nature,cao2020tunable,andrei2020graphene,balents2020superconductivity,zondiner2020cascade,wong2020cascade,bultinck2020ground,park2021tunable,khalaf2021charged} and twisted bilayer transition metal dichalcogenides (TMDs) \cite{wu2018hubbard,regan2020mott,tang2020simulation,xu2020correlated,wang2020correlated,kennes2021moire,jin2021stripe,huang2021correlated,chu2020nanoscale,pan2020quantum,morales2021metal,zhang2021electronic,bi2021excitonic,li2021imaging,angeli2021gamma} over the past several years has established moir\'e materials as an important platform for studies of strongly correlated electron physics.

In addition to modifying low-energy electronic properties, the moir\'e potentials in twisted bilayer 
TMDs also significantly modify interband excitonic collective modes,
and associated optical properties \cite{regan2022emerging,wu2017topological,wu2018theory,yu2017moire,tran2019evidence,jin2019observation,seyler2019signatures,alexeev2019resonantly,zhang2018moire,shimazaki2020strongly,zhang2020twist,baek2020highly,ruiz2020theory,liu2021signatures,brotons2021moir,baek2021optical,brem2020tunable}. 
The moir\'e modulation splits both intralayer and interlayer excitonic peaks 
in the optical spectrum into a series of subpeaks and modifies optical selection rules. 
Since the electrons, holes, and excitons are often strongly localized near minima of the periodic moir\'e 
potential, moir\'e superlattice 
systems can act as a new and more perfectly periodic realization of quantum dot arrays \cite{wu2018theory,yu2017moire},
and be used to engineer single-photon emitters and entangled photon 
sources \cite{baek2020highly,seyler2019signatures} with potential applications in quantum optics.

Theories of the excitonic properties of moir\'e materials
have progressed by treating excitons as point particles moving in a periodic moir\'e potential \cite{wu2017topological,wu2018theory,yu2017moire}. 
Recent numerical and experimental studies 
show that moir\'e modulation strengthens when strain is taken into account \cite{shabani2021deep,li2021imagingbands,li2021lattice,naik2018ultraflatbands}, and 
that the conduction band electrons and valence band holes that combine to form 
excitons in a moir\'e material can be attracted to different lateral positions within the moir\'e pattern \cite{guo2020shedding,naik2020origin,zhan2020tunability,maity2021reconstruction,kundu2021atomic}, 
distorting the excitonic modes.  In the strong modulation limit, the electron and hole moir\'e band widths 
become extremely small and the moir\'e material can be viewed as an accurately periodic array of quantum dots.  
In this article we study how strong confinement and lateral shifts between the electron and hole confining centers 
affect the properties of excitons and trions in the quantum dot array limit of a 
moir\'e material. In Sec.~\ref{sec:model} we introduce our model for electrons and holes 
in a moir\'e superlattice, and present results for 
Coulomb interaction matrix elements in the single-particle harmonic oscillator basis that we use for numerical calculations. 
In Secs.~\ref{sec:exciton} and \ref{sec:trion} we calculate the ground state energies and optical absorption strengths of 
excitons and trions respectively, by exact diagonalization (ED) of the corresponding few-body Hamiltonians in the 
harmonic oscillator basis. Finally, in Sec.~\ref{sec:discussion} we discuss the implications of our 
results, and point to future work that can be
approached with our method.  We find that the optical properties of systems in which electrons and holes
are trapped at the same position within the periodic moir\'e pattern are very different from those of systems in which electrons and holes are trapped at different positions; the former case is distinguished by fewer but stronger 
absorption features because of the approximate applicability of angular momentum conservation selection rules.  
Magnetic fields are potentially useful in characterizing moir\'e materials, but because of the strong electron-hole binding in two-dimensional material excitons and trions will yield substantial 
corrections that can be used to identify electron, hole, exciton, and trion configurations only at very strong magnetic fields.

\section{The model} \label{sec:model}

\begin{figure}
    \centering
    \includegraphics[width=0.49\linewidth]{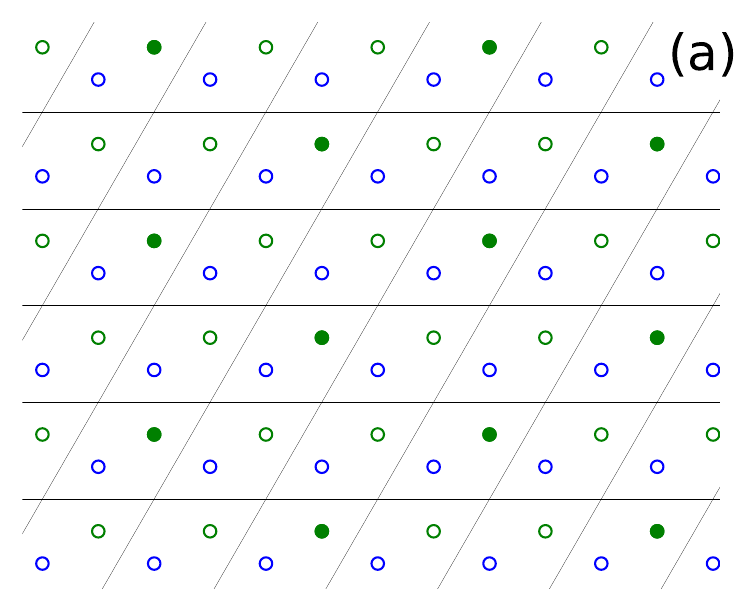}
    \includegraphics[width=0.49\linewidth]{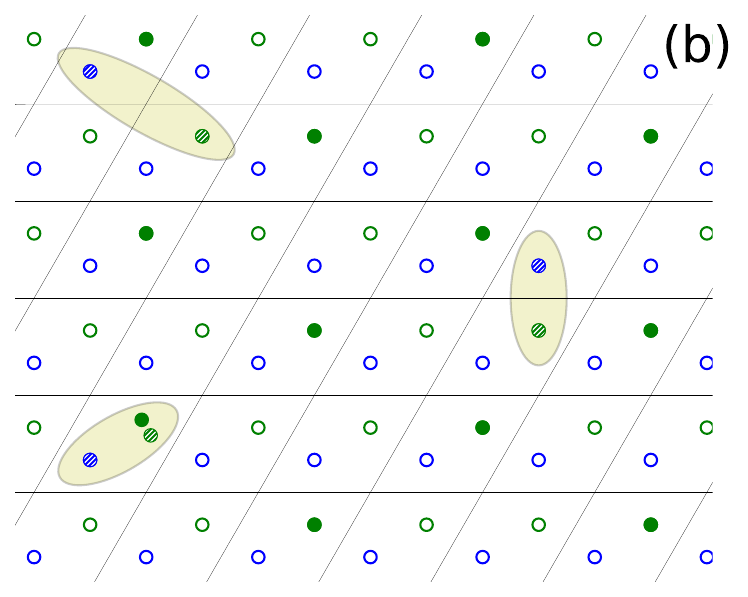}
    \includegraphics[width=0.49\linewidth]{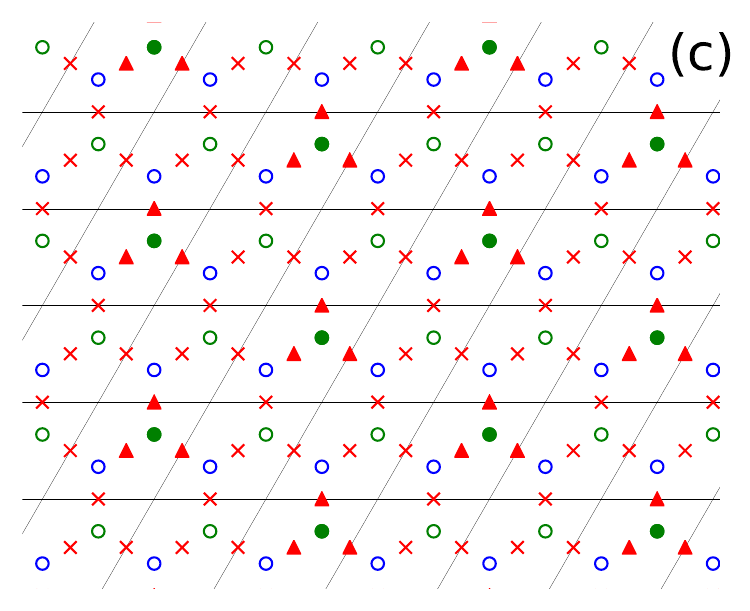}
    \includegraphics[width=0.49\linewidth]{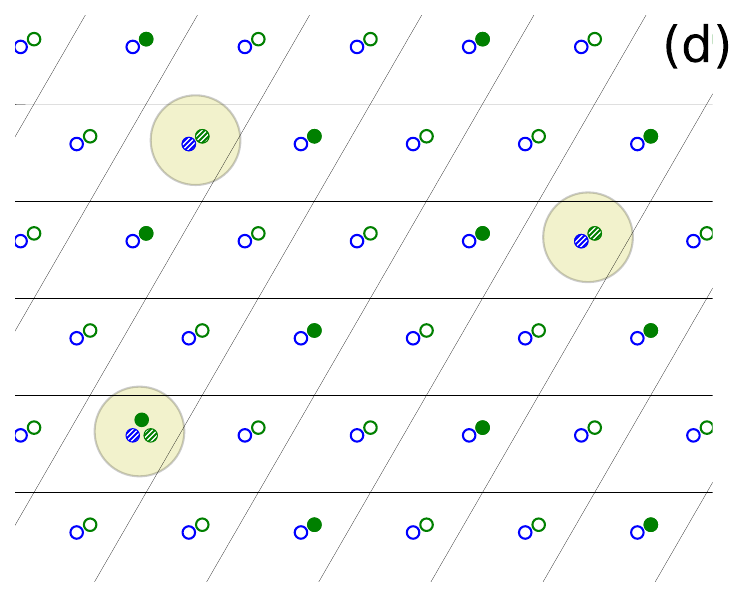}
    \caption{(a) A hole-doped moir\'e superlattice at filling factor $\nu=1/3$. The blue and green circles represent the electron and hole confining centers respectively, which are assumed here to be located at different high-symmetry stacking sites. 
    The filled and empty circles distinguish sites that are occupied by a 
    quasiparticle from those that are empty. The doped holes form a Wigner crystal in the ground state.
    (b) The moir\'e lattice with some electron-hole excitations. The yellow ellipses show three different types of electron-hole excitations: an exciton (right), a trion (lower left), and an example of a more widely separated electron-hole pair (top left). The optically excited electrons and holes are represented by the hatched blue and green circles.
    (c) If the center of an electron site and its neighboring hole site is associated with an exciton or trion excitation, as marked by the red crosses and triangles, together they form a kagome lattice.
    (d) Same as (b), except that the electrons and holes are localized at the same sites. The yellow circles show two excitons and a trion excitation.}
    \label{fig:eh_lattice}
\end{figure}

In a TMD heterobilayer moir\'e superlattice, electrons and holes tend to localize at one of three high-symmetry sites in a moir\'e unit cell \cite{shabani2021deep,li2021imagingbands,li2021lattice,naik2018ultraflatbands,guo2020shedding,naik2020origin,zhan2020tunability,maity2021reconstruction,kundu2021atomic} at which the local stacking places the metal atoms of one layer
above either metal, chalcogen, or empty sites of the layer below.  As summarized in Fig.~\ref{fig:eh_lattice}, there is 
an important distinction between the case in which electrons and hole are localized at the same positions in the 
moir\'e pattern or at different positions.  It is 
not yet confidently known which among the various TMD heterobilayer systems
fall into the former category and which fall into the latter category.
When the electron and hole sites are the same, they form a triangular 
lattice of sites at which excitons can be localized.  When the electron and hole sites are separated in the 2D plane, however,
their triangular lattices combine to form a hexagonal lattice like the one shown in Fig.~\ref{fig:eh_lattice}(a).  In a hole-doped system some of the hole sites are occupied, and at certain fractional fillings the ground state of the doped holes is a lattice Wigner crystal \cite{regan2020mott,tang2020simulation,xu2020correlated,li2021imaging}. Fig.~\ref{fig:eh_lattice} shows an example at filling factor $\nu=1/3$. Possible low-energy excitations of the system include electron-hole pairs at empty neighboring sites (an exciton), or an electron-hole pair at a filled hole site and its neighboring electron site (a trion). Excitations with electrons and holes farther away are also possible, but have higher energies and weaker optical absorptions. If we ignore these, and consider only near-neighbor exciton and trion excitations, each bond between an electron site and its neighboring hole site can be associated with a possible exciton or trion excitation (marked by red crosses and triangles in Fig.~\ref{fig:eh_lattice}(c)), depending on whether the hole site is occupied or not. Together these excitations form a kagome lattice.
Excitons and trions on different sites are coupled by intersite hopping of their constituents, 
and via Coulomb interactions.  We do not account for these complications in this paper, assuming that the excitons and trions are strongly localized at the moir\'e potential minima and 
that coupling between excitons and trions on different moir\'e lattice sites is therefore 
negligibly weak. Our approximation is valid in the strong modulation limit, which is 
always achieved at small twist angles, and in the limit of low exciton density that holds for most optical experiments.

\begin{figure}
\centering
\includegraphics[width=\linewidth]{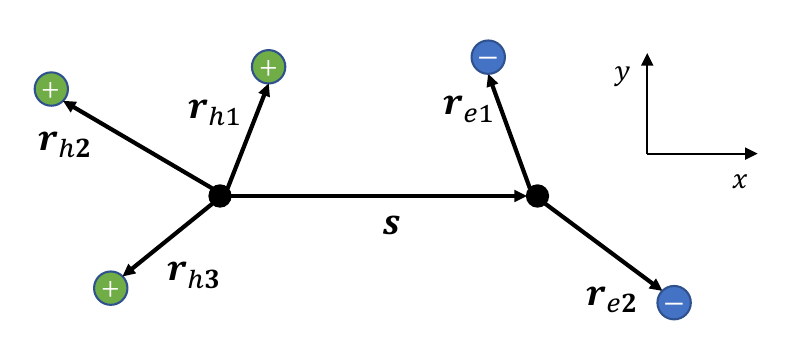}
\caption{Schematic illustration of the model system -- a few electrons and holes confined in a 2D quantum dot. The black dots stand for the confining potential minima for electrons and holes, and the blue and green dots represent electrons and holes respectively. 
The vectors $\bm r_{ei}$ and $\bm r_{hj}$ are defined relative to the potential minima for electrons and holes respectively, which are separated by the vector $\bm s = s\hat x$ when the electrons and holes are in the same layer and by 
$s\hat{x}+d\hat{z}$ when the electrons and holes are in layers separated by distance $d$.} \label{fig:qd_eh}
\end{figure}

In this article we consider both two-particle exciton and  three-particle trion electron-hole complexes, ignoring 
the possible role of coupling to other excitations of the system.
We study a system of $N_e$ electrons and $N_h$ holes confined in a 2D quantum dot (Fig.~\ref{fig:qd_eh}) with harmonic confining potentials for the electrons and holes, and a magnetic field $\bm B=B\hat z$ applied perpendicular to the 2D plane. 
We take advantage of the large spin-splitting of single-layer TMD bands at 
the Brillouin-zone corners by including only one conduction band and one 
valence band per spin.
We consider the general case in which the electrons and holes have different effective masses $m_e$ and $m_h$ and 
are localized at different lateral positions in the same or different layers by confining potentials that 
can have different oscillator frequencies $\omega_e$ and $\omega_h$.  For heterobilayer moir\'e superlattices,
states with electrons and holes in the same layer are relevant in absorption or reflection measurements 
and states with electrons and holes in different layers are relevant in luminescence experiments.
The system is described by Hamiltonian
\begin{equation} \label{eq:Hamiltonian}
H = H_e + H_h + V_{ee} + V_{hh} + V_{eh}.
\end{equation}
The electron and hole single-particle terms are (in cgs units)
\begin{align}
H_e &= \sum_{i=1}^{N_e} \left[\frac{1}{2m_e}(\bm p_{ei} + \frac ec \bm A_e(\bm r_{ei}))^2 + \frac 12 m_e\omega_e^2 r_{ei}^2 \right], \label{eq:H_e}\\
H_h &= \sum_{j=1}^{N_h} \left[\frac{1}{2m_h}(\bm p_{hj} - \frac ec \bm A_h(\bm r_{hj}))^2 + \frac 12 m_h\omega_h^2 r_{hj}^2 \right] \label{eq:H_h},
\end{align}
where $i$ and $j$ are labels for electrons and holes, and the origins of $\bm r_e$ and $\bm r_h$ are chosen to be 
at the minima of the corresponding confining potentials. 
We choose a gauge such that $\bm A_e(\bm r_e) = \frac 12 \bm B\times \bm r_e$ and $\bm A_h(\bm r_h) = \frac 12 \bm B\times \bm r_h$, i.e., the symmetric gauge, but with different origins for electrons and holes. 
We do not explicitly include in our calculations the valley Zeeman splitting \cite{li2014valley,macneill2015breaking,srivastava2015valley,aivazian2015magnetic}, which produces a constant shift in exciton energies but does not change exciton wavefunctions. Its effects can be easily added to our results. The non-trivial 
orbital effects of the magnetic field that we discuss in our work can be decoupled from the valley Zeeman splitting by optical measurements with circularly polarized light.
The interaction terms are
\begin{align}
V_{ee} &= \sum_{i_1<i_2}^{N_e} \frac{e^2}{\epsilon |\bm r_{ei_1} - \bm r_{ei_2}|}, \label{eq:V_ee}\\
V_{hh} &= \sum_{j_1<j_2}^{N_h} \frac{e^2}{\epsilon |\bm r_{hj_1} - \bm r_{hj_2}|}, \label{eq:V_hh}\\
V_{eh} &= \sum_{i=1}^{N_e} \sum_{j=1}^{N_h} \frac{-e^2}{\epsilon \sqrt{|\bm r_{ei} - \bm r_{hj} + \bm s|^2 + d^2}}, \label{eq:V_eh}
\end{align}
where $\bm s$ is the displacement between the confining potential minimum of electrons and holes.  
We choose the coordinate axes such that 
the 2D crystals are in the $\hat{x}$-$\hat{y}$ plane and the in-plane component of $\bm s$ is along the $\hat{x}$ direction (see Fig.~\ref{fig:qd_eh}). $d$ is the distance between the electron and hole layers ($d=0$ when the electrons and holes are in the same layer), and $\epsilon$ is the dielectric constant of the dielectric surrounding the system.  The most commonly employed 
surrounding dielectric is hexagonal boron nitride (hBN), which has an anisotropic dielectric tensor.  In this case 
$\epsilon= \sqrt{\epsilon_{zz} \epsilon_{\perp}}$ and the effective layer separation $d$ is related to the physical separation $d_0$ by
$d= d_0 \sqrt{\epsilon_{\perp}/\epsilon_{zz}}$, where $\epsilon_{\perp}$ and $\epsilon_{zz} $ are 
respectively the perpendicular-to-plane and in-plane components of the dielectric tensor.  

We diagonalize the few-particle Hamiltonian using a configuration interaction approach with 
basis state Slater determinants constructed from 
the eigenstates of $H_e$ and $H_h$.  Our calculations are similar to those reported in 
earlier papers \cite{wojs1995negatively,halonen1992excitons,que1992excitons,maksym1990quantum} on 
quantum dots in semiconductor quantum wells, 
except that we consider here the possibility that the electrons and holes are localized at different lateral positions. 
The harmonic oscillator (with mass $m$, charge $\pm e$, angular frequency $\omega$, in a perpendicular magnetic field $B$) eigenstates $\ket{n_+ n_-}$ are labelled by two quantum numbers $n_+$ and $n_-$, with energy eigenvalues
\begin{equation} \label{eq:sp_energy}
E_{n_+ n_-} = (n_+ + \frac 12)\hbar\Omega_+ + (n_- + \frac 12)\hbar\Omega_-,
\end{equation}
where $\Omega_{\pm} = \Omega \pm \omega_c/2$ with $\Omega = \sqrt{\omega^2 + (\omega_c/2)^2}$ and $\omega_c=eB/mc$. (See Appendix~\ref{app:ed_eqs} for derivations.) To diagonalize the Hamiltonian we choose basis states that are
antisymmetrized combination of electron-hole many particle product states
\begin{equation} \label{eq:basis_nosym}
\ket{n_{e1+} n_{e1-}, n_{e2+} n_{e2-}, \dots; n_{h1+} n_{h1-}, n_{h2+} n_{h2-}, \dots}.
\end{equation}
Antisymmetrization is not required between electrons or holes since these can be considered to be 
distinguishable.  The two-particle matrix elements for Coulomb interactions are evaluated by expanding the interactions in terms of ladder operators of the single-particle basis. For interactions among electrons or holes, the result is 
\begin{widetext}
\begin{equation}
\label{ee2particle}
\begin{split}
\bra{n_{1+}' n_{1-}', n_{2+}' n_{2-}'} V_{\alpha\alpha} \ket{n_{1+} n_{1-}, n_{2+} n_{2-}} &= \frac{\sqrt{\beta_{\alpha}}}{2} \frac{e^2}{\epsilon l_B} \left(\prod_{i\lambda} n_{i\lambda}'! n_{i\lambda}!\right)^{1/2} (-1)^{n_{1+}' + n_{1-} + n_{2+}' + n_{2-}} \delta_{l'l} \\
&\times \sum_{k_{i\lambda}=0}^{\min(n_{i\lambda}',n_{i\lambda})} \left(\prod_{i\lambda} \frac{(-1)^{k_{i\lambda}} 2^{-p} }{k_{i\lambda}! (n_{i\lambda}' - k_{i\lambda})! (n_{i\lambda} - k_{i\lambda})!} \right) \cdot \Gamma(p+\frac 12),
\end{split}
\end{equation}
where $\alpha=e,h$, the sums define separate factors for $\lambda=\pm$ and particle $i=1,2$,
the conserved angular momentum quantum numbers $l = n_{1+} - n_{1-} + n_{2+} - n_{2-}$ and $l' = n_{1+}' - n_{1-}' + n_{2+}' - n_{2-}'$, and $p = \sum_{i\lambda} (n_{i\lambda}' + n_{i\lambda} - 2k_{i\lambda})/2$.  In Eq.~\eqref{ee2particle}, the magnetic length $l_B = \sqrt{\hbar c/eB}$ and we have defined $\beta_{\alpha} = \sqrt{1+(2\omega_{\alpha}/\omega_{\alpha c})^2}$.

The two-particle matrix elements of $V_{eh}$ are similar but do not conserve
angular momentum because the electron and hole oscillators have different centers in general:
\begin{equation}
\begin{split}
\bra{n_{e+}' n_{e-}', n_{h+}' n_{h-}'} V_{eh} \ket{n_{e+} n_{e-}, n_{h+} n_{h-}} &= -\sqrt{\frac{\beta_{eh}}{2}} \frac{e^2}{\epsilon l_B} \left(\prod_{\alpha\lambda} n_{\alpha\lambda}'! n_{\alpha\lambda}!\right)^{1/2} (-1)^{n_{e+}' + n_{e-} + n_{h+}' + n_{h-}}  \\
&\times \sum_{k_{\alpha\lambda}=0}^{\min(n_{\alpha\lambda}',n_{\alpha\lambda})} \left(\prod_{\alpha\lambda} \frac{(-1)^{k_{\alpha\lambda}} (\frac{\beta_{eh}}{\beta_{\alpha}})^{\frac 12 (n_{\alpha\lambda}' + n_{\alpha\lambda}) - k_{\alpha\lambda}}}{k_{\alpha\lambda}! (n_{\alpha\lambda}' - k_{\alpha\lambda})! (n_{\alpha\lambda} - k_{\alpha\lambda})!} \right) \cdot I,
\end{split}
\end{equation}
where $\beta_{eh} = \beta_e \beta_h / (\beta_e+\beta_h)$, and the $(n',n,k)$-dependent integral
\begin{equation}
\label{eq:I}
I = \int_{0}^{\infty} d\tilde q^2 \, e^{-\tilde q^2} (\tilde q^2)^{p - \frac 12} e^{-\tilde q \tilde d} \cdot \int_{0}^{2\pi} \frac{d\phi}{2\pi} \, (ie^{i\phi})^{l-l'} e^{i\tilde q \tilde s \cos\phi},
\end{equation}
where we introduced the dimensionless lengths $\tilde d = \frac{\sqrt{2\beta_{eh}}}{l_B} d$, $\tilde s = \frac{\sqrt{2\beta_{eh}}}{l_B} s$, the total angular momentum quantum numbers $l = n_{e+} - n_{e-} - n_{h+} + n_{h-}$, $l' = n_{e+}' - n_{e-}' - n_{h+}' + n_{h-}'$, and $p = \sum_{\alpha\lambda} (n_{\alpha\lambda}' + n_{\alpha\lambda} - 2k_{\alpha\lambda})/2$. The radial part of the integral in Eq.~\eqref{eq:I} can be computed analytically and we obtain
\begin{equation}
I = 4^{-p}\, \Gamma(2p+1) \int_{0}^{2\pi} \frac{d\phi}{2\pi}\, (ie^{i\phi})^{l-l'}\, U(p+\frac 12, \frac 12, \frac 14 (\tilde d - i\tilde s \cos\phi)^2),
\end{equation}
\end{widetext}
where $U$ is the confluent hypergeometric function of the second kind. 
The angular integral must be evaluated numerically for $s \ne 0$.  For $s=0$,
\begin{equation}
\label{eq:Is=0}
I(s=0) = \delta_{l'l} \cdot 4^{-p}\, \Gamma(2p+1)\, U(p+\frac 12, \frac 12, \frac{\tilde d^2}{4}).
\end{equation}
When $d \to 0$, $I(s=0) \to \delta_{l'l} \cdot \Gamma (p + \frac 12)$. 
More detail on the derivations of these formulas is provided in Appendix~\ref{app:ed_eqs}. 
The generalization of these formulas to properly antisymmetrized multi-particle states basis with spin and/or valley 
degrees of freedom is straightforward.

\section{Excitons} \label{sec:exciton}

We first consider the case of excitons -- two-particle states with an electron and a hole that are distinguishable.
We neglect intervalley exchange interactions \cite{yu2014dirac,glazov2014exciton,yu2014valley,wu2015exciton} 
so that the excitonic states are valley independent. The $i$th eigenstate of the Hamiltonian can be written as
\begin{equation} \label{eq:psi_eh}
\ket{\Psi_{eh}^{(i)}} = \sum_{n_e n_h} C_{n_e n_h}^{(i)} \ket{n_e;n_h},
\end{equation}
where $n_e$ is a shorthand for the two indices $(n_{e+},n_{e-})$ that specify which electron basis state is occupied, and similarly for $n_h$. The coefficients $C_{n_e n_h}^{(i)}$ are obtained by numerical diagonalization of the Hamiltonian in the magnetic harmonic oscillator basis $\ket{n_e;n_h}$.  In our model, exciton energies are given relative to the band gap between potential extrema. 

In the weak-confinement limit, the relative-motion electron-hole pair state is very weakly distorted,
and the electron-hole center-of-mass moves in a weak modulation potential. 
The characteristic length scale for bound electron-hole pairs is the effective Bohr 
radius $a_B^*=\epsilon\hbar^2/me^2$ where $m = m_e m_h/(m_e+m_h)$ is the reduced mass of the electron-hole pair. 
The characteristic energy scale is the effective Rydberg ${\rm Ry}^* = e^2/2\epsilon a_B^* = me^4/2\epsilon^2 \hbar^2$. 
In our calculations we set $m_e=0.4m_{e0}$, $m_h=0.5m_{e0}$ where $m_{e0}$ is the free electron mass, and $\epsilon=5$, as approximate parameters for intralayer excitons in bilayer TMDs \cite{kormanyos2015k} surrounded by hBN. Then we find $a_B^*=\SI{1.19}{nm}$ and ${\rm Ry^*}=\SI{121}{meV}$.

In the strong-confinement and large-$s$ limit, the electron and hole are each confined in a harmonic potential, with length scale given by the oscillator length $l_{\alpha} = \sqrt{\hbar/m_{\alpha} \omega_{\alpha}}$ and energy scale $\hbar\omega_{\alpha}$. 
The magnetic field provides another length scale, the magnetic length $l_B=\sqrt{\hbar c/eB}$, and another energy scale, the cyclotron energy $\hbar\omega_{\alpha c}=\hbar eB/m_{\alpha}c$.
For our explicit calculations we choose $\hbar\omega_e = \SI{50}{meV}$, $\hbar\omega_h = \SI{60}{meV}$.  Then the harmonic confinement lengths $l_e=\SI{1.95}{nm}$ and $l_h=\SI{1.59}{nm}$.
At magnetic field  $B=\SI{10}{T}$, the magnetic length $l_B=\SI{8.11}{nm}$ is considerably longer and the cyclotron 
energies $\hbar\omega_{ec} = \SI{2.9}{meV}$, $\hbar\omega_{hc} = \SI{2.3}{meV}$ considerably smaller 
than other length and energy scales.  In a magnetic field, the oscillator length $L_{\alpha} = \sqrt{\hbar/m_{\alpha} \Omega_{\alpha}}$ with $\Omega_{\alpha} = \sqrt{\omega_{\alpha}^2 + (\omega_{\alpha c}/2)^2}$.  We therefore see that 
due to the large effective masses in the TMDs, the oscillator length is weakly perturbed by a
$B=\SI{10}{T}$ magnetic field ($L_{\alpha}\approx l_{\alpha} \ll l_B$).  The magnetic length and the 
oscillator length start to become comparable only for magnetic fields $\sim\SI{100}{T}$; below this field 
scale exciton states are weakly perturbed by magnetic fields. In the following we present results at zero magnetic field unless otherwise specified; the results differ very little when a moderate magnetic field is applied.

\begin{figure}
    \centering
    \includegraphics[width=\linewidth]{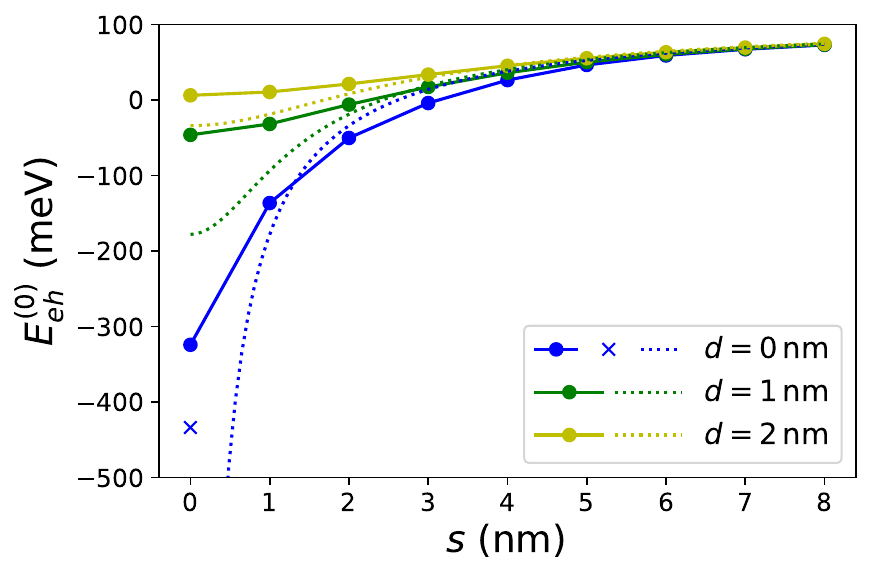}
    \caption{Ground-state exciton energy $E_{eh}^{(0)}$ as a function of the in-plane separation between the electron and hole confining potential minima $s$, for intralayer (blue $d=0$) and interlayer (green and yellow $d=1,\SI{2}{nm})$ excitons. The dots and solid lines are numerical results from ED calculations, and the dotted curves are the large-$s$ approximation of \eqref{eq:E_x_far}. The blue cross shows the weak-confinement approximation \eqref{eq:E_x_weak} expression which adds a zero-point confinement energy to the free exciton energy.}
    \label{fig:E_s_x}
\end{figure}

Fig.~\ref{fig:E_s_x} shows numerical results for the ground-state exciton energy as a function of the lateral electron-hole separation $s$. When $s\gg L_e,L_h$, the electron and hole are confined in widely separated harmonic potentials, electron-hole correlations are negligible, 
and interactions can be treated at lowest order in 
perturbation theory.  In this limit, the ground state energy is
\begin{equation} \label{eq:E_x_far}
E_{eh}^{(0)}(s) = E_e^{(0)} + E_h^{(0)} - \frac{e^2}{\epsilon \sqrt{s^2+d^2}},
\end{equation}
where $E_e^{(0)} = \hbar(\Omega_{e+}+\Omega_{e-})/2 = \hbar\Omega_{e}$ is the ground state of an electron in the quantum dot (and similarly for holes) with $\Omega_{\pm}$ and $\Omega$ defined after Eq.~\eqref{eq:sp_energy}. The dotted curves in Fig.~\ref{fig:E_s_x} show that the approximate ground state energy calculated from Eq.~\eqref{eq:E_x_far}, which entirely neglects electron-hole correlations,
agrees well with the numerical results when $s\gg L_e,L_h$. For intermediate $s\sim \SI{3}{nm}$, Coulomb attraction distorts the wavefunctions of the electron and hole, which effectively reduces the electron-hole separation and results in a lower ground-state energy than that given by the large-$s$ approximation \eqref{eq:E_x_far}. For small $s$, the electron and hole wavefunctions overlap in the 2D plane and the approximation of Eq.~\eqref{eq:E_x_far}, which becomes singular in this limit when $d=0$, breaks down. 
The singular behavior at small $s$ is cut off by the kinetic-energy pressure required by the Heisenberg uncertainly relation, 
and this results in a ground state energy that is higher than that given by Eq.~\eqref{eq:E_x_far}.
The approximate ground state energy of Eq.~\eqref{eq:E_x_far} is in reasonable agreement with numerical 
results when $s\gtrsim L_e,L_h$.

We see from Fig.~\ref{fig:E_s_x} that exciton properties depend qualitatively on the 
ratio of $s$ to confinement oscillator lengths.  As indicated in Fig.~\ref{fig:eh_lattice}, the experimentally relevant value of 
$s$ is either $s=0$, in which case electron-hole correlation is always important, or $s=a_M/\sqrt{3}$, where 
$a_M$ is the lattice constant of the triangular moir\'e Bravais lattice.
In the former case (see below) the large-$s$ limit is never relevant.  In the latter case, the large-$s$ limit is always achieved at 
large enough moir\'e period since the oscillator length of electrons or holes in moir\'e materials varies \cite{wu2018hubbard,angeli2021gamma} 
approximately as the square root of the moir\'e period.  The large-$s$ limit is simple not only for excitons, but for any finite-particle-number 
electron-hole complex, since the interactions between electrons and holes $V_{eh}$ (Eq.~\eqref{eq:V_eh}) simply add an electrostatic contribution to 
energies and do not alter electronic wavefunctions.  

If confinement is weak, the $s=0$ system can be viewed as a tightly bound exciton of mass $M=m_e+m_h$ 
moving in a weak confining potential with angular frequency 
$\omega = \sqrt{(m_e \omega_e^2 + m_h \omega_h^2)/(m_e+m_h)}$ \cite{halonen1992excitons}.
The ground state energy is then the sum of the ground state energies of the harmonic oscillator 
and the tightly bound exciton. For the intralayer exciton case $d=0$, the system is similar to a 2D hydrogen atom, and the total energy is given by the analytic result:
\begin{equation} \label{eq:E_x_weak}
E_{eh}^{(0)}(0) = \hbar\omega - 4\,{\rm Ry^*}.
\end{equation}
This approximate result is indicated by the blue cross in Fig.~\ref{fig:E_s_x},
where we see that the parameters chosen for this 
calculation are not in the weak confinement limit -- the exciton radius $\sim a_B^*$ is not small compared with the confining 
length of the harmonic potential $L = \sqrt{\hbar/M\omega}$ -- and Eq.~\eqref{eq:E_x_weak} is therefore inaccurate.
The stronger confinement weakens the correlations between electron and hole positions and 
therefore increases the total energy.

\begin{figure}
    \centering
    \includegraphics[width=\linewidth]{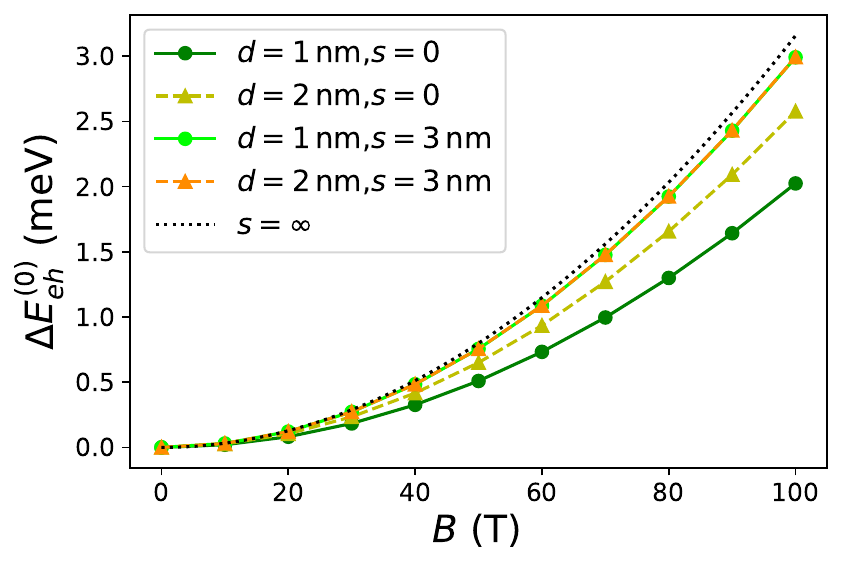}
    \caption{Diamagnetic exciton energy shift $\Delta E_{eh}^{(0)}(B) = E_{eh}^{(0)}(B) - E_{eh}^{(0)}(B=0)$ 
    of the excitonic ground state in a magnetic field $B$, for interlayer excitons at $d=1,\SI{2}{nm}$ and $s=0,\SI{3}{nm}$. The circular and triangular dots connected by solid or dashed lines are numerical results from ED calculations, and the dotted curve shows the energy increase obtained from the $s\to\infty$ limit \eqref{eq:E_x_far} expression in which only the zero-point confinement energy is shifted.}
    \label{fig:E_B_x}
\end{figure}

Fig.~\ref{fig:E_B_x} shows how the exciton ground state energy shifts in a magnetic field. In the large-$s$ limit it follows from Eq.~\eqref{eq:E_x_far} that the diamagnetic exciton energy shift comes from the single-particle energies of electrons and holes:
\begin{equation}
\begin{split}
&\Delta E_{eh}^{(0)}(B) \equiv E_{eh}^{(0)}(B) - E_{eh}^{(0)}(B=0) \\
&\xrightarrow{s\to\infty} \hbar(\Omega_{e} + \Omega_{h} - \omega_e - \omega_h) \\
&\approx \frac{e^2 B^2}{8c^2} \left(\frac{l_e^2}{m_e} + \frac{l_h^2}{m_h}\right).
\end{split}
\end{equation}
The last line of the above equation applies for a small magnetic field $B\ll 2m_{\alpha}\omega_{\alpha}c/e$. Fig.~\ref{fig:E_B_x} shows that for $s=\SI{3}{nm}$ the numerical results are already close to the $s\to\infty$ limit. The diamagnetic exciton energy shift grows quadratically with magnetic field both in the strong-binding case ($s=0$) and in the strong-confinement large-$s$ limit, as expected for the ground state with zero angular momentum. By first-order perturbation theory we see from Eqs.~\eqref{eq:H_e} and \eqref{eq:H_h} that the diamagnetic exciton energy shift for a zero-angular-mementum state is
\begin{equation}
\Delta E_{eh}^{(0)}(B) \approx \frac{e^2 B^2}{8c^2} \left(\frac{\langle \bm r_e^2 \rangle}{m_e} + \frac{\langle \bm r_h^2 \rangle}{m_h}\right),
\end{equation}
where the angle brackets denote ground state expectation values. 
When $s=0$ the electron and hole orbitals shrink due to strong Coulomb attraction, resulting in a weaker diamagnetic response. 
The diamagnetic response of excitons can in principle be used as an experimental probe of the binding strength of the electrons and 
holes and therefore distinguish the case in which they localized at the same site from the case in which they are 
localized at different sites.
In practice, however, this requires a very strong magnetic field because of the strong binding and large effective masses of electrons and holes in TMDs.

Next we study the optical absorption of the intralayer exciton states ($d=0$). In our model the current operator is given 
up to a constant factor by 
\begin{equation} \label{eq:current_op}
\begin{split}
\hat j \sim \sum_{\tau} \int d\bm r\, &[e_{\tau}^{\dagger}(\bm r) h_{\tau}^{\dagger}(\bm r) \exp(i\frac{eBs}{2\hbar c}y)  \\
&+ h_{\tau}(\bm r) e_{\tau}(\bm r) \exp(-i\frac{eBs}{2\hbar c}y)],
\end{split}
\end{equation}
where $e^{\dagger}$ and $h^{\dagger}$ are electron and hole creation operators, 
$\tau=\pm$ is the valley index, and the phase factors $\exp(\pm i\frac{eBs}{2\hbar c}y)$ come from our gauge choice (see Appendix~\ref{app:cur_op} for details). The matrix element for a transition from the vacuum state $\ket{\Psi_0}$ with no electrons or holes to an exciton state $\ket{\Psi_{eh}^{(i)}}$ (Eq.~\eqref{eq:psi_eh}) is proportional to
\begin{equation}
\bra{\Psi_0} \hat{j} \ket{\Psi_{eh}^{(i)}} \sim \sum_{n_e n_h} C_{n_e n_h}^{(i)} \int d\bm r\, \psi_{n_e}^{(e)}(\bm r) \psi_{n_h}^{(h)}(\bm r) e^{-i\frac{eBs}{2\hbar c}y},
\end{equation}
where $\psi_{n_e}^{(e)}(\bm r)$ is the wavefunction of an electron in state $n_e$, and similarly $\psi_{n_h}^{(h)}(\bm r)$ for holes. Then the absorption rate of state $i$ is proportional to the dimensionless quantity
\begin{equation} \label{eq:absorp_x}
A_{eh}^{(i)} = \Big| \sum_{n_e n_h} C_{n_e n_h}^{(i)} \int d\bm r\, \psi_{n_e}^{(e)}(\bm r) \psi_{n_h}^{(h)}(\bm r) e^{-i\frac{eBs}{2\hbar c}y} \Big|^2,
\end{equation}
which is, roughly speaking, the overlap of the electron and hole wavefunctions. The optical absorption strength of the ground state is calculated by numerical integration of the wavefunctions and shown in Fig.~\ref{fig:A_s_x}. The optical spectrum is a set of $\delta$-function peaks located at the energy eigenvalues of exciton states and with heights proportional to the corresponding optical absorption rates:
\begin{equation} \label{eq:spectrum_x}
A_{eh}(\omega) = \sum_i A_{eh}^{(i)} \,\delta(\hbar\omega-E_{eh}^{(i)}-E_g),
\end{equation}
where $E_g$ is the single-particle gap between the conduction and valence bands.

\begin{figure}
    \centering
    \includegraphics[width=\linewidth]{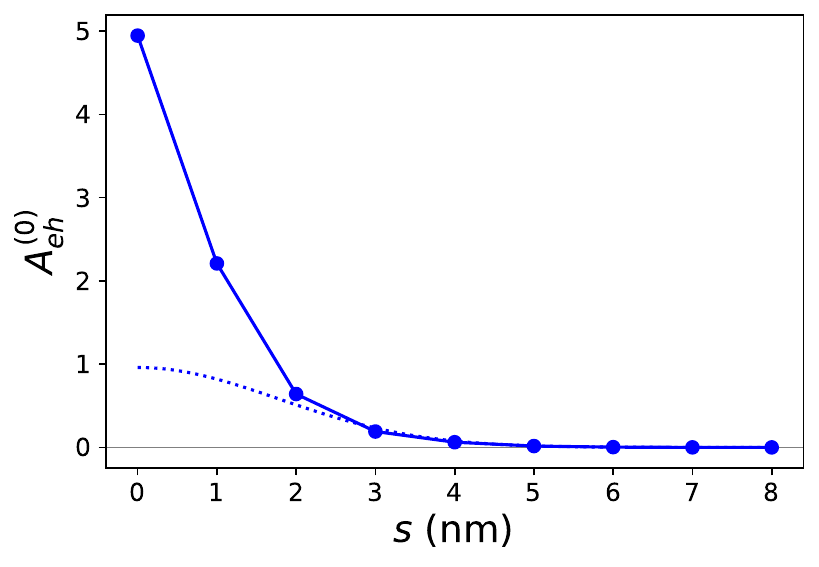}
    \caption{Optical absorption strength $A_{eh}^{(0)}$ of intralayer excitons as a function of the separation between the electron and hole confining potential minima $s$. The blue dots are the numerical results from ED calculations and the dotted curve is from the large-$s$ approximation \eqref{eq:A_x_far}.  Since the likely $s$ is either $0$ or $a_M/\sqrt{3}$, based on symmetry considerations, and 
    typical moir\'e periods are longer than \SI{5}{nm}, a substantial reduction in oscillator strength is expected in the $s \ne 0$ case.} \label{fig:A_s_x}
    \label{fig:A_s_x}
\end{figure}

We can obtain analytic expressions for the optical matrix elements in certain limiting cases. 
When $s\gg L_e,L_h$ and confinement is strong the electron and hole are in the ground states of their confining potentials 
and their wavefunctions are
\begin{equation}
\psi_0^{(\alpha)}(\bm r) = \frac{1}{\sqrt{\pi}} \frac{1}{L_{\alpha}} e^{-r^2/2L_{\alpha}^2},
\end{equation}
where $\alpha=e,h$ and the origin $\bm r=0$ is chosen at the bottom of the confining potential. The two-body wavefunction is approximately the product of $\psi_0^{(e)}$ and $\psi_0^{(h)}$. Then we can do the integral \eqref{eq:absorp_x} and find
\begin{equation} \label{eq:A_x_far}
A_{eh}^{(0)}(s) \approx \left(\frac{2L_e L_h}{L_e^2+L_h^2}\right)^2 \exp \left[-\frac{s^2}{L_e^2+L_h^2} \left(1+\frac{L_e^2 L_h^2}{4l_B^4}\right)\right].
\end{equation}
The dotted curve in Fig.~\ref{fig:A_s_x} shows that the above approximation is in good agreement with the numerical results at large $s$.

For $s=0$ and the weak-confinement limit, the wavefunction of the two-particle system can be 
separated into center-of-mass and relative motions
\begin{equation} \label{eq:wf_x_weak}
\Psi_{eh}^{(0)}(\bm r_e,\bm r_h) \approx \psi_{\rm cm}(\bm R) \psi_{\rm rel}(\bm r),
\end{equation}
where $\bm R = (m_e\bm r_e + m_h\bm r_h)/M$ and $\bm r = \bm r_e - \bm r_h$ are the center-of-mass and relative coordinates. The center-of-mass motion is the ground state of the harmonic oscillator:
\begin{equation}
\psi_{\rm cm}(\bm R) = \frac{1}{\sqrt{\pi}} \frac{1}{L} e^{-R^2/2L^2},
\end{equation}
where the confining length $L = \sqrt{\hbar/M\omega} = \SI{1.23}{nm}$. The relative motion is the ground state of the 2D hydrogen atom
\begin{equation} \label{eq:wf_2d_hydrogen}
\psi_{\rm rel}(\bm r) = \frac{1}{\sqrt{2\pi}} \frac{4}{a_B^*} e^{-2r/a_B^*}.
\end{equation}
With the wavefunction \eqref{eq:wf_x_weak} and the current operator \eqref{eq:current_op} we find
\begin{equation}
\bra{\Psi_0} \hat{j} \ket{\Psi_{eh}^{(0)}} = \int d\bm R\, \Psi_{eh}^{(0)}(\bm R,\bm R) \approx 4\sqrt{2} L/a_B^*,
\end{equation}
and therefore
\begin{equation} \label{eq:A_x_weak}
\begin{split}
A_{eh}^{(0)}(0) = |\bra{\Psi_0} \hat{j} \ket{\Psi_{eh}^{(0)}}|^2 \approx \frac{32L^2}{a_B^{*2}} \approx 34.
\end{split}
\end{equation}
This estimate is much larger than the result from ED calculation, as shown in Fig.~\ref{fig:A_s_x}. 
The difference can be understood as follows. 
Confinement adds a harmonic repuslive contribution to the relative motion which 
mixes the ground state of the 2D hydrogen atom \eqref{eq:wf_2d_hydrogen} with higher energy states. 
Since the wavefunctions of higher-energy hydrogenic states have smaller values at $\bm r=0$, the resulting optical absorption 
strength is reduced \cite{que1992excitons}. 
For the parameters used in the calculations illustrated in Fig.~\ref{fig:A_s_x},
the confining potential is far from weak, a lot of high energy states come into play, and the discrepancy 
between the exact numerical results and the approximation of \eqref{eq:A_x_weak} becomes large.

\section{Trions} \label{sec:trion}

Next we study a system of one electron and two holes -- a moir\'e-confined trion.
The three-particle state must be antisymmetric with respect to exchange of the two holes. 
We work in a representation in which the orbital and valley parts of the two-hole states separate,
and each is symmetric or antisymmetric with respect to the permutation. 
The orbital parts of the basis states are ($n_{h1}>n_{h2}$)
\begin{align}
&\ket{n_e; n_{h1} n_{h2}}_s = \frac{1}{\sqrt{2}} \ket{n_e}_e (\ket{n_{h1} n_{h2}}_h + \ket{n_{h2} n_{h1}}_h), \\
&\ket{n_e; n_{h1} n_{h2}}_a = \frac{1}{\sqrt{2}} \ket{n_e}_e (\ket{n_{h1} n_{h2}}_h - \ket{n_{h2} n_{h1}}_h), \\
&\ket{n_e; n_h n_h}_s = \ket{n_e}_e \ket{n_h n_h}_h,
\end{align}
where each $n$ is shorthand for the two indices $(n_+,n_-)$ that distinguish
two-dimensional oscillator single-particle states, 
the subscripts $s$ and $a$ stand for symmetric and antisymmetric states, and the kets with subscripts $e$ and $h$ are direct product states of electrons and holes without (anti)symmetrization. The valley parts of the basis states are
\begin{align}
&\ket{\pm;++}_s = \ket{\pm}_e \ket{++}_h, \\
&\ket{\pm;--}_s = \ket{\pm}_e \ket{--}_h, \\
&\ket{\pm;+-}_s = \frac{1}{\sqrt{2}} \ket{\pm}_e (\ket{+-}_h + \ket{-+}_h), \\
&\ket{\pm;+-}_a = \frac{1}{\sqrt{2}} \ket{\pm}_e (\ket{+-}_h - \ket{-+}_h),
\end{align}
where $+$ and $-$ stand for valley $+K$ and $-K$. 

We are interested in the optical excitations of a state with a single hole to a trion one-electron-two-hole state. 
The initial hole is in the ground state $n_h=0$ (by which we mean $n_{h+}=n_{h-}=0$), and we assume without loss of generality that it is in the $-K$ valley:
\begin{equation}
\ket{\Psi_h^{(0)}} = \ket{0}_h \ket{-}_h.
\end{equation}
In order for the optical matrix elements to be non-zero, the newly added electron and hole must be in the same valley.
There are three possibilities: $\ket{+;+-}_a$, $\ket{+;+-}_s$ and $\ket{-;--}_s$. Correspondingly the three-particle state has three possible forms:
\begin{widetext}
\begin{align}
\ket{\Psi_{ehh}^{(1,i)}} &= \sum_{n_e} \sum_{n_{h1} \ge n_{h2}} C_{n_e n_{h1} n_{h2}}^{(s,i)} \ket{n_e; n_{h1} n_{h2}}_s \ket{+;+-}_a, \label{eq:psi_ehh_1}\\
\ket{\Psi_{ehh}^{(2,i)}} &= \sum_{n_e} \sum_{n_{h1}>n_{h2}} C_{n_e n_{h1} n_{h2}}^{(a,i)} \ket{n_e; n_{h1} n_{h2}}_a \ket{+;+-}_s,  \label{eq:psi_ehh_2}\\
\ket{\Psi_{ehh}^{(3,i)}} &= \sum_{n_e} \sum_{n_{h1}>n_{h2}} C_{n_e n_{h1} n_{h2}}^{(a,i)} \ket{n_e; n_{h1} n_{h2}}_a \ket{-;--}_s.  \label{eq:psi_ehh_3}
\end{align}
\end{widetext}
The energy of the system depends only on the orbital part of the state, so the second and third forms of the three-particle state \eqref{eq:psi_ehh_2} and \eqref{eq:psi_ehh_3} with antisymmetric orbital parts are exactly degenerate in energy. Exchange statistics 
suggests that the ground state takes the first form \eqref{eq:psi_ehh_1}, since there is then no exclusion principle that forbids the two holes from occupying the same low-energy single-particle orbital ($n_{h1}=n_{h2}=0$). This expectation is confirmed by our numerical results.

Using the current operator \eqref{eq:current_op}, we find the optical absorption strength for intralayer trions
\begin{widetext}
\begin{align}
A_{ehh}^{(1,i)} &= \Big|\sum_{n_e} \int d\bm r \psi_{n_e}^{(e)}(\bm r) \left[\frac{1}{\sqrt{2}} \sum_{n_h>0} C_{n_e n_h 0}^{(s,i)} \psi_{n_h}^{(h)}(\bm r) + C_{n_e00}^{(s,i)} \psi_0^{(h)}(\bm r)\right] \exp(-i\frac{eBs}{2\hbar}y) \Big|^2 \equiv A_{ehh}^{(s,i)}, \label{eq:A_ehh_1}\\
A_{ehh}^{(2,i)} &= \frac{1}{2} \,\Big|\sum_{n_e} \sum_{n_h>0} C_{n_e n_h 0}^{(a,i)} \int d\bm r \psi_{n_e}^{(e)}(\bm r) \psi_{n_h}^{(h)}(\bm r) \exp(-i\frac{eBs}{2\hbar}y)\Big|^2 \equiv A_{ehh}^{(a,i)}, \\
A_{ehh}^{(3,i)} &= \Big|\sum_{n_e} \sum_{n_h>0} C_{n_e n_h 0}^{(a,i)} \int d\bm r \psi_{n_e}^{(e)}(\bm r) \psi_{n_h}^{(h)}(\bm r) \exp(-i\frac{eBs}{2\hbar}y) \Big|^2 = 2A_{ehh}^{(a,i)}.
\end{align}
\end{widetext}

When the polarization of the incident light is random (or equivalently, when the valley polarization of the initial hole is random), 
there is equal probability that an electron-hole pair is created in the same or opposite valley as the initial hole. 
Therefore the optical absorption spectrum is
\begin{equation} \label{eq:spectrum_t}
\begin{split}
A_{ehh}(\omega) = &\frac 12 \sum_{p=1}^3 \sum_i A_{ehh}^{(p,i)} \,\delta(\hbar\omega-E_{ehh}^{(p,i)}-E_g) \\
= &\frac 12 \sum_i A_{ehh}^{(s,i)} \,\delta(\hbar\omega-E_{eh}^{(s,i)}-E_g) \\
&+ \frac 32 \sum_i A_{ehh}^{(a,i)} \,\delta(\hbar\omega-E_{ehh}^{(a,i)}-E_g).
\end{split}
\end{equation}

\begin{figure}
    \centering
    \includegraphics[width=\linewidth]{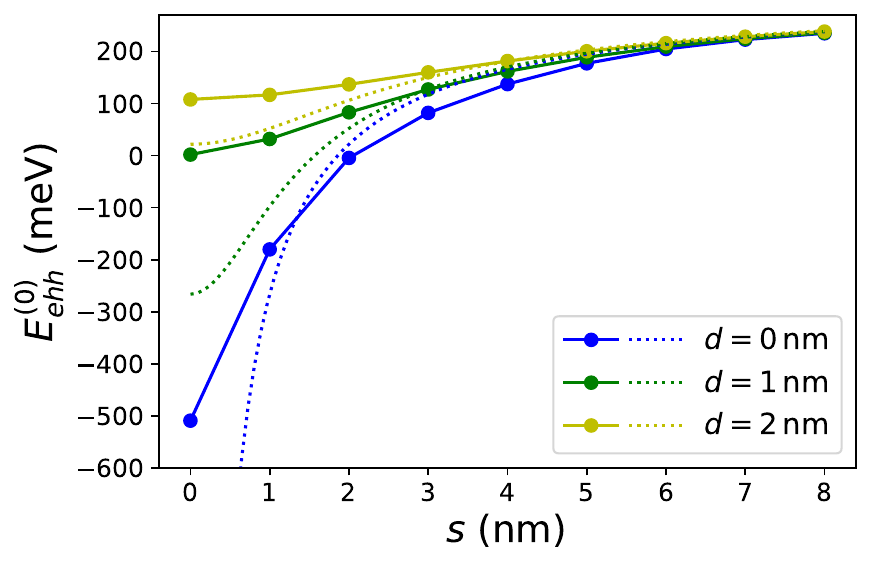}
    \caption{Ground-state trion energy $E_{ehh}^{(0)}$ as a function of the separation between the electron and hole confining potential minima $s$, for intralayer ($d=0$) and interlayer ($d=1,\SI{2}{nm}$) trions represented by different colors. The dots and solid lines are numerical results from ED calculations, and the dotted curves show results from the large-$s$ approximation \eqref{eq:E_s_t_far}.}
    \label{fig:E_s_t}
\end{figure}

Fig.~\ref{fig:E_s_t} shows the ground-state trion energy $E_{ehh}^{(0)}$ as a function of the electron-hole separation $s$. Other parameters of the calculations are chosen to be the same as for the calculations of excitons in Sec.~\ref{sec:exciton}. The dotted curve shows the large-$s$ approximation in which the motion of the electron (holes) is not affected by the presence of the holes (electron) in the other potential well, so the ground state energy is
\begin{equation} \label{eq:E_s_t_far}
E_{ehh}^{(0)}(s) \approx E_e^{(0)} + E_{hh}^{(0)} - \frac{2e^2}{\epsilon \sqrt{s^2+d^2}},
\end{equation}
where $E_{hh}^{(0)}$ is the ground state energy of two holes in the same quantum dot. 
A similar ED calculation with only holes present yields $E_{hh}^{(0)} = \SI{260}{meV}$
for the parameters of Fig.~\ref{fig:E_s_t}.  As shown in Fig.~\ref{fig:E_s_t}, the approximation \eqref{eq:E_s_t_far} is in good agreement with the numerical results for $s\gtrsim L_e,L_h$.

\begin{figure}
    \centering
    \includegraphics[width=\linewidth]{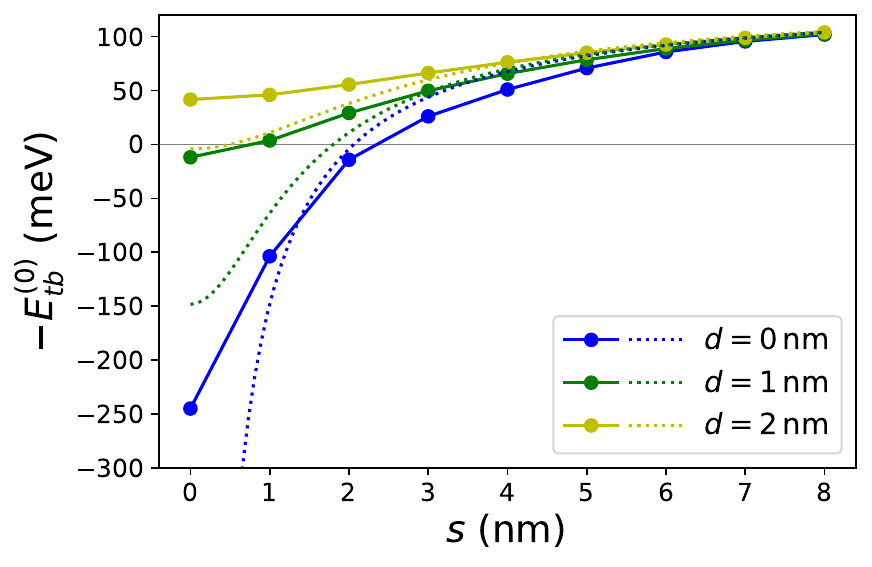}
    \caption{Ground state trion binding energy $E_{tb}$ as a function of the separation between the electron and hole confining potential minima $s$, for intralayer ($d=0$) and interlayer ($d=1,\SI{2}{nm}$) trions represented by different colors. The dots and solid lines are numerical results from ED calculations, 
    and the dotted curves show results from the large-$s$ approximation.}
    \label{fig:Etb_s}
\end{figure}

Fig.~\ref{fig:Etb_s} shows the trion binding energy, defined as 
\begin{equation}
E_{tb} = E_{eh}^{(0)} + E_h^{(0)} - E_{ehh}^{(0)},
\end{equation}
as a function of $s$. At small $s$, an extra hole binds with the electron-hole pair in such a way as to lower the energy. At large $s$, the repulsion between the two holes dominates and the energy increases when an extra hole is added to the system. 
The crossover from a bound trion to an unbound trion takes place between 2 to \SI{3}{nm} for an intralayer trion, and around \SI{1}{nm} for an interlayer trion at interlayer distance $d=\SI{1}{nm}$. An interlayer trion at $d=\SI{2}{nm}$ is always unbound even for $s=0$. 

The calculated trion binding energy at $d=s=0$ is much larger than the trion binding energy measured in monolayer TMDs \cite{mak2013tightly,ross2013electrical,jones2013optical}, indicating that strong moir\'e modulation significantly increases the trion binding strength.  This conclusion is in agreement with previous studies of trions in semiconductor quantum dots \cite{anisimovas2003excitonic,xie2000excitonic,bracker2005binding}. While we do
expect a clear increase of trion binding energy with confinement in experiment, our harmonic 
approximation overestimates the trion binding energy because anharmonic 
corrections become important outside a small region near the potential minimum. 
Anharmonic confinement is more important for trions than for excitons because of their larger size. 
The effective modulation strength in real moir\'e materials is weaker, resulting in a smaller trion binding energy.

\begin{figure}
    \centering
    \includegraphics[width=\linewidth]{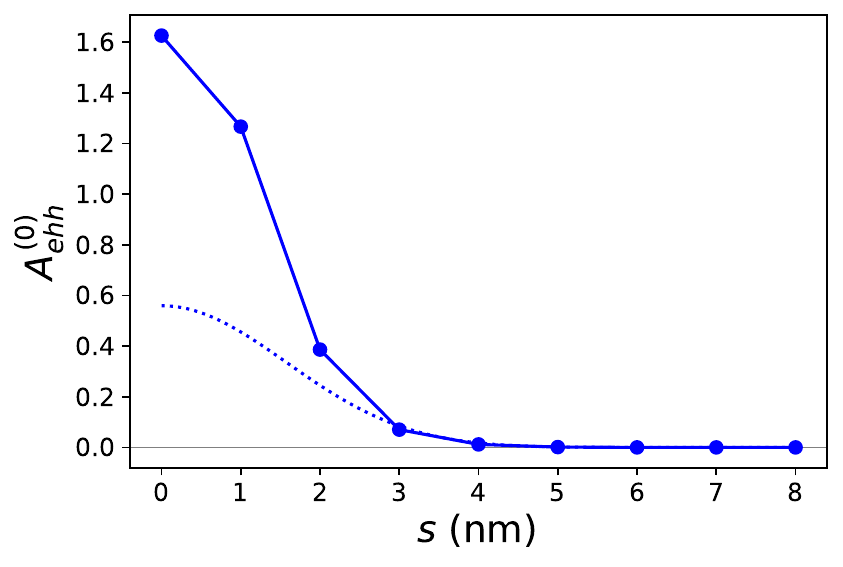}
    \caption{Optical absorption strength $A_{ehh}^{(0)}$ of intralayer trions as a function of the separation between the electron and hole confining potential minima $s$. The blue dots are numerical results from ED calculations, and the dotted curve is the large-$s$ approximation \eqref{eq:absorp_t}.}
    \label{fig:A_s_t}
\end{figure}

Fig.~\ref{fig:A_s_t} shows the optical absorption strength $A_{ehh}^{(0)}$ as a function of $s$. Comparison with Fig.~\ref{fig:A_s_x} 
shows that the optical absorption of an intralayer trion is always weaker than that of an intralayer exciton. This can be qualitatively understood by considering the $m_e\to\infty$ limit at $s=0$ for which the electron stays at the center of its quantum dot while the two holes move around it. 
The Coulomb repulsion between the two holes pushes them away from the electron into higher-energy states, so the overlap between the electron and hole wavefunctions, and therefore the optical absorption strength, decreases. For large $s\gg L_e,L_h$, the optical absorption strength
is small and can be calculated from Eq.~\eqref{eq:A_ehh_1} with
\begin{equation}
C^{(s,0)}_{n_e n_{h1} n_{h2}} = \delta_{n_e,0} C^{(s,0)}_{n_{h1} n_{h2}},
\end{equation}
where $C^{(s,0)}_{n_{h1} n_{h2}}$ (with $n_{h1}>n_{h2}$) are the expansion coefficients of the two-hole ground state in the symmetric two-hole basis. We notice from the ED calculation that the two-hole ground state comes predominantly from the single-particle ground state $\ket{00,00}_s$ and the lowest two excited states with zero total angular momentum $\ket{11,00}_s$ and $\ket{10,01}_s$:
\begin{equation}
\ket{\Psi_{hh}^{(0)}} \approx C_0 \ket{00,00}_s + C_1 \ket{11,00}_s + C_2 \ket{10,01}_s,
\end{equation}
with $C_0\approx 0.82$ and $C_1 = C_2 \approx 0.40$. Then from Eq.~\eqref{eq:A_ehh_1} we find an analytic expression for the optical absorption strength in the large-$s$ limit:
\begin{widetext}
\begin{equation} \label{eq:absorp_t}
A_{ehh}^{(0)}(s) \approx \Bigg| C_0 - \frac{C_1}{\sqrt{2}} \left[ \frac{L_e^2-L_h^2}{L_e^2+L_h^2} + \left(1-\frac{L_e^4}{4l_B^4}\right) \frac{L_h^2 s^2}{L_e^2+L_h^2} \right] \Bigg|^2
\left(\frac{2L_e L_h}{L_e^2+L_h^2}\right)^2 \exp \left[-\frac{s^2}{L_e^2+L_h^2} \left(1+\frac{L_e^2 L_h^2}{4l_B^4}\right)\right].
\end{equation}
\end{widetext}
This approximation is in good agreement with the numerical results at $s\gtrsim\SI{3}{nm}$ as shown by the dotted curve in Fig.~\ref{fig:A_s_t}.



\section{Summary and Discussion} \label{sec:discussion}
In this article we have numerically calculated the energies of exciton and trion states in moir\'e superlattices, 
and the matrix elements for interband absorption processes with exciton and trion final states.  We have modeled the 
moir\'e superlattice by harmonic confinement potentials -- an approximation that is valid when the 
confinement oscillator lengths are small compared to the moir\'e lattice constant.  Our calculations are similar to earlier 
ones \cite{wojs1995negatively,halonen1992excitons,que1992excitons,maksym1990quantum} that used harmonic confinement potentials to model quantum dots in semiconductor quantum wells, but differ
because of the relevance of the TMD valley degree of freedom, and by the possibility in the
moir\'e superlattice case that electrons and holes are localized 
at different lattice sites.

The relative positions of electron and hole confinement locations is very important for the optical properties of 
moir\'e superlattices.  As illustrated in Fig.~\ref{fig:eh_lattice}, the two most likely possibilities are that the 
confinement locations are at identical positions in the moir\'e unit cell, or that they are 
separated by a large fraction
of the moir\'e lattice constant.  It is not yet known whether one circumstance or the other is always realized, or whether 
this important distinction is interface-dependent.  
A recent experimental and computational study \cite{naik2022nature} of WSe$_2$/WS$_2$ heterostructures suggests that some of the optical absorption peaks come from excitons and trions in which the electrons and holes are confined at different lateral positions in the 2D plane, providing an example of the latter case. More experimental studies are needed to investigate how this property depends on the material platform.
Our calculations show that the two cases should in principle be distinguishable   
experimentally, since the rotational symmetries that apply in the common-confinement-location case, lead to a 
smaller number of very strong absorption peaks.  In the distinct-confinement-location case, the number of 
exciton and trion absorption peaks proliferates as angular-momentum-conservation selection rules are relaxed, but 
the overall absorption strength is still weakened due to electron-hole confinement.  
The situation is more complicated in realistic TMD materials due to the possible coexistence and mixing of both types of excitonic modes \cite{naik2022nature}. Nevertheless, we expect that our theory provides a useful guide to understand the nature of different excitonic peaks.
Because the Bohr radius length scale is 
so short in TMD two-dimensional semiconductors, the influence of magnetic fields on observable properties is weak at 
moderate magnetic field strengths.  In fact the results of our illustrative calculations at zero magnetic field are almost identical to the results at
$B\sim\SI{10}{T}$.  Magnetic fields are potentially very useful probes in 
sorting out the properties of electron-hole systems formed at particular interfaces, but will require extremely strong 
magnetic fields on the \SI{100}{T} scale.  

\begin{figure*}
    \centering
    \includegraphics[width=0.48\linewidth]{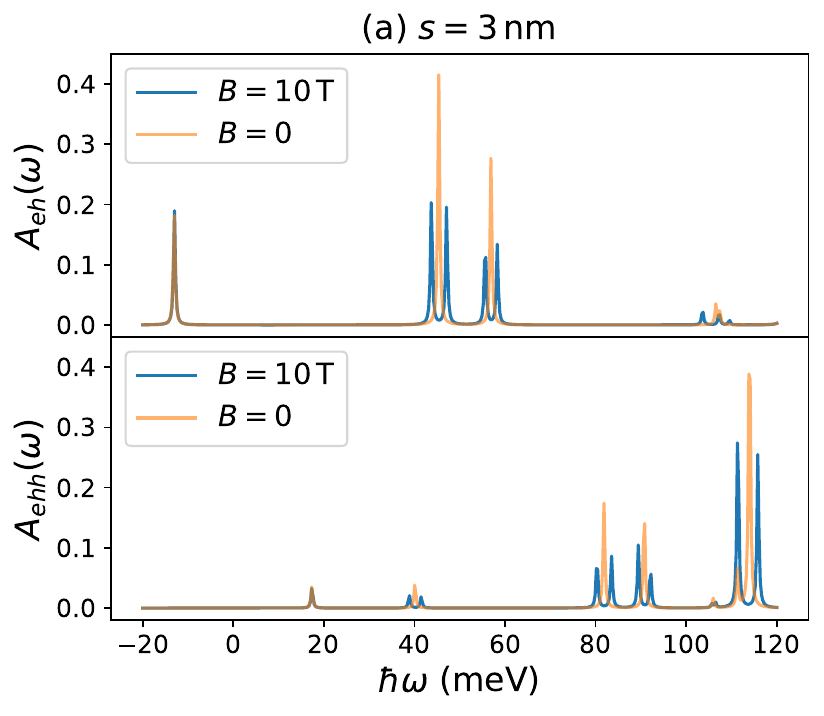}
    \includegraphics[width=0.48\linewidth]{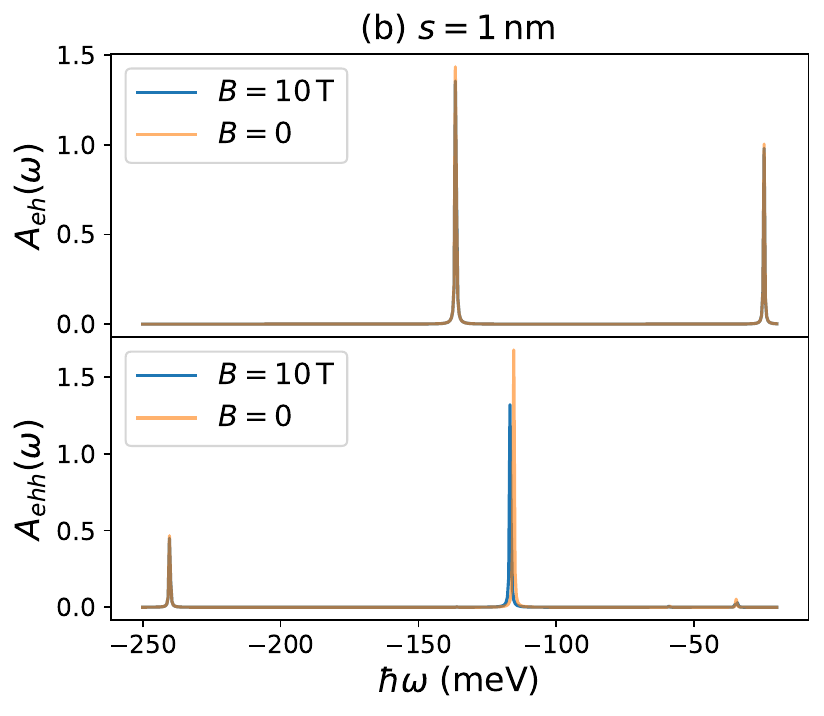}
    \caption{The optical absorption spectra of excitons ($A_{eh}$, upper panels) and trions ($A_{ehh}$, lower panels) at (a) $s=\SI{3}{nm}$; (b) $s=\SI{1}{nm}$. The $\delta$-functions in Eqs.~\eqref{eq:spectrum_x} and \eqref{eq:spectrum_t} are replaced with Lorentzian functions with height 1 and width \SI{0.2}{meV}.}
    \label{fig:A_nu}
\end{figure*}

A full theory of the optical absorption spectrum of a doped moir\'e system is complicated by importance of 
correlations among the doped charges present in the initial state
and correlations between electrons, holes, and trions located in different periods of the moir\'e superlattice \cite{baek2021optical}.
As a first approximation, we ignore these complications and calculate the optical absorption spectra of excitons and trions in our quantum dot model. In Fig.~\ref{fig:A_nu} we show the optical spectra of excitons \eqref{eq:spectrum_x} and trions \eqref{eq:spectrum_t} at a large $s=\SI{3}{nm}$ (a) and a small $s=\SI{1}{nm}$ (b), and compare the results at $B=0$ and $B=\SI{10}{T}$.
The band gap $E_g$ is set to zero in this figure since its only effect is a constant shift along the $\omega$-axis. The $\delta$-function is replaced with a Lorentzian function with height 1 and width \SI{0.2}{meV}. At $s=\SI{1}{nm}$ the electrons and holes are strongly bound and the two- or three-body eigenstates of the system have approximate rotational symmetry, so most of these states have optical absorption 
that is supressed by angular momentum conservation \cite{ruiz2020theory,wojs1995negatively,anisimovas2003excitonic}. In contrast, at $s=\SI{3}{nm}$ the electrons and holes are strongly localized in the harmonic potentials, and Coulomb interaction distorts the localized wavefunctions weakly. 
In this case the selection rules associated with angular momentum conservation are not relevant because 
rotational symmetry is strongly broken. The absorption peaks in Fig.~\ref{fig:A_nu}(a) are therefore much more closely spaced 
than in Fig.~\ref{fig:A_nu}(b). The peaks for states with nonzero angular momenta appear in the strong-confinement limit and are split in a magnetic field, while at small $s$ the optical spectrum changes very little in a magnetic field $B=\SI{10}{T}$. The valley Zeeman effect \cite{li2014valley,macneill2015breaking,srivastava2015valley,aivazian2015magnetic} leads to valley-splitting of excitonic peaks, whereas the level splittings in Fig.~\ref{fig:A_nu} are present even in valley-selected absorption measurements, because they are purely orbital effects.  The results do demonstrate that 
the absorption spectrum of moir\'e superlattices is strongly sensitive to electrical modulation of the carrier density.  


A more accurate description of the optical spectra of moir\'e materials would take account of correlations
between electrons and holes in different periods of the moir\'e pattern, but is beyond the scope of the present work.
These correlations will entangle photons emitted by nearby excitons, and promise the possibility of 
realizing unusual single and few photon light emitters.
One of the important contributions to inter-site exciton hopping processes come from the intervalley exchange interactions \cite{yu2014dirac,glazov2014exciton,yu2014valley,wu2015exciton}, which in the weak-modulation 
limit lead to topologically non-trivial exciton bands \cite{wu2017topological} in the presence of a finite Zeeman field.  We expect that this property will survive at strong modulation.
Our work has focused on excitons and trions, but our method also applies to more complex systems like biexcitons.  Because the moir\'e confinement length is relatively small, the transition energy between excitons and biexcitons localized at a single moir\'e site is expected to be large.
The method can also be extended to study a combination of localized and itinerant particles, which may help interpret recent experimental work \cite{xu2020correlated,gu2021dipolar} that used excitons in a TMD monolayer as an optical sensor of correlated insulating states in moir\'e materials.  We anticipate that close interaction between theory and experiment will be necessary to sort out all the richness of the optical properties of moir\'e superlattices.

\begin{acknowledgements}
Y.Z. and A.H.M. acknowledge many helpful conversations with Hui Deng and Feng Wang.  
This work was supported by Army Research Office (ARO) Grant \# W911NF-17-1-0312 (MURI).
\end{acknowledgements}

\appendix

\begin{widetext}
\section{Derivation of Coulomb matrix elements in the harmonic oscillator basis} \label{app:ed_eqs}

\subsection{Single-particle basis}
We consider an electron or a hole moving in a two-dimensional quantum dot, described by the Hamiltonian (in cgs units)
\begin{equation}
H = \frac{1}{2m} (\bm p\pm \frac ec \bm A) + \frac 12 m\omega^2\bm r^2,
\end{equation}
where the upper and lower signs are for electrons and holes respectively. We work in the symmetric gauge
\begin{equation}
A_x = -By/2, \quad A_y = Bx/2.
\end{equation}
To simplify notations we define the dimensionless quantities
\begin{equation}
\tilde{\bm r} = \bm r/l_B,\quad \tilde{\omega} = \omega/\omega_c,\quad \tilde H = H/\hbar\omega_c,
\end{equation}
where the magnetic length $l_B = \sqrt{\hbar c/eB}$ and the cyclotron frequency $\omega_c = eB/mc$. The dimensionless Hamiltonian can then be written as
\begin{equation}
\tilde H = \frac 12 (-i\partial_{\tilde x} \mp \frac{\tilde y}{2})^2 + \frac 12 (-i\partial_{\tilde y} \pm \frac{\tilde x}{2}) + \frac 12 \tilde{\omega}^2 (\tilde x^2 + \tilde y^2).
\end{equation}
Now we transform to the complex coordinates
\begin{equation}
z = x+iy,\quad z^* = x-iy,
\end{equation}
and their dimensionless counterparts $\tilde z,\tilde z^*$ defined in an obvious way. Then after some algebra the Hamiltonian can be written as
\begin{equation}
\tilde H = -2\partial_{\tilde z^*}\partial_{\tilde z} + (\frac 18 + \frac 12 \tilde{\omega}^2) \tilde z^* \tilde z \pm \frac 12 (\tilde z \partial_{\tilde z} - \tilde z^* \partial_{\tilde z^*}).
\end{equation}
Now we introduce the operators
\begin{equation}
\begin{split}
b_+ = \frac{1}{\sqrt 2} (\frac{\tilde z^*}{2} + 2\partial_{\tilde z}),\quad & b_+^{\dagger} = \frac{1}{\sqrt 2} (\frac{\tilde z}{2} - 2\partial_{\tilde z^*}), \\
b_- = \frac{1}{\sqrt 2} (\frac{\tilde z}{2} + 2\partial_{\tilde z^*}),\quad & b_-^{\dagger} = \frac{1}{\sqrt 2} (\frac{\tilde z^*}{2} - 2\partial_{\tilde z}),
\end{split}
\end{equation}
which satisfy the usual commutation relations of raising and lowering ladder operators. The Hamiltonian can be written in terms of these operators as
\begin{equation}
\tilde H = b_{\pm}^{\dagger} b_{\pm} + \frac 12 + \tilde{\omega}^2 (b_+^{\dagger} + b_-) (b_+ + b_-^{\dagger}),
\end{equation}
where as before the upper and lower signs (in the subscript) are for electrons and holes respectively. The Hamiltonian is quadratic in the ladder operators, and can be diagonalized by a Bogoliubov transformation
\begin{equation}
\begin{split}
a_+ &= b_{\pm} \cosh\frac{\eta}{2} + b_{\mp}^{\dagger} \sinh\frac{\eta}{2}, \\
a_- &= b_{\mp} \cosh\frac{\eta}{2} + b_{\pm}^{\dagger} \sinh\frac{\eta}{2},
\end{split}
\end{equation}
where
\begin{equation}
\tanh\eta = \frac{2\tilde{\omega}^2}{2\tilde{\omega}^2+1} = \frac{2\omega^2}{2\omega^2+\omega_c^2}.
\end{equation}
The Hamiltonian then takes the diagonal form
\begin{equation}
\tilde H = \tilde{\Omega}_+ (a_+^{\dagger}a_+ + \frac 12) + \tilde{\Omega}_- (a_-^{\dagger}a_- + \frac 12),
\end{equation}
where
\begin{equation}
\tilde{\Omega}_{\pm} = \frac 12 (\sqrt{1+4\tilde{\omega}^2} \pm 1).
\end{equation}
Restoring the physical dimensions, we have the diagonalized Hamiltonian
\begin{equation}
H = \hbar\Omega_+ (a_+^{\dagger}a_+ +\frac 12) + \hbar\Omega_- (a_-^{\dagger}a_- +\frac 12),
\end{equation}
with
\begin{equation}
\Omega_{\pm} = \sqrt{\omega^2 + (\frac{\omega_c}{2})^2} \pm \frac{\omega_c}{2} \equiv \Omega \pm \frac{\omega_c}{2}.
\end{equation}
The eigenstates of this Hamiltonian are
\begin{equation}
\ket{n_+ n_-} = (a_+^{\dagger})^{n_+} (a_-^{\dagger})^{n_-} \ket{0},
\end{equation}
with energy
\begin{equation}
E_{n_+ n_-} = (n_+ + \frac 12)\hbar\Omega_+ + (n_- + \frac 12)\hbar\Omega_-.
\end{equation}
For later use, the complex coordinate $z$ can be expressed in terms of the ladder operators as
\begin{equation} \label{eq:z_a}
z = \sqrt 2 l_B (b_+^{\dagger} + b_-) = \sqrt 2 l_B e^{-\eta/2} (a_{\pm}^{\dagger} + a_{\mp}) = L (a_{\pm}^{\dagger} + a_{\mp}),
\end{equation}
where $L = \sqrt{2}l_B e^{-\eta/2} = \sqrt{\hbar/m\Omega}$ is the total confining length. The normalized real-space wavefunction of the energy eigenstates is
\begin{equation}
\psi_{n_+ n_-}(r,\theta) = \left[\frac{n!}{(n+|l|)!} \frac{1}{\pi} \right]^{1/2} \frac 1L \left(\frac{r}{L}\right)^{|l|} e^{-r^2/2L^2} L_n^{|l|}(r^2/L^2)\cdot e^{il\theta},
\end{equation}
where $n=\min(n_+,n_-)$, $l=\pm(n_+-n_-)$, $\Omega=\sqrt{\omega^2+(\omega_c/2)^2}$, and $L_n^{|l|}$ is the associated Laguerre polynomial.

\subsection{Coulomb matrix elements}
Now we calculate the Coulomb matrix elements in the above basis (generalized to multiple particles, see Eq.~\eqref{eq:basis_nosym}). We start with the most complicated one, $V_{eh}$ (Eq.~\eqref{eq:V_eh}), the electron-hole interaction. The calculations for $V_{ee}$ and $V_{hh}$ are similar and simpler. We first write $V_{eh}$ in Fourier form (consider only one electron and one hole for simplicity):
\begin{equation}
V_{eh} = -\frac{2\pi e^2}{\epsilon} \int \frac{d^2 q}{(2\pi)^2}\, \frac{1}{|q|} e^{-|q|d} e^{iq_x s} e^{i(q^* z_e + qz_e^*)/2} e^{-i(q^* z_h + qz_h^*)/2},
\end{equation}
where $q=q_x+iq_y$ is a complex integration variable. By Eq.~\eqref{eq:z_a} we can write the complex coordinates $z$ in terms of ladder operators $a_{\pm}, a_{\pm}^{\dagger}$, then the matrix elements of $V_{eh}$ can be written as
\begin{equation} \label{eq:Veh_long}
\begin{split}
\bra{n_{e+}' n_{e-}', n_{h+}' n_{h-}'} V_{eh} \ket{n_{e+} n_{e-}, n_{h+} n_{h-}} &= -\frac{2\pi e^2}{\epsilon} \int \frac{d^2 q}{(2\pi)^2}\, \frac{1}{|q|} e^{-|q|d} e^{iq_x s} \bra{n_{e+}'} e^{iL_e (q^* a_{e+}^{\dagger} + qa_{e+})} \ket{n_{e+}} \\
\times \bra{n_{e-}'} e^{iL_e (q^* a_{e-}^{\dagger} + qa_{e-})} \ket{n_{e-}}  &\bra{n_{h+}'} e^{-iL_h (qa_{h+}^{\dagger} + q^* a_{h+})} \ket{n_{h+}} \bra{n_{h-}'} e^{-iL_h (qa_{h-}^{\dagger} + q^* a_{h-})} \ket{n_{h-}},
\end{split}
\end{equation}
Then we need to evaluate matrix elements of the form $\bra{n'} e^{i(fa^{\dagger} + f^* a)} \ket{n}$. To this end we make use of the operator identity
\begin{equation}
e^{i(fa^{\dagger} + f^* a)} = e^{-|f|^2/2} e^{ifa^{\dagger}} e^{if^* a},
\end{equation}
which then leads to
\begin{equation}
\begin{split}
\bra{n'} e^{i(fa^{\dagger} + f^* a)} \ket{n} &= e^{-|f|^2/2} \bra{n'} e^{ifa^{\dagger}} e^{if^* a} \ket{n} \\
&= e^{-|f|^2/2} \sum_{k'=0}^{n'} \sum_{k=0}^{n} \frac{(if)^{k'} (if^*)^k}{k'!k!} \bra{n'} (a^{\dagger})^{k'} a^k \ket{n} \\
&= e^{-|f|^2/2} (ie^{i\varphi})^{n'-n} (n'!n!)^{1/2} \sum_{k=0}^{\min(n',n)} \frac{(-1)^{n-k} |f|^{n'+n-2k}}{k!(n'-k)!(n-k)!},
\end{split}
\end{equation}
where $\varphi$ is the argument of the complex number $f=|f|e^{i\varphi}$. With this formula we go back to Eq.~\eqref{eq:Veh_long} and get
\begin{equation}
\begin{split}
\bra{n_{e+}' n_{e-}', n_{h+}' n_{h-}'} V_{eh} \ket{n_{e+} n_{e-}, n_{h+} n_{h-}} &= -\sqrt{\frac{\beta_{eh}}{2}} \frac{e^2}{\epsilon l_B} \left(\prod_{\alpha\lambda} n_{\alpha\lambda}'! n_{\alpha\lambda}!\right)^{1/2} (-1)^{n_{e+}' + n_{e-} + n_{h+}' + n_{h-}}  \\
&\times \sum_{k_{\alpha\lambda}=0}^{\min(n_{\alpha\lambda}',n_{\alpha\lambda})} \left(\prod_{\alpha\lambda} \frac{(-1)^{k_{\alpha\lambda}} (\frac{\beta_{eh}}{\beta_{\alpha}})^{\frac 12 (n_{\alpha\lambda}' + n_{\alpha\lambda}) - k_{\alpha\lambda}}}{k_{\alpha\lambda}! (n_{\alpha\lambda}' - k_{\alpha\lambda})! (n_{\alpha\lambda} - k_{\alpha\lambda})!} \right) \cdot I,
\end{split}
\end{equation}
where the summation indices $\alpha=e,h$ and $\lambda=\pm$ (the summation sign really means four summations), and we defined $\beta_{\alpha} = e^{\eta_{\alpha}} = \sqrt{1+(2\omega_{\alpha}/\omega_{\alpha c})^2}$ and $\beta_{eh} = (\beta_e^{-1} + \beta_h^{-1})^{-1}$, and the $(n',n,k)$-dependent integral
\begin{equation} \label{eq:int_Veh}
I = \int_{0}^{\infty} d\tilde q^2 \, e^{-\tilde q^2} (\tilde q^2)^{p - \frac 12} e^{-\tilde q \tilde d} \cdot \int_{0}^{2\pi} \frac{d\phi}{2\pi} \, (ie^{i\phi})^{l-l'} e^{i\tilde q \tilde s \cos\phi},
\end{equation}
with the dimensionless variables
\begin{equation}
\tilde q = \frac{l_B}{\sqrt{2\beta_{eh}}} |q|, \quad \tilde d = \frac{\sqrt{2\beta_{eh}}}{l_B} d, \quad \tilde s = \frac{\sqrt{2\beta_{eh}}}{l_B} s,
\end{equation}
the index
\begin{equation}
p = \sum_{\alpha\lambda} (\frac{n_{\alpha\lambda}' + n_{\alpha\lambda}}{2} - k_{\alpha\lambda})
\end{equation}
and the total angular momentum quantum numbers
\begin{equation}
l = n_{e+} - n_{e-} - n_{h+} + n_{h-}, \quad l' = n_{e+}' - n_{e-}' - n_{h+}' + n_{h-}'.
\end{equation}
The integral \eqref{eq:int_Veh} can be simplified in special cases. When $s=0$, the angular integral is simply $\delta_{l'l}$. When $d$ is also zero, the radial integral yields a gamma function and the result is
\begin{equation}
I = \delta_{l'l} \cdot \Gamma (p + \frac 12).
\end{equation}
When $s=0$ and $d\ne 0$, the radial integral yields the product of a gamma function and a (Tricomi's) confluent hypergeometric function
\begin{equation}
I = \delta_{l'l} \cdot 4^{-p}\, \Gamma(2p+1)\, U(p+\frac 12, \frac 12, \frac{\tilde d^2}{4}).
\end{equation}
However, in the general case where $s\ne 0$, only the radial part can be integrated in closed form and the angular integral remains:
\begin{equation}
I = 4^{-p}\, \Gamma(2p+1) \int_{0}^{2\pi} \frac{d\phi}{2\pi}\, (ie^{i\phi})^{l-l'}\, U(p+\frac 12, \frac 12, \frac 14 (\tilde d - i\tilde s \cos\phi)^2).
\end{equation}
In practical calculations the integral has to be done numerically. The matrix elements of $V_{ee}$ and $V_{hh}$ can be calculated in the same way, although the result looks simpler since $d=s=0$ for the same type of particles. The result is (without antisymmetrization)
\begin{equation}
\begin{split}
\bra{n_{1+}' n_{1-}', n_{2+}' n_{2-}'} V_{\alpha\alpha} \ket{n_{1+} n_{1-}, n_{2+} n_{2-}} &= \frac{\sqrt{\beta_{\alpha}}}{2} \frac{e^2}{\epsilon l_B} \left(\prod_{i\lambda} n_{i\lambda}'! n_{i\lambda}!\right)^{1/2} (-1)^{n_{1+}' + n_{1-} + n_{2+}' + n_{2-}} \delta_{l'l} \\
&\times \sum_{k_{i\lambda}=0}^{\min(n_{i\lambda}',n_{i\lambda})} \left(\prod_{i\lambda} \frac{(-1)^{k_{i\lambda}} 2^{-p} }{k_{i\lambda}! (n_{i\lambda}' - k_{i\lambda})! (n_{i\lambda} - k_{i\lambda})!} \right) \cdot \Gamma(p+\frac 12),
\end{split}
\end{equation}
where $\alpha=e$ or $h$, the summation indices $i=1,2$ is the particle label (the $e/h$ label is omitted), $\lambda=\pm$, and we defined
\begin{equation}
p = \sum_{i\lambda} (\frac{n_{i\lambda}' + n_{i\lambda}}{2} - k_{i\lambda}),
\end{equation}
\begin{equation}
l = n_{1+} - n_{1-} + n_{2+} - n_{2-}, \quad l' = n_{1+}' - n_{1-}' + n_{2+}' - n_{2-}'.
\end{equation}

\section{The current operator} \label{app:cur_op}
In the two-band model the current operator is (naively) proportional to $\sum_{\bm k} (e_{\bm k}^{\dagger} h_{-\bm k}^{\dagger} + h_{-\bm k} e_{\bm k})$, which in real space takes the form
\begin{equation}
\hat j \sim \int d\bm r [e^{\dagger}(\bm r) h^{\dagger}(\bm r) + h(\bm r) e(\bm r)],
\end{equation}
where $e$ and $h$ are annihilation operators of electron and hole states. Note however that in writing Eqs.~\eqref{eq:H_e} and \eqref{eq:H_h} with $\bm A_{\alpha}(\bm r_{\alpha}) = \frac 12 \bm B\times \bm r_{\alpha}$ for both $\alpha=e$ and $h$, and with $\bm r_e$ and $\bm r_h$ measured from different origins (see Fig.~\ref{fig:qd_eh}), we have implicitly performed a gauge transformation
\begin{equation} \label{eq:gauge_transform}
\bm A_e(\bm r_e) \to \bm A_e(\bm r_e) - \frac{Bs}{2}\hat y,\quad \psi(\{\bm r_{ei}\}, \{\bm r_{hj}\}) \to \psi(\{\bm r_{ei}\}, \{\bm r_{hj}\}) \exp\left(i\frac{eBs}{2\hbar c} \sum_i y_{ei} \right),
\end{equation}
compared to the gauge where $\bm A(\bm r) = \frac 12 \bm B\times \bm r$ for both electrons and holes, with the origin set at the confining potential minimum for holes. After the gauge transformation the field operator $e(\bm r)$ gains an $\bm r$-dependent phase factor, so in this gauge the current operator becomes
\begin{equation}
\hat j \sim \int d\bm r \left[e^{\dagger}(\bm r) h^{\dagger}(\bm r) \exp\left(i\frac{eBs}{2\hbar c}y \right)  + h(\bm r) e(\bm r) \exp\left(-i\frac{eBs}{2\hbar c}y \right) \right].
\end{equation}

\section{Basis cutoff}

\begin{table}[h!]
\centering
\begin{tabular}{|c|c|c|c|c|c|c|c|c|c|}
\hline
$s$/nm & 0 & 1 & 2 & 3 & 4 & 5 & 6 & 7 & 8 \\\hline
$N_{x0}$ & 150000 & 3000 & 3000 & 2000 & 2000 & 500 & 200 & 100 & 100 \\\hline
$N_{t0}$ & 500000 & 10000 & 8000 & 8000 & 8000 & 4000 & 4000 & 4000 & 4000 \\\hline
$N_{x1}$ & 20000 & 2000 & 1000 & 500 & 200 & 100 & 100 & 100 & 100 \\\hline
$N_{t1}$ & 50000 & 10000 & 5000 & 5000 & 4000 & 4000 & 4000 & 4000 & 4000 \\\hline
$N_{x2}$ & 10000 & 2000 & 1000 & 500 & 200 & 100 & 100 & 100 & 100 \\\hline
$N_{t2}$ & 20000 & 4000 & 4000 & 4000 & 4000 & 4000 & 4000 & 4000 & 4000 \\\hline
\end{tabular}
\caption{Number of basis states for the ED calculations for excitons ($N_{x0},N_{x1},N_{x2}$ for $d=0,1,\SI{2}{nm}$) and trions ($N_{t0},N_{t1},N_{t2}$) at different in-plane electron-hole separations $s$.}
\label{tab:num_basis}
\end{table}

The energies and optical absorption strengths are obtained by numerically diagonalizing the Hamiltonian \eqref{eq:Hamiltonian} as a matrix in the harmonic oscillator basis, which has to be cut off under some energy scale. The number of basis states used in the ED calculations are shown in Table~\ref{tab:num_basis}. With the exception of the $s=0$ case, the number of basis states is chosen such that the ground-state energy is accurate within \SI{1}{meV} (in fact the error is much smaller for large $s$), and the optical absorption strength (for $d=0$) is accurate within 1\%. In the magnetic harmonic oscillator basis, convergence is faster for larger $s$. For $s=0$, the ED calculations suffer from poor convergence with the number of basis states, as already noted in Ref.~\cite{halonen1992excitons} where the exciton problem is solved instead in the center-of-mass and relative coordinates. However, since we are mostly interested in the $s\ne 0$ case, and for its easier generalization to more complicated systems like trions and biexcitons, we solve the $s=0$ case also in the harmonic oscillator basis, but with a huge number of basis states. The calculation is manageable when we take advantage of the rotational symmetry of the system and only diagonalize the block with total angular momentum zero (which contains the ground state and is optically active).

\end{widetext}

\bibliography{qdxt.bib}

\begin{thebibliography}{84}%
\makeatletter
\providecommand \@ifxundefined [1]{%
 \@ifx{#1\undefined}
}%
\providecommand \@ifnum [1]{%
 \ifnum #1\expandafter \@firstoftwo
 \else \expandafter \@secondoftwo
 \fi
}%
\providecommand \@ifx [1]{%
 \ifx #1\expandafter \@firstoftwo
 \else \expandafter \@secondoftwo
 \fi
}%
\providecommand \natexlab [1]{#1}%
\providecommand \enquote  [1]{``#1''}%
\providecommand \bibnamefont  [1]{#1}%
\providecommand \bibfnamefont [1]{#1}%
\providecommand \citenamefont [1]{#1}%
\providecommand \href@noop [0]{\@secondoftwo}%
\providecommand \href [0]{\begingroup \@sanitize@url \@href}%
\providecommand \@href[1]{\@@startlink{#1}\@@href}%
\providecommand \@@href[1]{\endgroup#1\@@endlink}%
\providecommand \@sanitize@url [0]{\catcode `\\12\catcode `\$12\catcode
  `\&12\catcode `\#12\catcode `\^12\catcode `\_12\catcode `\%12\relax}%
\providecommand \@@startlink[1]{}%
\providecommand \@@endlink[0]{}%
\providecommand \url  [0]{\begingroup\@sanitize@url \@url }%
\providecommand \@url [1]{\endgroup\@href {#1}{\urlprefix }}%
\providecommand \urlprefix  [0]{URL }%
\providecommand \Eprint [0]{\href }%
\providecommand \doibase [0]{https://doi.org/}%
\providecommand \selectlanguage [0]{\@gobble}%
\providecommand \bibinfo  [0]{\@secondoftwo}%
\providecommand \bibfield  [0]{\@secondoftwo}%
\providecommand \translation [1]{[#1]}%
\providecommand \BibitemOpen [0]{}%
\providecommand \bibitemStop [0]{}%
\providecommand \bibitemNoStop [0]{.\EOS\space}%
\providecommand \EOS [0]{\spacefactor3000\relax}%
\providecommand \BibitemShut  [1]{\csname bibitem#1\endcsname}%
\let\auto@bib@innerbib\@empty
\bibitem [{\citenamefont {Bistritzer}\ and\ \citenamefont
  {MacDonald}(2011)}]{bistritzer2011moire}%
  \BibitemOpen
  \bibfield  {author} {\bibinfo {author} {\bibfnamefont {R.}~\bibnamefont
  {Bistritzer}}\ and\ \bibinfo {author} {\bibfnamefont {A.~H.}\ \bibnamefont
  {MacDonald}},\ }\bibfield  {title} {\bibinfo {title} {Moir{\'e} bands in
  twisted double-layer graphene},\ }\href@noop {} {\bibfield  {journal}
  {\bibinfo  {journal} {Proceedings of the National Academy of Sciences}\
  }\textbf {\bibinfo {volume} {108}},\ \bibinfo {pages} {12233} (\bibinfo
  {year} {2011})}\BibitemShut {NoStop}%
\bibitem [{\citenamefont {Cao}\ \emph {et~al.}(2018{\natexlab{a}})\citenamefont
  {Cao}, \citenamefont {Fatemi}, \citenamefont {Demir}, \citenamefont {Fang},
  \citenamefont {Tomarken}, \citenamefont {Luo}, \citenamefont
  {Sanchez-Yamagishi}, \citenamefont {Watanabe}, \citenamefont {Taniguchi},
  \citenamefont {Kaxiras} \emph {et~al.}}]{cao2018correlated}%
  \BibitemOpen
  \bibfield  {author} {\bibinfo {author} {\bibfnamefont {Y.}~\bibnamefont
  {Cao}}, \bibinfo {author} {\bibfnamefont {V.}~\bibnamefont {Fatemi}},
  \bibinfo {author} {\bibfnamefont {A.}~\bibnamefont {Demir}}, \bibinfo
  {author} {\bibfnamefont {S.}~\bibnamefont {Fang}}, \bibinfo {author}
  {\bibfnamefont {S.~L.}\ \bibnamefont {Tomarken}}, \bibinfo {author}
  {\bibfnamefont {J.~Y.}\ \bibnamefont {Luo}}, \bibinfo {author} {\bibfnamefont
  {J.~D.}\ \bibnamefont {Sanchez-Yamagishi}}, \bibinfo {author} {\bibfnamefont
  {K.}~\bibnamefont {Watanabe}}, \bibinfo {author} {\bibfnamefont
  {T.}~\bibnamefont {Taniguchi}}, \bibinfo {author} {\bibfnamefont
  {E.}~\bibnamefont {Kaxiras}}, \emph {et~al.},\ }\bibfield  {title} {\bibinfo
  {title} {Correlated insulator behaviour at half-filling in magic-angle
  graphene superlattices},\ }\href@noop {} {\bibfield  {journal} {\bibinfo
  {journal} {Nature}\ }\textbf {\bibinfo {volume} {556}},\ \bibinfo {pages}
  {80} (\bibinfo {year} {2018}{\natexlab{a}})}\BibitemShut {NoStop}%
\bibitem [{\citenamefont {Cao}\ \emph {et~al.}(2018{\natexlab{b}})\citenamefont
  {Cao}, \citenamefont {Fatemi}, \citenamefont {Fang}, \citenamefont
  {Watanabe}, \citenamefont {Taniguchi}, \citenamefont {Kaxiras},\ and\
  \citenamefont {Jarillo-Herrero}}]{cao2018unconventional}%
  \BibitemOpen
  \bibfield  {author} {\bibinfo {author} {\bibfnamefont {Y.}~\bibnamefont
  {Cao}}, \bibinfo {author} {\bibfnamefont {V.}~\bibnamefont {Fatemi}},
  \bibinfo {author} {\bibfnamefont {S.}~\bibnamefont {Fang}}, \bibinfo {author}
  {\bibfnamefont {K.}~\bibnamefont {Watanabe}}, \bibinfo {author}
  {\bibfnamefont {T.}~\bibnamefont {Taniguchi}}, \bibinfo {author}
  {\bibfnamefont {E.}~\bibnamefont {Kaxiras}},\ and\ \bibinfo {author}
  {\bibfnamefont {P.}~\bibnamefont {Jarillo-Herrero}},\ }\bibfield  {title}
  {\bibinfo {title} {Unconventional superconductivity in magic-angle graphene
  superlattices},\ }\href@noop {} {\bibfield  {journal} {\bibinfo  {journal}
  {Nature}\ }\textbf {\bibinfo {volume} {556}},\ \bibinfo {pages} {43}
  (\bibinfo {year} {2018}{\natexlab{b}})}\BibitemShut {NoStop}%
\bibitem [{\citenamefont {Lu}\ \emph {et~al.}(2019)\citenamefont {Lu},
  \citenamefont {Stepanov}, \citenamefont {Yang}, \citenamefont {Xie},
  \citenamefont {Aamir}, \citenamefont {Das}, \citenamefont {Urgell},
  \citenamefont {Watanabe}, \citenamefont {Taniguchi}, \citenamefont {Zhang}
  \emph {et~al.}}]{lu2019superconductors}%
  \BibitemOpen
  \bibfield  {author} {\bibinfo {author} {\bibfnamefont {X.}~\bibnamefont
  {Lu}}, \bibinfo {author} {\bibfnamefont {P.}~\bibnamefont {Stepanov}},
  \bibinfo {author} {\bibfnamefont {W.}~\bibnamefont {Yang}}, \bibinfo {author}
  {\bibfnamefont {M.}~\bibnamefont {Xie}}, \bibinfo {author} {\bibfnamefont
  {M.~A.}\ \bibnamefont {Aamir}}, \bibinfo {author} {\bibfnamefont
  {I.}~\bibnamefont {Das}}, \bibinfo {author} {\bibfnamefont {C.}~\bibnamefont
  {Urgell}}, \bibinfo {author} {\bibfnamefont {K.}~\bibnamefont {Watanabe}},
  \bibinfo {author} {\bibfnamefont {T.}~\bibnamefont {Taniguchi}}, \bibinfo
  {author} {\bibfnamefont {G.}~\bibnamefont {Zhang}}, \emph {et~al.},\
  }\bibfield  {title} {\bibinfo {title} {Superconductors, orbital magnets and
  correlated states in magic-angle bilayer graphene},\ }\href@noop {}
  {\bibfield  {journal} {\bibinfo  {journal} {Nature}\ }\textbf {\bibinfo
  {volume} {574}},\ \bibinfo {pages} {653} (\bibinfo {year}
  {2019})}\BibitemShut {NoStop}%
\bibitem [{\citenamefont {Yankowitz}\ \emph {et~al.}(2019)\citenamefont
  {Yankowitz}, \citenamefont {Chen}, \citenamefont {Polshyn}, \citenamefont
  {Zhang}, \citenamefont {Watanabe}, \citenamefont {Taniguchi}, \citenamefont
  {Graf}, \citenamefont {Young},\ and\ \citenamefont
  {Dean}}]{yankowitz2019tuning}%
  \BibitemOpen
  \bibfield  {author} {\bibinfo {author} {\bibfnamefont {M.}~\bibnamefont
  {Yankowitz}}, \bibinfo {author} {\bibfnamefont {S.}~\bibnamefont {Chen}},
  \bibinfo {author} {\bibfnamefont {H.}~\bibnamefont {Polshyn}}, \bibinfo
  {author} {\bibfnamefont {Y.}~\bibnamefont {Zhang}}, \bibinfo {author}
  {\bibfnamefont {K.}~\bibnamefont {Watanabe}}, \bibinfo {author}
  {\bibfnamefont {T.}~\bibnamefont {Taniguchi}}, \bibinfo {author}
  {\bibfnamefont {D.}~\bibnamefont {Graf}}, \bibinfo {author} {\bibfnamefont
  {A.~F.}\ \bibnamefont {Young}},\ and\ \bibinfo {author} {\bibfnamefont
  {C.~R.}\ \bibnamefont {Dean}},\ }\bibfield  {title} {\bibinfo {title} {Tuning
  superconductivity in twisted bilayer graphene},\ }\href@noop {} {\bibfield
  {journal} {\bibinfo  {journal} {Science}\ }\textbf {\bibinfo {volume}
  {363}},\ \bibinfo {pages} {1059} (\bibinfo {year} {2019})}\BibitemShut
  {NoStop}%
\bibitem [{\citenamefont {Po}\ \emph {et~al.}(2018)\citenamefont {Po},
  \citenamefont {Zou}, \citenamefont {Vishwanath},\ and\ \citenamefont
  {Senthil}}]{po2018origin}%
  \BibitemOpen
  \bibfield  {author} {\bibinfo {author} {\bibfnamefont {H.~C.}\ \bibnamefont
  {Po}}, \bibinfo {author} {\bibfnamefont {L.}~\bibnamefont {Zou}}, \bibinfo
  {author} {\bibfnamefont {A.}~\bibnamefont {Vishwanath}},\ and\ \bibinfo
  {author} {\bibfnamefont {T.}~\bibnamefont {Senthil}},\ }\bibfield  {title}
  {\bibinfo {title} {Origin of mott insulating behavior and superconductivity
  in twisted bilayer graphene},\ }\href@noop {} {\bibfield  {journal} {\bibinfo
   {journal} {Physical Review X}\ }\textbf {\bibinfo {volume} {8}},\ \bibinfo
  {pages} {031089} (\bibinfo {year} {2018})}\BibitemShut {NoStop}%
\bibitem [{\citenamefont {Koshino}\ \emph {et~al.}(2018)\citenamefont
  {Koshino}, \citenamefont {Yuan}, \citenamefont {Koretsune}, \citenamefont
  {Ochi}, \citenamefont {Kuroki},\ and\ \citenamefont
  {Fu}}]{koshino2018maximally}%
  \BibitemOpen
  \bibfield  {author} {\bibinfo {author} {\bibfnamefont {M.}~\bibnamefont
  {Koshino}}, \bibinfo {author} {\bibfnamefont {N.~F.}\ \bibnamefont {Yuan}},
  \bibinfo {author} {\bibfnamefont {T.}~\bibnamefont {Koretsune}}, \bibinfo
  {author} {\bibfnamefont {M.}~\bibnamefont {Ochi}}, \bibinfo {author}
  {\bibfnamefont {K.}~\bibnamefont {Kuroki}},\ and\ \bibinfo {author}
  {\bibfnamefont {L.}~\bibnamefont {Fu}},\ }\bibfield  {title} {\bibinfo
  {title} {Maximally localized {Wannier} orbitals and the extended {Hubbard}
  model for twisted bilayer graphene},\ }\href@noop {} {\bibfield  {journal}
  {\bibinfo  {journal} {Physical Review X}\ }\textbf {\bibinfo {volume} {8}},\
  \bibinfo {pages} {031087} (\bibinfo {year} {2018})}\BibitemShut {NoStop}%
\bibitem [{\citenamefont {Kerelsky}\ \emph {et~al.}(2019)\citenamefont
  {Kerelsky}, \citenamefont {McGilly}, \citenamefont {Kennes}, \citenamefont
  {Xian}, \citenamefont {Yankowitz}, \citenamefont {Chen}, \citenamefont
  {Watanabe}, \citenamefont {Taniguchi}, \citenamefont {Hone}, \citenamefont
  {Dean} \emph {et~al.}}]{kerelsky2019maximized}%
  \BibitemOpen
  \bibfield  {author} {\bibinfo {author} {\bibfnamefont {A.}~\bibnamefont
  {Kerelsky}}, \bibinfo {author} {\bibfnamefont {L.~J.}\ \bibnamefont
  {McGilly}}, \bibinfo {author} {\bibfnamefont {D.~M.}\ \bibnamefont {Kennes}},
  \bibinfo {author} {\bibfnamefont {L.}~\bibnamefont {Xian}}, \bibinfo {author}
  {\bibfnamefont {M.}~\bibnamefont {Yankowitz}}, \bibinfo {author}
  {\bibfnamefont {S.}~\bibnamefont {Chen}}, \bibinfo {author} {\bibfnamefont
  {K.}~\bibnamefont {Watanabe}}, \bibinfo {author} {\bibfnamefont
  {T.}~\bibnamefont {Taniguchi}}, \bibinfo {author} {\bibfnamefont
  {J.}~\bibnamefont {Hone}}, \bibinfo {author} {\bibfnamefont {C.}~\bibnamefont
  {Dean}}, \emph {et~al.},\ }\bibfield  {title} {\bibinfo {title} {Maximized
  electron interactions at the magic angle in twisted bilayer graphene},\
  }\href@noop {} {\bibfield  {journal} {\bibinfo  {journal} {Nature}\ }\textbf
  {\bibinfo {volume} {572}},\ \bibinfo {pages} {95} (\bibinfo {year}
  {2019})}\BibitemShut {NoStop}%
\bibitem [{\citenamefont {Serlin}\ \emph {et~al.}(2020)\citenamefont {Serlin},
  \citenamefont {Tschirhart}, \citenamefont {Polshyn}, \citenamefont {Zhang},
  \citenamefont {Zhu}, \citenamefont {Watanabe}, \citenamefont {Taniguchi},
  \citenamefont {Balents},\ and\ \citenamefont {Young}}]{serlin2020intrinsic}%
  \BibitemOpen
  \bibfield  {author} {\bibinfo {author} {\bibfnamefont {M.}~\bibnamefont
  {Serlin}}, \bibinfo {author} {\bibfnamefont {C.}~\bibnamefont {Tschirhart}},
  \bibinfo {author} {\bibfnamefont {H.}~\bibnamefont {Polshyn}}, \bibinfo
  {author} {\bibfnamefont {Y.}~\bibnamefont {Zhang}}, \bibinfo {author}
  {\bibfnamefont {J.}~\bibnamefont {Zhu}}, \bibinfo {author} {\bibfnamefont
  {K.}~\bibnamefont {Watanabe}}, \bibinfo {author} {\bibfnamefont
  {T.}~\bibnamefont {Taniguchi}}, \bibinfo {author} {\bibfnamefont
  {L.}~\bibnamefont {Balents}},\ and\ \bibinfo {author} {\bibfnamefont
  {A.}~\bibnamefont {Young}},\ }\bibfield  {title} {\bibinfo {title} {Intrinsic
  quantized anomalous hall effect in a moir{\'e} heterostructure},\ }\href@noop
  {} {\bibfield  {journal} {\bibinfo  {journal} {Science}\ }\textbf {\bibinfo
  {volume} {367}},\ \bibinfo {pages} {900} (\bibinfo {year}
  {2020})}\BibitemShut {NoStop}%
\bibitem [{\citenamefont {Kang}\ and\ \citenamefont
  {Vafek}(2018)}]{kang2018symmetry}%
  \BibitemOpen
  \bibfield  {author} {\bibinfo {author} {\bibfnamefont {J.}~\bibnamefont
  {Kang}}\ and\ \bibinfo {author} {\bibfnamefont {O.}~\bibnamefont {Vafek}},\
  }\bibfield  {title} {\bibinfo {title} {Symmetry, maximally localized
  {Wannier} states, and a low-energy model for twisted bilayer graphene narrow
  bands},\ }\href@noop {} {\bibfield  {journal} {\bibinfo  {journal} {Physical
  Review X}\ }\textbf {\bibinfo {volume} {8}},\ \bibinfo {pages} {031088}
  (\bibinfo {year} {2018})}\BibitemShut {NoStop}%
\bibitem [{\citenamefont {Isobe}\ \emph {et~al.}(2018)\citenamefont {Isobe},
  \citenamefont {Yuan},\ and\ \citenamefont {Fu}}]{isobe2018unconventional}%
  \BibitemOpen
  \bibfield  {author} {\bibinfo {author} {\bibfnamefont {H.}~\bibnamefont
  {Isobe}}, \bibinfo {author} {\bibfnamefont {N.~F.}\ \bibnamefont {Yuan}},\
  and\ \bibinfo {author} {\bibfnamefont {L.}~\bibnamefont {Fu}},\ }\bibfield
  {title} {\bibinfo {title} {Unconventional superconductivity and density waves
  in twisted bilayer graphene},\ }\href@noop {} {\bibfield  {journal} {\bibinfo
   {journal} {Physical Review X}\ }\textbf {\bibinfo {volume} {8}},\ \bibinfo
  {pages} {041041} (\bibinfo {year} {2018})}\BibitemShut {NoStop}%
\bibitem [{\citenamefont {Wu}\ \emph {et~al.}(2018{\natexlab{a}})\citenamefont
  {Wu}, \citenamefont {MacDonald},\ and\ \citenamefont
  {Martin}}]{wu2018superconductivity}%
  \BibitemOpen
  \bibfield  {author} {\bibinfo {author} {\bibfnamefont {F.}~\bibnamefont
  {Wu}}, \bibinfo {author} {\bibfnamefont {A.}~\bibnamefont {MacDonald}},\ and\
  \bibinfo {author} {\bibfnamefont {I.}~\bibnamefont {Martin}},\ }\bibfield
  {title} {\bibinfo {title} {Theory of phonon-mediated superconductivity in
  twisted bilayer graphene},\ }\href@noop {} {\bibfield  {journal} {\bibinfo
  {journal} {Physical review letters}\ }\textbf {\bibinfo {volume} {121}},\
  \bibinfo {pages} {257001} (\bibinfo {year} {2018}{\natexlab{a}})}\BibitemShut
  {NoStop}%
\bibitem [{\citenamefont {Lian}\ \emph {et~al.}(2019)\citenamefont {Lian},
  \citenamefont {Wang},\ and\ \citenamefont {Bernevig}}]{lian2019twisted}%
  \BibitemOpen
  \bibfield  {author} {\bibinfo {author} {\bibfnamefont {B.}~\bibnamefont
  {Lian}}, \bibinfo {author} {\bibfnamefont {Z.}~\bibnamefont {Wang}},\ and\
  \bibinfo {author} {\bibfnamefont {B.~A.}\ \bibnamefont {Bernevig}},\
  }\bibfield  {title} {\bibinfo {title} {Twisted bilayer graphene: a
  phonon-driven superconductor},\ }\href@noop {} {\bibfield  {journal}
  {\bibinfo  {journal} {Physical review letters}\ }\textbf {\bibinfo {volume}
  {122}},\ \bibinfo {pages} {257002} (\bibinfo {year} {2019})}\BibitemShut
  {NoStop}%
\bibitem [{\citenamefont {Xie}\ and\ \citenamefont
  {MacDonald}(2020)}]{xie2020nature}%
  \BibitemOpen
  \bibfield  {author} {\bibinfo {author} {\bibfnamefont {M.}~\bibnamefont
  {Xie}}\ and\ \bibinfo {author} {\bibfnamefont {A.~H.}\ \bibnamefont
  {MacDonald}},\ }\bibfield  {title} {\bibinfo {title} {Nature of the
  correlated insulator states in twisted bilayer graphene},\ }\href@noop {}
  {\bibfield  {journal} {\bibinfo  {journal} {Physical review letters}\
  }\textbf {\bibinfo {volume} {124}},\ \bibinfo {pages} {097601} (\bibinfo
  {year} {2020})}\BibitemShut {NoStop}%
\bibitem [{\citenamefont {Cao}\ \emph {et~al.}(2020)\citenamefont {Cao},
  \citenamefont {Rodan-Legrain}, \citenamefont {Rubies-Bigorda}, \citenamefont
  {Park}, \citenamefont {Watanabe}, \citenamefont {Taniguchi},\ and\
  \citenamefont {Jarillo-Herrero}}]{cao2020tunable}%
  \BibitemOpen
  \bibfield  {author} {\bibinfo {author} {\bibfnamefont {Y.}~\bibnamefont
  {Cao}}, \bibinfo {author} {\bibfnamefont {D.}~\bibnamefont {Rodan-Legrain}},
  \bibinfo {author} {\bibfnamefont {O.}~\bibnamefont {Rubies-Bigorda}},
  \bibinfo {author} {\bibfnamefont {J.~M.}\ \bibnamefont {Park}}, \bibinfo
  {author} {\bibfnamefont {K.}~\bibnamefont {Watanabe}}, \bibinfo {author}
  {\bibfnamefont {T.}~\bibnamefont {Taniguchi}},\ and\ \bibinfo {author}
  {\bibfnamefont {P.}~\bibnamefont {Jarillo-Herrero}},\ }\bibfield  {title}
  {\bibinfo {title} {Tunable correlated states and spin-polarized phases in
  twisted bilayer--bilayer graphene},\ }\href@noop {} {\bibfield  {journal}
  {\bibinfo  {journal} {Nature}\ }\textbf {\bibinfo {volume} {583}},\ \bibinfo
  {pages} {215} (\bibinfo {year} {2020})}\BibitemShut {NoStop}%
\bibitem [{\citenamefont {Andrei}\ and\ \citenamefont
  {MacDonald}(2020)}]{andrei2020graphene}%
  \BibitemOpen
  \bibfield  {author} {\bibinfo {author} {\bibfnamefont {E.~Y.}\ \bibnamefont
  {Andrei}}\ and\ \bibinfo {author} {\bibfnamefont {A.~H.}\ \bibnamefont
  {MacDonald}},\ }\bibfield  {title} {\bibinfo {title} {Graphene bilayers with
  a twist},\ }\href@noop {} {\bibfield  {journal} {\bibinfo  {journal} {Nature
  Materials}\ }\textbf {\bibinfo {volume} {19}},\ \bibinfo {pages} {1265}
  (\bibinfo {year} {2020})}\BibitemShut {NoStop}%
\bibitem [{\citenamefont {Balents}\ \emph {et~al.}(2020)\citenamefont
  {Balents}, \citenamefont {Dean}, \citenamefont {Efetov},\ and\ \citenamefont
  {Young}}]{balents2020superconductivity}%
  \BibitemOpen
  \bibfield  {author} {\bibinfo {author} {\bibfnamefont {L.}~\bibnamefont
  {Balents}}, \bibinfo {author} {\bibfnamefont {C.~R.}\ \bibnamefont {Dean}},
  \bibinfo {author} {\bibfnamefont {D.~K.}\ \bibnamefont {Efetov}},\ and\
  \bibinfo {author} {\bibfnamefont {A.~F.}\ \bibnamefont {Young}},\ }\bibfield
  {title} {\bibinfo {title} {Superconductivity and strong correlations in
  moir{\'e} flat bands},\ }\href@noop {} {\bibfield  {journal} {\bibinfo
  {journal} {Nature Physics}\ }\textbf {\bibinfo {volume} {16}},\ \bibinfo
  {pages} {725} (\bibinfo {year} {2020})}\BibitemShut {NoStop}%
\bibitem [{\citenamefont {Zondiner}\ \emph {et~al.}(2020)\citenamefont
  {Zondiner}, \citenamefont {Rozen}, \citenamefont {Rodan-Legrain},
  \citenamefont {Cao}, \citenamefont {Queiroz}, \citenamefont {Taniguchi},
  \citenamefont {Watanabe}, \citenamefont {Oreg}, \citenamefont {von Oppen},
  \citenamefont {Stern} \emph {et~al.}}]{zondiner2020cascade}%
  \BibitemOpen
  \bibfield  {author} {\bibinfo {author} {\bibfnamefont {U.}~\bibnamefont
  {Zondiner}}, \bibinfo {author} {\bibfnamefont {A.}~\bibnamefont {Rozen}},
  \bibinfo {author} {\bibfnamefont {D.}~\bibnamefont {Rodan-Legrain}}, \bibinfo
  {author} {\bibfnamefont {Y.}~\bibnamefont {Cao}}, \bibinfo {author}
  {\bibfnamefont {R.}~\bibnamefont {Queiroz}}, \bibinfo {author} {\bibfnamefont
  {T.}~\bibnamefont {Taniguchi}}, \bibinfo {author} {\bibfnamefont
  {K.}~\bibnamefont {Watanabe}}, \bibinfo {author} {\bibfnamefont
  {Y.}~\bibnamefont {Oreg}}, \bibinfo {author} {\bibfnamefont {F.}~\bibnamefont
  {von Oppen}}, \bibinfo {author} {\bibfnamefont {A.}~\bibnamefont {Stern}},
  \emph {et~al.},\ }\bibfield  {title} {\bibinfo {title} {Cascade of phase
  transitions and dirac revivals in magic-angle graphene},\ }\href@noop {}
  {\bibfield  {journal} {\bibinfo  {journal} {Nature}\ }\textbf {\bibinfo
  {volume} {582}},\ \bibinfo {pages} {203} (\bibinfo {year}
  {2020})}\BibitemShut {NoStop}%
\bibitem [{\citenamefont {Wong}\ \emph {et~al.}(2020)\citenamefont {Wong},
  \citenamefont {Nuckolls}, \citenamefont {Oh}, \citenamefont {Lian},
  \citenamefont {Xie}, \citenamefont {Jeon}, \citenamefont {Watanabe},
  \citenamefont {Taniguchi}, \citenamefont {Bernevig},\ and\ \citenamefont
  {Yazdani}}]{wong2020cascade}%
  \BibitemOpen
  \bibfield  {author} {\bibinfo {author} {\bibfnamefont {D.}~\bibnamefont
  {Wong}}, \bibinfo {author} {\bibfnamefont {K.~P.}\ \bibnamefont {Nuckolls}},
  \bibinfo {author} {\bibfnamefont {M.}~\bibnamefont {Oh}}, \bibinfo {author}
  {\bibfnamefont {B.}~\bibnamefont {Lian}}, \bibinfo {author} {\bibfnamefont
  {Y.}~\bibnamefont {Xie}}, \bibinfo {author} {\bibfnamefont {S.}~\bibnamefont
  {Jeon}}, \bibinfo {author} {\bibfnamefont {K.}~\bibnamefont {Watanabe}},
  \bibinfo {author} {\bibfnamefont {T.}~\bibnamefont {Taniguchi}}, \bibinfo
  {author} {\bibfnamefont {B.~A.}\ \bibnamefont {Bernevig}},\ and\ \bibinfo
  {author} {\bibfnamefont {A.}~\bibnamefont {Yazdani}},\ }\bibfield  {title}
  {\bibinfo {title} {Cascade of electronic transitions in magic-angle twisted
  bilayer graphene},\ }\href@noop {} {\bibfield  {journal} {\bibinfo  {journal}
  {Nature}\ }\textbf {\bibinfo {volume} {582}},\ \bibinfo {pages} {198}
  (\bibinfo {year} {2020})}\BibitemShut {NoStop}%
\bibitem [{\citenamefont {Bultinck}\ \emph {et~al.}(2020)\citenamefont
  {Bultinck}, \citenamefont {Khalaf}, \citenamefont {Liu}, \citenamefont
  {Chatterjee}, \citenamefont {Vishwanath},\ and\ \citenamefont
  {Zaletel}}]{bultinck2020ground}%
  \BibitemOpen
  \bibfield  {author} {\bibinfo {author} {\bibfnamefont {N.}~\bibnamefont
  {Bultinck}}, \bibinfo {author} {\bibfnamefont {E.}~\bibnamefont {Khalaf}},
  \bibinfo {author} {\bibfnamefont {S.}~\bibnamefont {Liu}}, \bibinfo {author}
  {\bibfnamefont {S.}~\bibnamefont {Chatterjee}}, \bibinfo {author}
  {\bibfnamefont {A.}~\bibnamefont {Vishwanath}},\ and\ \bibinfo {author}
  {\bibfnamefont {M.~P.}\ \bibnamefont {Zaletel}},\ }\bibfield  {title}
  {\bibinfo {title} {Ground state and hidden symmetry of magic-angle graphene
  at even integer filling},\ }\href@noop {} {\bibfield  {journal} {\bibinfo
  {journal} {Physical Review X}\ }\textbf {\bibinfo {volume} {10}},\ \bibinfo
  {pages} {031034} (\bibinfo {year} {2020})}\BibitemShut {NoStop}%
\bibitem [{\citenamefont {Park}\ \emph {et~al.}(2021)\citenamefont {Park},
  \citenamefont {Cao}, \citenamefont {Watanabe}, \citenamefont {Taniguchi},\
  and\ \citenamefont {Jarillo-Herrero}}]{park2021tunable}%
  \BibitemOpen
  \bibfield  {author} {\bibinfo {author} {\bibfnamefont {J.~M.}\ \bibnamefont
  {Park}}, \bibinfo {author} {\bibfnamefont {Y.}~\bibnamefont {Cao}}, \bibinfo
  {author} {\bibfnamefont {K.}~\bibnamefont {Watanabe}}, \bibinfo {author}
  {\bibfnamefont {T.}~\bibnamefont {Taniguchi}},\ and\ \bibinfo {author}
  {\bibfnamefont {P.}~\bibnamefont {Jarillo-Herrero}},\ }\bibfield  {title}
  {\bibinfo {title} {Tunable strongly coupled superconductivity in magic-angle
  twisted trilayer graphene},\ }\href@noop {} {\bibfield  {journal} {\bibinfo
  {journal} {Nature}\ }\textbf {\bibinfo {volume} {590}},\ \bibinfo {pages}
  {249} (\bibinfo {year} {2021})}\BibitemShut {NoStop}%
\bibitem [{\citenamefont {Khalaf}\ \emph {et~al.}(2021)\citenamefont {Khalaf},
  \citenamefont {Chatterjee}, \citenamefont {Bultinck}, \citenamefont
  {Zaletel},\ and\ \citenamefont {Vishwanath}}]{khalaf2021charged}%
  \BibitemOpen
  \bibfield  {author} {\bibinfo {author} {\bibfnamefont {E.}~\bibnamefont
  {Khalaf}}, \bibinfo {author} {\bibfnamefont {S.}~\bibnamefont {Chatterjee}},
  \bibinfo {author} {\bibfnamefont {N.}~\bibnamefont {Bultinck}}, \bibinfo
  {author} {\bibfnamefont {M.~P.}\ \bibnamefont {Zaletel}},\ and\ \bibinfo
  {author} {\bibfnamefont {A.}~\bibnamefont {Vishwanath}},\ }\bibfield  {title}
  {\bibinfo {title} {Charged skyrmions and topological origin of
  superconductivity in magic-angle graphene},\ }\href@noop {} {\bibfield
  {journal} {\bibinfo  {journal} {Science advances}\ }\textbf {\bibinfo
  {volume} {7}},\ \bibinfo {pages} {eabf5299} (\bibinfo {year}
  {2021})}\BibitemShut {NoStop}%
\bibitem [{\citenamefont {Wu}\ \emph {et~al.}(2018{\natexlab{b}})\citenamefont
  {Wu}, \citenamefont {Lovorn}, \citenamefont {Tutuc},\ and\ \citenamefont
  {MacDonald}}]{wu2018hubbard}%
  \BibitemOpen
  \bibfield  {author} {\bibinfo {author} {\bibfnamefont {F.}~\bibnamefont
  {Wu}}, \bibinfo {author} {\bibfnamefont {T.}~\bibnamefont {Lovorn}}, \bibinfo
  {author} {\bibfnamefont {E.}~\bibnamefont {Tutuc}},\ and\ \bibinfo {author}
  {\bibfnamefont {A.~H.}\ \bibnamefont {MacDonald}},\ }\bibfield  {title}
  {\bibinfo {title} {Hubbard model physics in transition metal dichalcogenide
  moir{\'e} bands},\ }\href@noop {} {\bibfield  {journal} {\bibinfo  {journal}
  {Physical review letters}\ }\textbf {\bibinfo {volume} {121}},\ \bibinfo
  {pages} {026402} (\bibinfo {year} {2018}{\natexlab{b}})}\BibitemShut
  {NoStop}%
\bibitem [{\citenamefont {Regan}\ \emph {et~al.}(2020)\citenamefont {Regan},
  \citenamefont {Wang}, \citenamefont {Jin}, \citenamefont {Utama},
  \citenamefont {Gao}, \citenamefont {Wei}, \citenamefont {Zhao}, \citenamefont
  {Zhao}, \citenamefont {Zhang}, \citenamefont {Yumigeta} \emph
  {et~al.}}]{regan2020mott}%
  \BibitemOpen
  \bibfield  {author} {\bibinfo {author} {\bibfnamefont {E.~C.}\ \bibnamefont
  {Regan}}, \bibinfo {author} {\bibfnamefont {D.}~\bibnamefont {Wang}},
  \bibinfo {author} {\bibfnamefont {C.}~\bibnamefont {Jin}}, \bibinfo {author}
  {\bibfnamefont {M.~I.~B.}\ \bibnamefont {Utama}}, \bibinfo {author}
  {\bibfnamefont {B.}~\bibnamefont {Gao}}, \bibinfo {author} {\bibfnamefont
  {X.}~\bibnamefont {Wei}}, \bibinfo {author} {\bibfnamefont {S.}~\bibnamefont
  {Zhao}}, \bibinfo {author} {\bibfnamefont {W.}~\bibnamefont {Zhao}}, \bibinfo
  {author} {\bibfnamefont {Z.}~\bibnamefont {Zhang}}, \bibinfo {author}
  {\bibfnamefont {K.}~\bibnamefont {Yumigeta}}, \emph {et~al.},\ }\bibfield
  {title} {\bibinfo {title} {Mott and generalized {Wigner crystal states in
  WSe$_2$/WS$_2$ moir{\'e} superlattices}},\ }\href@noop {} {\bibfield
  {journal} {\bibinfo  {journal} {Nature}\ }\textbf {\bibinfo {volume} {579}},\
  \bibinfo {pages} {359} (\bibinfo {year} {2020})}\BibitemShut {NoStop}%
\bibitem [{\citenamefont {Tang}\ \emph {et~al.}(2020)\citenamefont {Tang},
  \citenamefont {Li}, \citenamefont {Li}, \citenamefont {Xu}, \citenamefont
  {Liu}, \citenamefont {Barmak}, \citenamefont {Watanabe}, \citenamefont
  {Taniguchi}, \citenamefont {MacDonald}, \citenamefont {Shan} \emph
  {et~al.}}]{tang2020simulation}%
  \BibitemOpen
  \bibfield  {author} {\bibinfo {author} {\bibfnamefont {Y.}~\bibnamefont
  {Tang}}, \bibinfo {author} {\bibfnamefont {L.}~\bibnamefont {Li}}, \bibinfo
  {author} {\bibfnamefont {T.}~\bibnamefont {Li}}, \bibinfo {author}
  {\bibfnamefont {Y.}~\bibnamefont {Xu}}, \bibinfo {author} {\bibfnamefont
  {S.}~\bibnamefont {Liu}}, \bibinfo {author} {\bibfnamefont {K.}~\bibnamefont
  {Barmak}}, \bibinfo {author} {\bibfnamefont {K.}~\bibnamefont {Watanabe}},
  \bibinfo {author} {\bibfnamefont {T.}~\bibnamefont {Taniguchi}}, \bibinfo
  {author} {\bibfnamefont {A.~H.}\ \bibnamefont {MacDonald}}, \bibinfo {author}
  {\bibfnamefont {J.}~\bibnamefont {Shan}}, \emph {et~al.},\ }\bibfield
  {title} {\bibinfo {title} {Simulation of hubbard model physics in
  {WSe$_2$/WS$_2$} moir{\'e} superlattices},\ }\href@noop {} {\bibfield
  {journal} {\bibinfo  {journal} {Nature}\ }\textbf {\bibinfo {volume} {579}},\
  \bibinfo {pages} {353} (\bibinfo {year} {2020})}\BibitemShut {NoStop}%
\bibitem [{\citenamefont {Xu}\ \emph {et~al.}(2020)\citenamefont {Xu},
  \citenamefont {Liu}, \citenamefont {Rhodes}, \citenamefont {Watanabe},
  \citenamefont {Taniguchi}, \citenamefont {Hone}, \citenamefont {Elser},
  \citenamefont {Mak},\ and\ \citenamefont {Shan}}]{xu2020correlated}%
  \BibitemOpen
  \bibfield  {author} {\bibinfo {author} {\bibfnamefont {Y.}~\bibnamefont
  {Xu}}, \bibinfo {author} {\bibfnamefont {S.}~\bibnamefont {Liu}}, \bibinfo
  {author} {\bibfnamefont {D.~A.}\ \bibnamefont {Rhodes}}, \bibinfo {author}
  {\bibfnamefont {K.}~\bibnamefont {Watanabe}}, \bibinfo {author}
  {\bibfnamefont {T.}~\bibnamefont {Taniguchi}}, \bibinfo {author}
  {\bibfnamefont {J.}~\bibnamefont {Hone}}, \bibinfo {author} {\bibfnamefont
  {V.}~\bibnamefont {Elser}}, \bibinfo {author} {\bibfnamefont {K.~F.}\
  \bibnamefont {Mak}},\ and\ \bibinfo {author} {\bibfnamefont {J.}~\bibnamefont
  {Shan}},\ }\bibfield  {title} {\bibinfo {title} {Correlated insulating states
  at fractional fillings of moir{\'e} superlattices},\ }\href@noop {}
  {\bibfield  {journal} {\bibinfo  {journal} {Nature}\ }\textbf {\bibinfo
  {volume} {587}},\ \bibinfo {pages} {214} (\bibinfo {year}
  {2020})}\BibitemShut {NoStop}%
\bibitem [{\citenamefont {Wang}\ \emph {et~al.}(2020)\citenamefont {Wang},
  \citenamefont {Shih}, \citenamefont {Ghiotto}, \citenamefont {Xian},
  \citenamefont {Rhodes}, \citenamefont {Tan}, \citenamefont {Claassen},
  \citenamefont {Kennes}, \citenamefont {Bai}, \citenamefont {Kim} \emph
  {et~al.}}]{wang2020correlated}%
  \BibitemOpen
  \bibfield  {author} {\bibinfo {author} {\bibfnamefont {L.}~\bibnamefont
  {Wang}}, \bibinfo {author} {\bibfnamefont {E.-M.}\ \bibnamefont {Shih}},
  \bibinfo {author} {\bibfnamefont {A.}~\bibnamefont {Ghiotto}}, \bibinfo
  {author} {\bibfnamefont {L.}~\bibnamefont {Xian}}, \bibinfo {author}
  {\bibfnamefont {D.~A.}\ \bibnamefont {Rhodes}}, \bibinfo {author}
  {\bibfnamefont {C.}~\bibnamefont {Tan}}, \bibinfo {author} {\bibfnamefont
  {M.}~\bibnamefont {Claassen}}, \bibinfo {author} {\bibfnamefont {D.~M.}\
  \bibnamefont {Kennes}}, \bibinfo {author} {\bibfnamefont {Y.}~\bibnamefont
  {Bai}}, \bibinfo {author} {\bibfnamefont {B.}~\bibnamefont {Kim}}, \emph
  {et~al.},\ }\bibfield  {title} {\bibinfo {title} {Correlated electronic
  phases in twisted bilayer transition metal dichalcogenides},\ }\href@noop {}
  {\bibfield  {journal} {\bibinfo  {journal} {Nature materials}\ }\textbf
  {\bibinfo {volume} {19}},\ \bibinfo {pages} {861} (\bibinfo {year}
  {2020})}\BibitemShut {NoStop}%
\bibitem [{\citenamefont {Kennes}\ \emph {et~al.}(2021)\citenamefont {Kennes},
  \citenamefont {Claassen}, \citenamefont {Xian}, \citenamefont {Georges},
  \citenamefont {Millis}, \citenamefont {Hone}, \citenamefont {Dean},
  \citenamefont {Basov}, \citenamefont {Pasupathy},\ and\ \citenamefont
  {Rubio}}]{kennes2021moire}%
  \BibitemOpen
  \bibfield  {author} {\bibinfo {author} {\bibfnamefont {D.~M.}\ \bibnamefont
  {Kennes}}, \bibinfo {author} {\bibfnamefont {M.}~\bibnamefont {Claassen}},
  \bibinfo {author} {\bibfnamefont {L.}~\bibnamefont {Xian}}, \bibinfo {author}
  {\bibfnamefont {A.}~\bibnamefont {Georges}}, \bibinfo {author} {\bibfnamefont
  {A.~J.}\ \bibnamefont {Millis}}, \bibinfo {author} {\bibfnamefont
  {J.}~\bibnamefont {Hone}}, \bibinfo {author} {\bibfnamefont {C.~R.}\
  \bibnamefont {Dean}}, \bibinfo {author} {\bibfnamefont {D.}~\bibnamefont
  {Basov}}, \bibinfo {author} {\bibfnamefont {A.~N.}\ \bibnamefont
  {Pasupathy}},\ and\ \bibinfo {author} {\bibfnamefont {A.}~\bibnamefont
  {Rubio}},\ }\bibfield  {title} {\bibinfo {title} {Moir{\'e} heterostructures
  as a condensed-matter quantum simulator},\ }\href@noop {} {\bibfield
  {journal} {\bibinfo  {journal} {Nature Physics}\ }\textbf {\bibinfo {volume}
  {17}},\ \bibinfo {pages} {155} (\bibinfo {year} {2021})}\BibitemShut
  {NoStop}%
\bibitem [{\citenamefont {Jin}\ \emph {et~al.}(2021)\citenamefont {Jin},
  \citenamefont {Tao}, \citenamefont {Li}, \citenamefont {Xu}, \citenamefont
  {Tang}, \citenamefont {Zhu}, \citenamefont {Liu}, \citenamefont {Watanabe},
  \citenamefont {Taniguchi}, \citenamefont {Hone} \emph
  {et~al.}}]{jin2021stripe}%
  \BibitemOpen
  \bibfield  {author} {\bibinfo {author} {\bibfnamefont {C.}~\bibnamefont
  {Jin}}, \bibinfo {author} {\bibfnamefont {Z.}~\bibnamefont {Tao}}, \bibinfo
  {author} {\bibfnamefont {T.}~\bibnamefont {Li}}, \bibinfo {author}
  {\bibfnamefont {Y.}~\bibnamefont {Xu}}, \bibinfo {author} {\bibfnamefont
  {Y.}~\bibnamefont {Tang}}, \bibinfo {author} {\bibfnamefont {J.}~\bibnamefont
  {Zhu}}, \bibinfo {author} {\bibfnamefont {S.}~\bibnamefont {Liu}}, \bibinfo
  {author} {\bibfnamefont {K.}~\bibnamefont {Watanabe}}, \bibinfo {author}
  {\bibfnamefont {T.}~\bibnamefont {Taniguchi}}, \bibinfo {author}
  {\bibfnamefont {J.~C.}\ \bibnamefont {Hone}}, \emph {et~al.},\ }\bibfield
  {title} {\bibinfo {title} {Stripe phases in $\rm{WSe_2/WS_2}$ moir{\'e}
  superlattices},\ }\href@noop {} {\bibfield  {journal} {\bibinfo  {journal}
  {Nature Materials}\ ,\ \bibinfo {pages} {1}} (\bibinfo {year}
  {2021})}\BibitemShut {NoStop}%
\bibitem [{\citenamefont {Huang}\ \emph {et~al.}(2021)\citenamefont {Huang},
  \citenamefont {Wang}, \citenamefont {Miao}, \citenamefont {Wang},
  \citenamefont {Li}, \citenamefont {Lian}, \citenamefont {Taniguchi},
  \citenamefont {Watanabe}, \citenamefont {Okamoto}, \citenamefont {Xiao} \emph
  {et~al.}}]{huang2021correlated}%
  \BibitemOpen
  \bibfield  {author} {\bibinfo {author} {\bibfnamefont {X.}~\bibnamefont
  {Huang}}, \bibinfo {author} {\bibfnamefont {T.}~\bibnamefont {Wang}},
  \bibinfo {author} {\bibfnamefont {S.}~\bibnamefont {Miao}}, \bibinfo {author}
  {\bibfnamefont {C.}~\bibnamefont {Wang}}, \bibinfo {author} {\bibfnamefont
  {Z.}~\bibnamefont {Li}}, \bibinfo {author} {\bibfnamefont {Z.}~\bibnamefont
  {Lian}}, \bibinfo {author} {\bibfnamefont {T.}~\bibnamefont {Taniguchi}},
  \bibinfo {author} {\bibfnamefont {K.}~\bibnamefont {Watanabe}}, \bibinfo
  {author} {\bibfnamefont {S.}~\bibnamefont {Okamoto}}, \bibinfo {author}
  {\bibfnamefont {D.}~\bibnamefont {Xiao}}, \emph {et~al.},\ }\bibfield
  {title} {\bibinfo {title} {Correlated insulating states at fractional
  fillings of the $\rm{WS_2/WSe_2}$ moir{\'e} lattice},\ }\href@noop {}
  {\bibfield  {journal} {\bibinfo  {journal} {Nature Physics}\ }\textbf
  {\bibinfo {volume} {17}},\ \bibinfo {pages} {715} (\bibinfo {year}
  {2021})}\BibitemShut {NoStop}%
\bibitem [{\citenamefont {Chu}\ \emph {et~al.}(2020)\citenamefont {Chu},
  \citenamefont {Regan}, \citenamefont {Ma}, \citenamefont {Wang},
  \citenamefont {Xu}, \citenamefont {Utama}, \citenamefont {Yumigeta},
  \citenamefont {Blei}, \citenamefont {Watanabe}, \citenamefont {Taniguchi}
  \emph {et~al.}}]{chu2020nanoscale}%
  \BibitemOpen
  \bibfield  {author} {\bibinfo {author} {\bibfnamefont {Z.}~\bibnamefont
  {Chu}}, \bibinfo {author} {\bibfnamefont {E.~C.}\ \bibnamefont {Regan}},
  \bibinfo {author} {\bibfnamefont {X.}~\bibnamefont {Ma}}, \bibinfo {author}
  {\bibfnamefont {D.}~\bibnamefont {Wang}}, \bibinfo {author} {\bibfnamefont
  {Z.}~\bibnamefont {Xu}}, \bibinfo {author} {\bibfnamefont {M.~I.~B.}\
  \bibnamefont {Utama}}, \bibinfo {author} {\bibfnamefont {K.}~\bibnamefont
  {Yumigeta}}, \bibinfo {author} {\bibfnamefont {M.}~\bibnamefont {Blei}},
  \bibinfo {author} {\bibfnamefont {K.}~\bibnamefont {Watanabe}}, \bibinfo
  {author} {\bibfnamefont {T.}~\bibnamefont {Taniguchi}}, \emph {et~al.},\
  }\bibfield  {title} {\bibinfo {title} {Nanoscale conductivity imaging of
  correlated electronic states in $\rm{Wse_2/WS_2}$ moir{\'e} superlattices},\
  }\href@noop {} {\bibfield  {journal} {\bibinfo  {journal} {Physical review
  letters}\ }\textbf {\bibinfo {volume} {125}},\ \bibinfo {pages} {186803}
  (\bibinfo {year} {2020})}\BibitemShut {NoStop}%
\bibitem [{\citenamefont {Pan}\ \emph {et~al.}(2020)\citenamefont {Pan},
  \citenamefont {Wu},\ and\ \citenamefont {Sarma}}]{pan2020quantum}%
  \BibitemOpen
  \bibfield  {author} {\bibinfo {author} {\bibfnamefont {H.}~\bibnamefont
  {Pan}}, \bibinfo {author} {\bibfnamefont {F.}~\bibnamefont {Wu}},\ and\
  \bibinfo {author} {\bibfnamefont {S.~D.}\ \bibnamefont {Sarma}},\ }\bibfield
  {title} {\bibinfo {title} {Quantum phase diagram of a {Moir{\'e}-Hubbard}
  model},\ }\href@noop {} {\bibfield  {journal} {\bibinfo  {journal} {Physical
  Review B}\ }\textbf {\bibinfo {volume} {102}},\ \bibinfo {pages} {201104}
  (\bibinfo {year} {2020})}\BibitemShut {NoStop}%
\bibitem [{\citenamefont {Morales-Dur{\'a}n}\ \emph {et~al.}(2021)\citenamefont
  {Morales-Dur{\'a}n}, \citenamefont {MacDonald},\ and\ \citenamefont
  {Potasz}}]{morales2021metal}%
  \BibitemOpen
  \bibfield  {author} {\bibinfo {author} {\bibfnamefont {N.}~\bibnamefont
  {Morales-Dur{\'a}n}}, \bibinfo {author} {\bibfnamefont {A.~H.}\ \bibnamefont
  {MacDonald}},\ and\ \bibinfo {author} {\bibfnamefont {P.}~\bibnamefont
  {Potasz}},\ }\bibfield  {title} {\bibinfo {title} {Metal-insulator transition
  in transition metal dichalcogenide heterobilayer moir{\'e} superlattices},\
  }\href@noop {} {\bibfield  {journal} {\bibinfo  {journal} {Physical Review
  B}\ }\textbf {\bibinfo {volume} {103}},\ \bibinfo {pages} {L241110} (\bibinfo
  {year} {2021})}\BibitemShut {NoStop}%
\bibitem [{\citenamefont {Zhang}\ \emph {et~al.}(2021)\citenamefont {Zhang},
  \citenamefont {Liu},\ and\ \citenamefont {Fu}}]{zhang2021electronic}%
  \BibitemOpen
  \bibfield  {author} {\bibinfo {author} {\bibfnamefont {Y.}~\bibnamefont
  {Zhang}}, \bibinfo {author} {\bibfnamefont {T.}~\bibnamefont {Liu}},\ and\
  \bibinfo {author} {\bibfnamefont {L.}~\bibnamefont {Fu}},\ }\bibfield
  {title} {\bibinfo {title} {Electronic structures, charge transfer, and charge
  order in twisted transition metal dichalcogenide bilayers},\ }\href@noop {}
  {\bibfield  {journal} {\bibinfo  {journal} {Physical Review B}\ }\textbf
  {\bibinfo {volume} {103}},\ \bibinfo {pages} {155142} (\bibinfo {year}
  {2021})}\BibitemShut {NoStop}%
\bibitem [{\citenamefont {Bi}\ and\ \citenamefont
  {Fu}(2021)}]{bi2021excitonic}%
  \BibitemOpen
  \bibfield  {author} {\bibinfo {author} {\bibfnamefont {Z.}~\bibnamefont
  {Bi}}\ and\ \bibinfo {author} {\bibfnamefont {L.}~\bibnamefont {Fu}},\
  }\bibfield  {title} {\bibinfo {title} {Excitonic density wave and spin-valley
  superfluid in bilayer transition metal dichalcogenide},\ }\href@noop {}
  {\bibfield  {journal} {\bibinfo  {journal} {Nature communications}\ }\textbf
  {\bibinfo {volume} {12}},\ \bibinfo {pages} {1} (\bibinfo {year}
  {2021})}\BibitemShut {NoStop}%
\bibitem [{\citenamefont {Li}\ \emph {et~al.}(2021{\natexlab{a}})\citenamefont
  {Li}, \citenamefont {Li}, \citenamefont {Regan}, \citenamefont {Wang},
  \citenamefont {Zhao}, \citenamefont {Kahn}, \citenamefont {Yumigeta},
  \citenamefont {Blei}, \citenamefont {Taniguchi}, \citenamefont {Watanabe}
  \emph {et~al.}}]{li2021imaging}%
  \BibitemOpen
  \bibfield  {author} {\bibinfo {author} {\bibfnamefont {H.}~\bibnamefont
  {Li}}, \bibinfo {author} {\bibfnamefont {S.}~\bibnamefont {Li}}, \bibinfo
  {author} {\bibfnamefont {E.~C.}\ \bibnamefont {Regan}}, \bibinfo {author}
  {\bibfnamefont {D.}~\bibnamefont {Wang}}, \bibinfo {author} {\bibfnamefont
  {W.}~\bibnamefont {Zhao}}, \bibinfo {author} {\bibfnamefont {S.}~\bibnamefont
  {Kahn}}, \bibinfo {author} {\bibfnamefont {K.}~\bibnamefont {Yumigeta}},
  \bibinfo {author} {\bibfnamefont {M.}~\bibnamefont {Blei}}, \bibinfo {author}
  {\bibfnamefont {T.}~\bibnamefont {Taniguchi}}, \bibinfo {author}
  {\bibfnamefont {K.}~\bibnamefont {Watanabe}}, \emph {et~al.},\ }\bibfield
  {title} {\bibinfo {title} {Imaging generalized wigner crystal states in a
  {WSe$_2$/WS$_2$} moir\'e superlattice},\ }\href@noop {} {\bibfield  {journal}
  {\bibinfo  {journal} {arXiv preprint arXiv:2106.10599}\ } (\bibinfo {year}
  {2021}{\natexlab{a}})}\BibitemShut {NoStop}%
\bibitem [{\citenamefont {Angeli}\ and\ \citenamefont
  {MacDonald}(2021)}]{angeli2021gamma}%
  \BibitemOpen
  \bibfield  {author} {\bibinfo {author} {\bibfnamefont {M.}~\bibnamefont
  {Angeli}}\ and\ \bibinfo {author} {\bibfnamefont {A.~H.}\ \bibnamefont
  {MacDonald}},\ }\bibfield  {title} {\bibinfo {title} {{$\Gamma$} valley
  transition metal dichalcogenide moir{\'e} bands},\ }\href@noop {} {\bibfield
  {journal} {\bibinfo  {journal} {Proceedings of the National Academy of
  Sciences}\ }\textbf {\bibinfo {volume} {118}} (\bibinfo {year}
  {2021})}\BibitemShut {NoStop}%
\bibitem [{\citenamefont {Regan}\ \emph {et~al.}(2022)\citenamefont {Regan},
  \citenamefont {Wang}, \citenamefont {Paik}, \citenamefont {Zeng},
  \citenamefont {Zhang}, \citenamefont {Zhu}, \citenamefont {MacDonald},
  \citenamefont {Deng},\ and\ \citenamefont {Wang}}]{regan2022emerging}%
  \BibitemOpen
  \bibfield  {author} {\bibinfo {author} {\bibfnamefont {E.~C.}\ \bibnamefont
  {Regan}}, \bibinfo {author} {\bibfnamefont {D.}~\bibnamefont {Wang}},
  \bibinfo {author} {\bibfnamefont {E.~Y.}\ \bibnamefont {Paik}}, \bibinfo
  {author} {\bibfnamefont {Y.}~\bibnamefont {Zeng}}, \bibinfo {author}
  {\bibfnamefont {L.}~\bibnamefont {Zhang}}, \bibinfo {author} {\bibfnamefont
  {J.}~\bibnamefont {Zhu}}, \bibinfo {author} {\bibfnamefont {A.~H.}\
  \bibnamefont {MacDonald}}, \bibinfo {author} {\bibfnamefont {H.}~\bibnamefont
  {Deng}},\ and\ \bibinfo {author} {\bibfnamefont {F.}~\bibnamefont {Wang}},\
  }\bibfield  {title} {\bibinfo {title} {Emerging exciton physics in transition
  metal dichalcogenide heterobilayers},\ }\href@noop {} {\bibfield  {journal}
  {\bibinfo  {journal} {Nature Reviews Materials}\ ,\ \bibinfo {pages} {1}}
  (\bibinfo {year} {2022})}\BibitemShut {NoStop}%
\bibitem [{\citenamefont {Wu}\ \emph {et~al.}(2017)\citenamefont {Wu},
  \citenamefont {Lovorn},\ and\ \citenamefont {MacDonald}}]{wu2017topological}%
  \BibitemOpen
  \bibfield  {author} {\bibinfo {author} {\bibfnamefont {F.}~\bibnamefont
  {Wu}}, \bibinfo {author} {\bibfnamefont {T.}~\bibnamefont {Lovorn}},\ and\
  \bibinfo {author} {\bibfnamefont {A.~H.}\ \bibnamefont {MacDonald}},\
  }\bibfield  {title} {\bibinfo {title} {Topological exciton bands in moir{\'e}
  heterojunctions},\ }\href@noop {} {\bibfield  {journal} {\bibinfo  {journal}
  {Physical review letters}\ }\textbf {\bibinfo {volume} {118}},\ \bibinfo
  {pages} {147401} (\bibinfo {year} {2017})}\BibitemShut {NoStop}%
\bibitem [{\citenamefont {Wu}\ \emph {et~al.}(2018{\natexlab{c}})\citenamefont
  {Wu}, \citenamefont {Lovorn},\ and\ \citenamefont
  {MacDonald}}]{wu2018theory}%
  \BibitemOpen
  \bibfield  {author} {\bibinfo {author} {\bibfnamefont {F.}~\bibnamefont
  {Wu}}, \bibinfo {author} {\bibfnamefont {T.}~\bibnamefont {Lovorn}},\ and\
  \bibinfo {author} {\bibfnamefont {A.}~\bibnamefont {MacDonald}},\ }\bibfield
  {title} {\bibinfo {title} {Theory of optical absorption by interlayer
  excitons in transition metal dichalcogenide heterobilayers},\ }\href@noop {}
  {\bibfield  {journal} {\bibinfo  {journal} {Physical Review B}\ }\textbf
  {\bibinfo {volume} {97}},\ \bibinfo {pages} {035306} (\bibinfo {year}
  {2018}{\natexlab{c}})}\BibitemShut {NoStop}%
\bibitem [{\citenamefont {Yu}\ \emph {et~al.}(2017)\citenamefont {Yu},
  \citenamefont {Liu}, \citenamefont {Tang}, \citenamefont {Xu},\ and\
  \citenamefont {Yao}}]{yu2017moire}%
  \BibitemOpen
  \bibfield  {author} {\bibinfo {author} {\bibfnamefont {H.}~\bibnamefont
  {Yu}}, \bibinfo {author} {\bibfnamefont {G.-B.}\ \bibnamefont {Liu}},
  \bibinfo {author} {\bibfnamefont {J.}~\bibnamefont {Tang}}, \bibinfo {author}
  {\bibfnamefont {X.}~\bibnamefont {Xu}},\ and\ \bibinfo {author}
  {\bibfnamefont {W.}~\bibnamefont {Yao}},\ }\bibfield  {title} {\bibinfo
  {title} {Moir{\'e} excitons: From programmable quantum emitter arrays to
  spin-orbit--coupled artificial lattices},\ }\href@noop {} {\bibfield
  {journal} {\bibinfo  {journal} {Science advances}\ }\textbf {\bibinfo
  {volume} {3}},\ \bibinfo {pages} {e1701696} (\bibinfo {year}
  {2017})}\BibitemShut {NoStop}%
\bibitem [{\citenamefont {Tran}\ \emph {et~al.}(2019)\citenamefont {Tran},
  \citenamefont {Moody}, \citenamefont {Wu}, \citenamefont {Lu}, \citenamefont
  {Choi}, \citenamefont {Kim}, \citenamefont {Rai}, \citenamefont {Sanchez},
  \citenamefont {Quan}, \citenamefont {Singh} \emph
  {et~al.}}]{tran2019evidence}%
  \BibitemOpen
  \bibfield  {author} {\bibinfo {author} {\bibfnamefont {K.}~\bibnamefont
  {Tran}}, \bibinfo {author} {\bibfnamefont {G.}~\bibnamefont {Moody}},
  \bibinfo {author} {\bibfnamefont {F.}~\bibnamefont {Wu}}, \bibinfo {author}
  {\bibfnamefont {X.}~\bibnamefont {Lu}}, \bibinfo {author} {\bibfnamefont
  {J.}~\bibnamefont {Choi}}, \bibinfo {author} {\bibfnamefont {K.}~\bibnamefont
  {Kim}}, \bibinfo {author} {\bibfnamefont {A.}~\bibnamefont {Rai}}, \bibinfo
  {author} {\bibfnamefont {D.~A.}\ \bibnamefont {Sanchez}}, \bibinfo {author}
  {\bibfnamefont {J.}~\bibnamefont {Quan}}, \bibinfo {author} {\bibfnamefont
  {A.}~\bibnamefont {Singh}}, \emph {et~al.},\ }\bibfield  {title} {\bibinfo
  {title} {Evidence for moir{\'e} excitons in van der {Waals}
  heterostructures},\ }\href@noop {} {\bibfield  {journal} {\bibinfo  {journal}
  {Nature}\ }\textbf {\bibinfo {volume} {567}},\ \bibinfo {pages} {71}
  (\bibinfo {year} {2019})}\BibitemShut {NoStop}%
\bibitem [{\citenamefont {Jin}\ \emph {et~al.}(2019)\citenamefont {Jin},
  \citenamefont {Regan}, \citenamefont {Yan}, \citenamefont {Utama},
  \citenamefont {Wang}, \citenamefont {Zhao}, \citenamefont {Qin},
  \citenamefont {Yang}, \citenamefont {Zheng}, \citenamefont {Shi} \emph
  {et~al.}}]{jin2019observation}%
  \BibitemOpen
  \bibfield  {author} {\bibinfo {author} {\bibfnamefont {C.}~\bibnamefont
  {Jin}}, \bibinfo {author} {\bibfnamefont {E.~C.}\ \bibnamefont {Regan}},
  \bibinfo {author} {\bibfnamefont {A.}~\bibnamefont {Yan}}, \bibinfo {author}
  {\bibfnamefont {M.~I.~B.}\ \bibnamefont {Utama}}, \bibinfo {author}
  {\bibfnamefont {D.}~\bibnamefont {Wang}}, \bibinfo {author} {\bibfnamefont
  {S.}~\bibnamefont {Zhao}}, \bibinfo {author} {\bibfnamefont {Y.}~\bibnamefont
  {Qin}}, \bibinfo {author} {\bibfnamefont {S.}~\bibnamefont {Yang}}, \bibinfo
  {author} {\bibfnamefont {Z.}~\bibnamefont {Zheng}}, \bibinfo {author}
  {\bibfnamefont {S.}~\bibnamefont {Shi}}, \emph {et~al.},\ }\bibfield  {title}
  {\bibinfo {title} {Observation of moir{\'e} excitons in {WSe$_2$/WS$_2$}
  heterostructure superlattices},\ }\href@noop {} {\bibfield  {journal}
  {\bibinfo  {journal} {Nature}\ }\textbf {\bibinfo {volume} {567}},\ \bibinfo
  {pages} {76} (\bibinfo {year} {2019})}\BibitemShut {NoStop}%
\bibitem [{\citenamefont {Seyler}\ \emph {et~al.}(2019)\citenamefont {Seyler},
  \citenamefont {Rivera}, \citenamefont {Yu}, \citenamefont {Wilson},
  \citenamefont {Ray}, \citenamefont {Mandrus}, \citenamefont {Yan},
  \citenamefont {Yao},\ and\ \citenamefont {Xu}}]{seyler2019signatures}%
  \BibitemOpen
  \bibfield  {author} {\bibinfo {author} {\bibfnamefont {K.~L.}\ \bibnamefont
  {Seyler}}, \bibinfo {author} {\bibfnamefont {P.}~\bibnamefont {Rivera}},
  \bibinfo {author} {\bibfnamefont {H.}~\bibnamefont {Yu}}, \bibinfo {author}
  {\bibfnamefont {N.~P.}\ \bibnamefont {Wilson}}, \bibinfo {author}
  {\bibfnamefont {E.~L.}\ \bibnamefont {Ray}}, \bibinfo {author} {\bibfnamefont
  {D.~G.}\ \bibnamefont {Mandrus}}, \bibinfo {author} {\bibfnamefont
  {J.}~\bibnamefont {Yan}}, \bibinfo {author} {\bibfnamefont {W.}~\bibnamefont
  {Yao}},\ and\ \bibinfo {author} {\bibfnamefont {X.}~\bibnamefont {Xu}},\
  }\bibfield  {title} {\bibinfo {title} {Signatures of moir{\'e}-trapped valley
  excitons in {MoSe$_2$/WSe$_2$} heterobilayers},\ }\href@noop {} {\bibfield
  {journal} {\bibinfo  {journal} {Nature}\ }\textbf {\bibinfo {volume} {567}},\
  \bibinfo {pages} {66} (\bibinfo {year} {2019})}\BibitemShut {NoStop}%
\bibitem [{\citenamefont {Alexeev}\ \emph {et~al.}(2019)\citenamefont
  {Alexeev}, \citenamefont {Ruiz-Tijerina}, \citenamefont {Danovich},
  \citenamefont {Hamer}, \citenamefont {Terry}, \citenamefont {Nayak},
  \citenamefont {Ahn}, \citenamefont {Pak}, \citenamefont {Lee}, \citenamefont
  {Sohn} \emph {et~al.}}]{alexeev2019resonantly}%
  \BibitemOpen
  \bibfield  {author} {\bibinfo {author} {\bibfnamefont {E.~M.}\ \bibnamefont
  {Alexeev}}, \bibinfo {author} {\bibfnamefont {D.~A.}\ \bibnamefont
  {Ruiz-Tijerina}}, \bibinfo {author} {\bibfnamefont {M.}~\bibnamefont
  {Danovich}}, \bibinfo {author} {\bibfnamefont {M.~J.}\ \bibnamefont {Hamer}},
  \bibinfo {author} {\bibfnamefont {D.~J.}\ \bibnamefont {Terry}}, \bibinfo
  {author} {\bibfnamefont {P.~K.}\ \bibnamefont {Nayak}}, \bibinfo {author}
  {\bibfnamefont {S.}~\bibnamefont {Ahn}}, \bibinfo {author} {\bibfnamefont
  {S.}~\bibnamefont {Pak}}, \bibinfo {author} {\bibfnamefont {J.}~\bibnamefont
  {Lee}}, \bibinfo {author} {\bibfnamefont {J.~I.}\ \bibnamefont {Sohn}}, \emph
  {et~al.},\ }\bibfield  {title} {\bibinfo {title} {Resonantly hybridized
  excitons in moir{\'e} superlattices in van der {Waals} heterostructures},\
  }\href@noop {} {\bibfield  {journal} {\bibinfo  {journal} {Nature}\ }\textbf
  {\bibinfo {volume} {567}},\ \bibinfo {pages} {81} (\bibinfo {year}
  {2019})}\BibitemShut {NoStop}%
\bibitem [{\citenamefont {Zhang}\ \emph {et~al.}(2018)\citenamefont {Zhang},
  \citenamefont {Surrente}, \citenamefont {Baranowski}, \citenamefont {Maude},
  \citenamefont {Gant}, \citenamefont {Castellanos-Gomez},\ and\ \citenamefont
  {Plochocka}}]{zhang2018moire}%
  \BibitemOpen
  \bibfield  {author} {\bibinfo {author} {\bibfnamefont {N.}~\bibnamefont
  {Zhang}}, \bibinfo {author} {\bibfnamefont {A.}~\bibnamefont {Surrente}},
  \bibinfo {author} {\bibfnamefont {M.}~\bibnamefont {Baranowski}}, \bibinfo
  {author} {\bibfnamefont {D.~K.}\ \bibnamefont {Maude}}, \bibinfo {author}
  {\bibfnamefont {P.}~\bibnamefont {Gant}}, \bibinfo {author} {\bibfnamefont
  {A.}~\bibnamefont {Castellanos-Gomez}},\ and\ \bibinfo {author}
  {\bibfnamefont {P.}~\bibnamefont {Plochocka}},\ }\bibfield  {title} {\bibinfo
  {title} {Moir{\'e} intralayer excitons in a $\rm{MoSe_2/MoS_2}$
  heterostructure},\ }\href@noop {} {\bibfield  {journal} {\bibinfo  {journal}
  {Nano letters}\ }\textbf {\bibinfo {volume} {18}},\ \bibinfo {pages} {7651}
  (\bibinfo {year} {2018})}\BibitemShut {NoStop}%
\bibitem [{\citenamefont {Shimazaki}\ \emph {et~al.}(2020)\citenamefont
  {Shimazaki}, \citenamefont {Schwartz}, \citenamefont {Watanabe},
  \citenamefont {Taniguchi}, \citenamefont {Kroner},\ and\ \citenamefont
  {Imamo{\u{g}}lu}}]{shimazaki2020strongly}%
  \BibitemOpen
  \bibfield  {author} {\bibinfo {author} {\bibfnamefont {Y.}~\bibnamefont
  {Shimazaki}}, \bibinfo {author} {\bibfnamefont {I.}~\bibnamefont {Schwartz}},
  \bibinfo {author} {\bibfnamefont {K.}~\bibnamefont {Watanabe}}, \bibinfo
  {author} {\bibfnamefont {T.}~\bibnamefont {Taniguchi}}, \bibinfo {author}
  {\bibfnamefont {M.}~\bibnamefont {Kroner}},\ and\ \bibinfo {author}
  {\bibfnamefont {A.}~\bibnamefont {Imamo{\u{g}}lu}},\ }\bibfield  {title}
  {\bibinfo {title} {Strongly correlated electrons and hybrid excitons in a
  moir{\'e} heterostructure},\ }\href@noop {} {\bibfield  {journal} {\bibinfo
  {journal} {Nature}\ }\textbf {\bibinfo {volume} {580}},\ \bibinfo {pages}
  {472} (\bibinfo {year} {2020})}\BibitemShut {NoStop}%
\bibitem [{\citenamefont {Zhang}\ \emph {et~al.}(2020)\citenamefont {Zhang},
  \citenamefont {Zhang}, \citenamefont {Wu}, \citenamefont {Wang},
  \citenamefont {Gogna}, \citenamefont {Hou}, \citenamefont {Watanabe},
  \citenamefont {Taniguchi}, \citenamefont {Kulkarni}, \citenamefont {Kuo}
  \emph {et~al.}}]{zhang2020twist}%
  \BibitemOpen
  \bibfield  {author} {\bibinfo {author} {\bibfnamefont {L.}~\bibnamefont
  {Zhang}}, \bibinfo {author} {\bibfnamefont {Z.}~\bibnamefont {Zhang}},
  \bibinfo {author} {\bibfnamefont {F.}~\bibnamefont {Wu}}, \bibinfo {author}
  {\bibfnamefont {D.}~\bibnamefont {Wang}}, \bibinfo {author} {\bibfnamefont
  {R.}~\bibnamefont {Gogna}}, \bibinfo {author} {\bibfnamefont
  {S.}~\bibnamefont {Hou}}, \bibinfo {author} {\bibfnamefont {K.}~\bibnamefont
  {Watanabe}}, \bibinfo {author} {\bibfnamefont {T.}~\bibnamefont {Taniguchi}},
  \bibinfo {author} {\bibfnamefont {K.}~\bibnamefont {Kulkarni}}, \bibinfo
  {author} {\bibfnamefont {T.}~\bibnamefont {Kuo}}, \emph {et~al.},\ }\bibfield
   {title} {\bibinfo {title} {Twist-angle dependence of moir{\'e} excitons in
  {WS$_2$/MoSe$_2$} heterobilayers},\ }\href@noop {} {\bibfield  {journal}
  {\bibinfo  {journal} {Nature communications}\ }\textbf {\bibinfo {volume}
  {11}},\ \bibinfo {pages} {1} (\bibinfo {year} {2020})}\BibitemShut {NoStop}%
\bibitem [{\citenamefont {Baek}\ \emph {et~al.}(2020)\citenamefont {Baek},
  \citenamefont {Brotons-Gisbert}, \citenamefont {Koong}, \citenamefont
  {Campbell}, \citenamefont {Rambach}, \citenamefont {Watanabe}, \citenamefont
  {Taniguchi},\ and\ \citenamefont {Gerardot}}]{baek2020highly}%
  \BibitemOpen
  \bibfield  {author} {\bibinfo {author} {\bibfnamefont {H.}~\bibnamefont
  {Baek}}, \bibinfo {author} {\bibfnamefont {M.}~\bibnamefont
  {Brotons-Gisbert}}, \bibinfo {author} {\bibfnamefont {Z.}~\bibnamefont
  {Koong}}, \bibinfo {author} {\bibfnamefont {A.}~\bibnamefont {Campbell}},
  \bibinfo {author} {\bibfnamefont {M.}~\bibnamefont {Rambach}}, \bibinfo
  {author} {\bibfnamefont {K.}~\bibnamefont {Watanabe}}, \bibinfo {author}
  {\bibfnamefont {T.}~\bibnamefont {Taniguchi}},\ and\ \bibinfo {author}
  {\bibfnamefont {B.~D.}\ \bibnamefont {Gerardot}},\ }\bibfield  {title}
  {\bibinfo {title} {Highly energy-tunable quantum light from moir{\'e}-trapped
  excitons},\ }\href@noop {} {\bibfield  {journal} {\bibinfo  {journal}
  {Science advances}\ }\textbf {\bibinfo {volume} {6}},\ \bibinfo {pages}
  {eaba8526} (\bibinfo {year} {2020})}\BibitemShut {NoStop}%
\bibitem [{\citenamefont {Ruiz-Tijerina}\ \emph {et~al.}(2020)\citenamefont
  {Ruiz-Tijerina}, \citenamefont {Soltero},\ and\ \citenamefont
  {Mireles}}]{ruiz2020theory}%
  \BibitemOpen
  \bibfield  {author} {\bibinfo {author} {\bibfnamefont {D.~A.}\ \bibnamefont
  {Ruiz-Tijerina}}, \bibinfo {author} {\bibfnamefont {I.}~\bibnamefont
  {Soltero}},\ and\ \bibinfo {author} {\bibfnamefont {F.}~\bibnamefont
  {Mireles}},\ }\bibfield  {title} {\bibinfo {title} {Theory of moir{\'e}
  localized excitons in transition metal dichalcogenide heterobilayers},\
  }\href@noop {} {\bibfield  {journal} {\bibinfo  {journal} {Physical Review
  B}\ }\textbf {\bibinfo {volume} {102}},\ \bibinfo {pages} {195403} (\bibinfo
  {year} {2020})}\BibitemShut {NoStop}%
\bibitem [{\citenamefont {Liu}\ \emph {et~al.}(2021)\citenamefont {Liu},
  \citenamefont {Barr{\'e}}, \citenamefont {van Baren}, \citenamefont {Wilson},
  \citenamefont {Taniguchi}, \citenamefont {Watanabe}, \citenamefont {Cui},
  \citenamefont {Gabor}, \citenamefont {Heinz}, \citenamefont {Chang} \emph
  {et~al.}}]{liu2021signatures}%
  \BibitemOpen
  \bibfield  {author} {\bibinfo {author} {\bibfnamefont {E.}~\bibnamefont
  {Liu}}, \bibinfo {author} {\bibfnamefont {E.}~\bibnamefont {Barr{\'e}}},
  \bibinfo {author} {\bibfnamefont {J.}~\bibnamefont {van Baren}}, \bibinfo
  {author} {\bibfnamefont {M.}~\bibnamefont {Wilson}}, \bibinfo {author}
  {\bibfnamefont {T.}~\bibnamefont {Taniguchi}}, \bibinfo {author}
  {\bibfnamefont {K.}~\bibnamefont {Watanabe}}, \bibinfo {author}
  {\bibfnamefont {Y.-T.}\ \bibnamefont {Cui}}, \bibinfo {author} {\bibfnamefont
  {N.~M.}\ \bibnamefont {Gabor}}, \bibinfo {author} {\bibfnamefont {T.~F.}\
  \bibnamefont {Heinz}}, \bibinfo {author} {\bibfnamefont {Y.-C.}\ \bibnamefont
  {Chang}}, \emph {et~al.},\ }\bibfield  {title} {\bibinfo {title} {Signatures
  of moir{\'e} trions in {WSe$_2$/MoSe$_2$} heterobilayers},\ }\href@noop {}
  {\bibfield  {journal} {\bibinfo  {journal} {Nature}\ }\textbf {\bibinfo
  {volume} {594}},\ \bibinfo {pages} {46} (\bibinfo {year} {2021})}\BibitemShut
  {NoStop}%
\bibitem [{\citenamefont {Brotons-Gisbert}\ \emph {et~al.}(2021)\citenamefont
  {Brotons-Gisbert}, \citenamefont {Baek}, \citenamefont {Campbell},
  \citenamefont {Watanabe}, \citenamefont {Taniguchi},\ and\ \citenamefont
  {Gerardot}}]{brotons2021moir}%
  \BibitemOpen
  \bibfield  {author} {\bibinfo {author} {\bibfnamefont {M.}~\bibnamefont
  {Brotons-Gisbert}}, \bibinfo {author} {\bibfnamefont {H.}~\bibnamefont
  {Baek}}, \bibinfo {author} {\bibfnamefont {A.}~\bibnamefont {Campbell}},
  \bibinfo {author} {\bibfnamefont {K.}~\bibnamefont {Watanabe}}, \bibinfo
  {author} {\bibfnamefont {T.}~\bibnamefont {Taniguchi}},\ and\ \bibinfo
  {author} {\bibfnamefont {B.~D.}\ \bibnamefont {Gerardot}},\ }\bibfield
  {title} {\bibinfo {title} {Moir\'e-trapped interlayer trions in a
  charge-tunable {WSe$_2$/MoSe$_2$} heterobilayer},\ }\href@noop {} {\bibfield
  {journal} {\bibinfo  {journal} {arXiv preprint arXiv:2101.07747}\ } (\bibinfo
  {year} {2021})}\BibitemShut {NoStop}%
\bibitem [{\citenamefont {Baek}\ \emph {et~al.}(2021)\citenamefont {Baek},
  \citenamefont {Brotons-Gisbert}, \citenamefont {Campbell}, \citenamefont
  {Watanabe}, \citenamefont {Taniguchi},\ and\ \citenamefont
  {Gerardot}}]{baek2021optical}%
  \BibitemOpen
  \bibfield  {author} {\bibinfo {author} {\bibfnamefont {H.}~\bibnamefont
  {Baek}}, \bibinfo {author} {\bibfnamefont {M.}~\bibnamefont
  {Brotons-Gisbert}}, \bibinfo {author} {\bibfnamefont {A.}~\bibnamefont
  {Campbell}}, \bibinfo {author} {\bibfnamefont {K.}~\bibnamefont {Watanabe}},
  \bibinfo {author} {\bibfnamefont {T.}~\bibnamefont {Taniguchi}},\ and\
  \bibinfo {author} {\bibfnamefont {B.~D.}\ \bibnamefont {Gerardot}},\
  }\bibfield  {title} {\bibinfo {title} {Optical read-out of {Coulomb}
  staircases in a moir\'e superlattice via trapped interlayer trions},\
  }\href@noop {} {\bibfield  {journal} {\bibinfo  {journal} {arXiv preprint
  arXiv:2102.01358}\ } (\bibinfo {year} {2021})}\BibitemShut {NoStop}%
\bibitem [{\citenamefont {Brem}\ \emph {et~al.}(2020)\citenamefont {Brem},
  \citenamefont {Linder{\"a}lv}, \citenamefont {Erhart},\ and\ \citenamefont
  {Malic}}]{brem2020tunable}%
  \BibitemOpen
  \bibfield  {author} {\bibinfo {author} {\bibfnamefont {S.}~\bibnamefont
  {Brem}}, \bibinfo {author} {\bibfnamefont {C.}~\bibnamefont {Linder{\"a}lv}},
  \bibinfo {author} {\bibfnamefont {P.}~\bibnamefont {Erhart}},\ and\ \bibinfo
  {author} {\bibfnamefont {E.}~\bibnamefont {Malic}},\ }\bibfield  {title}
  {\bibinfo {title} {Tunable phases of moir{\'e} excitons in van der {Waals}
  heterostructures},\ }\href@noop {} {\bibfield  {journal} {\bibinfo  {journal}
  {Nano letters}\ }\textbf {\bibinfo {volume} {20}},\ \bibinfo {pages} {8534}
  (\bibinfo {year} {2020})}\BibitemShut {NoStop}%
\bibitem [{\citenamefont {Shabani}\ \emph {et~al.}(2021)\citenamefont
  {Shabani}, \citenamefont {Halbertal}, \citenamefont {Wu}, \citenamefont
  {Chen}, \citenamefont {Liu}, \citenamefont {Hone}, \citenamefont {Yao},
  \citenamefont {Basov}, \citenamefont {Zhu},\ and\ \citenamefont
  {Pasupathy}}]{shabani2021deep}%
  \BibitemOpen
  \bibfield  {author} {\bibinfo {author} {\bibfnamefont {S.}~\bibnamefont
  {Shabani}}, \bibinfo {author} {\bibfnamefont {D.}~\bibnamefont {Halbertal}},
  \bibinfo {author} {\bibfnamefont {W.}~\bibnamefont {Wu}}, \bibinfo {author}
  {\bibfnamefont {M.}~\bibnamefont {Chen}}, \bibinfo {author} {\bibfnamefont
  {S.}~\bibnamefont {Liu}}, \bibinfo {author} {\bibfnamefont {J.}~\bibnamefont
  {Hone}}, \bibinfo {author} {\bibfnamefont {W.}~\bibnamefont {Yao}}, \bibinfo
  {author} {\bibfnamefont {D.~N.}\ \bibnamefont {Basov}}, \bibinfo {author}
  {\bibfnamefont {X.}~\bibnamefont {Zhu}},\ and\ \bibinfo {author}
  {\bibfnamefont {A.~N.}\ \bibnamefont {Pasupathy}},\ }\bibfield  {title}
  {\bibinfo {title} {Deep moir{\'e} potentials in twisted transition metal
  dichalcogenide bilayers},\ }\href@noop {} {\bibfield  {journal} {\bibinfo
  {journal} {Nature Physics}\ }\textbf {\bibinfo {volume} {17}},\ \bibinfo
  {pages} {720} (\bibinfo {year} {2021})}\BibitemShut {NoStop}%
\bibitem [{\citenamefont {Li}\ \emph {et~al.}(2021{\natexlab{b}})\citenamefont
  {Li}, \citenamefont {Li}, \citenamefont {Naik}, \citenamefont {Xie},
  \citenamefont {Li}, \citenamefont {Wang}, \citenamefont {Regan},
  \citenamefont {Wang}, \citenamefont {Zhao}, \citenamefont {Zhao} \emph
  {et~al.}}]{li2021imagingbands}%
  \BibitemOpen
  \bibfield  {author} {\bibinfo {author} {\bibfnamefont {H.}~\bibnamefont
  {Li}}, \bibinfo {author} {\bibfnamefont {S.}~\bibnamefont {Li}}, \bibinfo
  {author} {\bibfnamefont {M.~H.}\ \bibnamefont {Naik}}, \bibinfo {author}
  {\bibfnamefont {J.}~\bibnamefont {Xie}}, \bibinfo {author} {\bibfnamefont
  {X.}~\bibnamefont {Li}}, \bibinfo {author} {\bibfnamefont {J.}~\bibnamefont
  {Wang}}, \bibinfo {author} {\bibfnamefont {E.}~\bibnamefont {Regan}},
  \bibinfo {author} {\bibfnamefont {D.}~\bibnamefont {Wang}}, \bibinfo {author}
  {\bibfnamefont {W.}~\bibnamefont {Zhao}}, \bibinfo {author} {\bibfnamefont
  {S.}~\bibnamefont {Zhao}}, \emph {et~al.},\ }\bibfield  {title} {\bibinfo
  {title} {Imaging moir{\'e} flat bands in three-dimensional reconstructed wse
  2/ws 2 superlattices},\ }\href@noop {} {\bibfield  {journal} {\bibinfo
  {journal} {Nature materials}\ ,\ \bibinfo {pages} {1}} (\bibinfo {year}
  {2021}{\natexlab{b}})}\BibitemShut {NoStop}%
\bibitem [{\citenamefont {Li}\ \emph {et~al.}(2021{\natexlab{c}})\citenamefont
  {Li}, \citenamefont {Hu}, \citenamefont {Feng}, \citenamefont {Zhou},
  \citenamefont {An}, \citenamefont {Law}, \citenamefont {Wang},\ and\
  \citenamefont {Lin}}]{li2021lattice}%
  \BibitemOpen
  \bibfield  {author} {\bibinfo {author} {\bibfnamefont {E.}~\bibnamefont
  {Li}}, \bibinfo {author} {\bibfnamefont {J.-X.}\ \bibnamefont {Hu}}, \bibinfo
  {author} {\bibfnamefont {X.}~\bibnamefont {Feng}}, \bibinfo {author}
  {\bibfnamefont {Z.}~\bibnamefont {Zhou}}, \bibinfo {author} {\bibfnamefont
  {L.}~\bibnamefont {An}}, \bibinfo {author} {\bibfnamefont {K.~T.}\
  \bibnamefont {Law}}, \bibinfo {author} {\bibfnamefont {N.}~\bibnamefont
  {Wang}},\ and\ \bibinfo {author} {\bibfnamefont {N.}~\bibnamefont {Lin}},\
  }\bibfield  {title} {\bibinfo {title} {Lattice reconstruction induced
  multiple ultra-flat bands in twisted bilayer {WSe$_2$}},\ }\href@noop {}
  {\bibfield  {journal} {\bibinfo  {journal} {arXiv preprint arXiv:2103.06479}\
  } (\bibinfo {year} {2021}{\natexlab{c}})}\BibitemShut {NoStop}%
\bibitem [{\citenamefont {Naik}\ and\ \citenamefont
  {Jain}(2018)}]{naik2018ultraflatbands}%
  \BibitemOpen
  \bibfield  {author} {\bibinfo {author} {\bibfnamefont {M.~H.}\ \bibnamefont
  {Naik}}\ and\ \bibinfo {author} {\bibfnamefont {M.}~\bibnamefont {Jain}},\
  }\bibfield  {title} {\bibinfo {title} {Ultraflatbands and shear solitons in
  moir{\'e} patterns of twisted bilayer transition metal dichalcogenides},\
  }\href@noop {} {\bibfield  {journal} {\bibinfo  {journal} {Physical review
  letters}\ }\textbf {\bibinfo {volume} {121}},\ \bibinfo {pages} {266401}
  (\bibinfo {year} {2018})}\BibitemShut {NoStop}%
\bibitem [{\citenamefont {Guo}\ \emph {et~al.}(2020)\citenamefont {Guo},
  \citenamefont {Zhang},\ and\ \citenamefont {Lu}}]{guo2020shedding}%
  \BibitemOpen
  \bibfield  {author} {\bibinfo {author} {\bibfnamefont {H.}~\bibnamefont
  {Guo}}, \bibinfo {author} {\bibfnamefont {X.}~\bibnamefont {Zhang}},\ and\
  \bibinfo {author} {\bibfnamefont {G.}~\bibnamefont {Lu}},\ }\bibfield
  {title} {\bibinfo {title} {Shedding light on moir{\'e} excitons: A
  first-principles perspective},\ }\href@noop {} {\bibfield  {journal}
  {\bibinfo  {journal} {Science advances}\ }\textbf {\bibinfo {volume} {6}},\
  \bibinfo {pages} {eabc5638} (\bibinfo {year} {2020})}\BibitemShut {NoStop}%
\bibitem [{\citenamefont {Naik}\ \emph {et~al.}(2020)\citenamefont {Naik},
  \citenamefont {Kundu}, \citenamefont {Maity},\ and\ \citenamefont
  {Jain}}]{naik2020origin}%
  \BibitemOpen
  \bibfield  {author} {\bibinfo {author} {\bibfnamefont {M.~H.}\ \bibnamefont
  {Naik}}, \bibinfo {author} {\bibfnamefont {S.}~\bibnamefont {Kundu}},
  \bibinfo {author} {\bibfnamefont {I.}~\bibnamefont {Maity}},\ and\ \bibinfo
  {author} {\bibfnamefont {M.}~\bibnamefont {Jain}},\ }\bibfield  {title}
  {\bibinfo {title} {Origin and evolution of ultraflat bands in twisted bilayer
  transition metal dichalcogenides: Realization of triangular quantum dots},\
  }\href@noop {} {\bibfield  {journal} {\bibinfo  {journal} {Physical Review
  B}\ }\textbf {\bibinfo {volume} {102}},\ \bibinfo {pages} {075413} (\bibinfo
  {year} {2020})}\BibitemShut {NoStop}%
\bibitem [{\citenamefont {Zhan}\ \emph {et~al.}(2020)\citenamefont {Zhan},
  \citenamefont {Zhang}, \citenamefont {Lv}, \citenamefont {Zhong},
  \citenamefont {Yu}, \citenamefont {Guinea}, \citenamefont
  {Silva-Guill{\'e}n},\ and\ \citenamefont {Yuan}}]{zhan2020tunability}%
  \BibitemOpen
  \bibfield  {author} {\bibinfo {author} {\bibfnamefont {Z.}~\bibnamefont
  {Zhan}}, \bibinfo {author} {\bibfnamefont {Y.}~\bibnamefont {Zhang}},
  \bibinfo {author} {\bibfnamefont {P.}~\bibnamefont {Lv}}, \bibinfo {author}
  {\bibfnamefont {H.}~\bibnamefont {Zhong}}, \bibinfo {author} {\bibfnamefont
  {G.}~\bibnamefont {Yu}}, \bibinfo {author} {\bibfnamefont {F.}~\bibnamefont
  {Guinea}}, \bibinfo {author} {\bibfnamefont {J.~{\'A}.}\ \bibnamefont
  {Silva-Guill{\'e}n}},\ and\ \bibinfo {author} {\bibfnamefont
  {S.}~\bibnamefont {Yuan}},\ }\bibfield  {title} {\bibinfo {title} {Tunability
  of multiple ultraflat bands and effect of spin-orbit coupling in twisted
  bilayer transition metal dichalcogenides},\ }\href@noop {} {\bibfield
  {journal} {\bibinfo  {journal} {Physical Review B}\ }\textbf {\bibinfo
  {volume} {102}},\ \bibinfo {pages} {241106} (\bibinfo {year}
  {2020})}\BibitemShut {NoStop}%
\bibitem [{\citenamefont {Maity}\ \emph {et~al.}(2021)\citenamefont {Maity},
  \citenamefont {Maiti}, \citenamefont {Krishnamurthy},\ and\ \citenamefont
  {Jain}}]{maity2021reconstruction}%
  \BibitemOpen
  \bibfield  {author} {\bibinfo {author} {\bibfnamefont {I.}~\bibnamefont
  {Maity}}, \bibinfo {author} {\bibfnamefont {P.~K.}\ \bibnamefont {Maiti}},
  \bibinfo {author} {\bibfnamefont {H.}~\bibnamefont {Krishnamurthy}},\ and\
  \bibinfo {author} {\bibfnamefont {M.}~\bibnamefont {Jain}},\ }\bibfield
  {title} {\bibinfo {title} {Reconstruction of moir{\'e} lattices in twisted
  transition metal dichalcogenide bilayers},\ }\href@noop {} {\bibfield
  {journal} {\bibinfo  {journal} {Physical Review B}\ }\textbf {\bibinfo
  {volume} {103}},\ \bibinfo {pages} {L121102} (\bibinfo {year}
  {2021})}\BibitemShut {NoStop}%
\bibitem [{\citenamefont {Kundu}\ \emph {et~al.}(2021)\citenamefont {Kundu},
  \citenamefont {Maity}, \citenamefont {Bajaj}, \citenamefont {Krishnamurthy},\
  and\ \citenamefont {Jain}}]{kundu2021atomic}%
  \BibitemOpen
  \bibfield  {author} {\bibinfo {author} {\bibfnamefont {S.}~\bibnamefont
  {Kundu}}, \bibinfo {author} {\bibfnamefont {I.}~\bibnamefont {Maity}},
  \bibinfo {author} {\bibfnamefont {R.}~\bibnamefont {Bajaj}}, \bibinfo
  {author} {\bibfnamefont {H.}~\bibnamefont {Krishnamurthy}},\ and\ \bibinfo
  {author} {\bibfnamefont {M.}~\bibnamefont {Jain}},\ }\bibfield  {title}
  {\bibinfo {title} {Atomic reconstruction and flat bands in strain engineered
  transition metal dichalcogenide bilayer moir\'e systems},\ }\href@noop {}
  {\bibfield  {journal} {\bibinfo  {journal} {arXiv preprint arXiv:2108.01112}\
  } (\bibinfo {year} {2021})}\BibitemShut {NoStop}%
\bibitem [{\citenamefont {Li}\ \emph {et~al.}(2014)\citenamefont {Li},
  \citenamefont {Ludwig}, \citenamefont {Low}, \citenamefont {Chernikov},
  \citenamefont {Cui}, \citenamefont {Arefe}, \citenamefont {Kim},
  \citenamefont {Van Der~Zande}, \citenamefont {Rigosi}, \citenamefont {Hill}
  \emph {et~al.}}]{li2014valley}%
  \BibitemOpen
  \bibfield  {author} {\bibinfo {author} {\bibfnamefont {Y.}~\bibnamefont
  {Li}}, \bibinfo {author} {\bibfnamefont {J.}~\bibnamefont {Ludwig}}, \bibinfo
  {author} {\bibfnamefont {T.}~\bibnamefont {Low}}, \bibinfo {author}
  {\bibfnamefont {A.}~\bibnamefont {Chernikov}}, \bibinfo {author}
  {\bibfnamefont {X.}~\bibnamefont {Cui}}, \bibinfo {author} {\bibfnamefont
  {G.}~\bibnamefont {Arefe}}, \bibinfo {author} {\bibfnamefont {Y.~D.}\
  \bibnamefont {Kim}}, \bibinfo {author} {\bibfnamefont {A.~M.}\ \bibnamefont
  {Van Der~Zande}}, \bibinfo {author} {\bibfnamefont {A.}~\bibnamefont
  {Rigosi}}, \bibinfo {author} {\bibfnamefont {H.~M.}\ \bibnamefont {Hill}},
  \emph {et~al.},\ }\bibfield  {title} {\bibinfo {title} {{Valley splitting and
  polarization by the Zeeman effect in monolayer MoSe$_2$}},\ }\href@noop {}
  {\bibfield  {journal} {\bibinfo  {journal} {Physical review letters}\
  }\textbf {\bibinfo {volume} {113}},\ \bibinfo {pages} {266804} (\bibinfo
  {year} {2014})}\BibitemShut {NoStop}%
\bibitem [{\citenamefont {MacNeill}\ \emph {et~al.}(2015)\citenamefont
  {MacNeill}, \citenamefont {Heikes}, \citenamefont {Mak}, \citenamefont
  {Anderson}, \citenamefont {Korm{\'a}nyos}, \citenamefont {Z{\'o}lyomi},
  \citenamefont {Park},\ and\ \citenamefont {Ralph}}]{macneill2015breaking}%
  \BibitemOpen
  \bibfield  {author} {\bibinfo {author} {\bibfnamefont {D.}~\bibnamefont
  {MacNeill}}, \bibinfo {author} {\bibfnamefont {C.}~\bibnamefont {Heikes}},
  \bibinfo {author} {\bibfnamefont {K.~F.}\ \bibnamefont {Mak}}, \bibinfo
  {author} {\bibfnamefont {Z.}~\bibnamefont {Anderson}}, \bibinfo {author}
  {\bibfnamefont {A.}~\bibnamefont {Korm{\'a}nyos}}, \bibinfo {author}
  {\bibfnamefont {V.}~\bibnamefont {Z{\'o}lyomi}}, \bibinfo {author}
  {\bibfnamefont {J.}~\bibnamefont {Park}},\ and\ \bibinfo {author}
  {\bibfnamefont {D.~C.}\ \bibnamefont {Ralph}},\ }\bibfield  {title} {\bibinfo
  {title} {Breaking of valley degeneracy by magnetic field in monolayer
  {MoSe$_2$}},\ }\href@noop {} {\bibfield  {journal} {\bibinfo  {journal}
  {Physical review letters}\ }\textbf {\bibinfo {volume} {114}},\ \bibinfo
  {pages} {037401} (\bibinfo {year} {2015})}\BibitemShut {NoStop}%
\bibitem [{\citenamefont {Srivastava}\ \emph {et~al.}(2015)\citenamefont
  {Srivastava}, \citenamefont {Sidler}, \citenamefont {Allain}, \citenamefont
  {Lembke}, \citenamefont {Kis},\ and\ \citenamefont
  {Imamo{\u{g}}lu}}]{srivastava2015valley}%
  \BibitemOpen
  \bibfield  {author} {\bibinfo {author} {\bibfnamefont {A.}~\bibnamefont
  {Srivastava}}, \bibinfo {author} {\bibfnamefont {M.}~\bibnamefont {Sidler}},
  \bibinfo {author} {\bibfnamefont {A.~V.}\ \bibnamefont {Allain}}, \bibinfo
  {author} {\bibfnamefont {D.~S.}\ \bibnamefont {Lembke}}, \bibinfo {author}
  {\bibfnamefont {A.}~\bibnamefont {Kis}},\ and\ \bibinfo {author}
  {\bibfnamefont {A.}~\bibnamefont {Imamo{\u{g}}lu}},\ }\bibfield  {title}
  {\bibinfo {title} {{Valley Zeeman effect in elementary optical excitations of
  monolayer WSe$_2$}},\ }\href@noop {} {\bibfield  {journal} {\bibinfo
  {journal} {Nature Physics}\ }\textbf {\bibinfo {volume} {11}},\ \bibinfo
  {pages} {141} (\bibinfo {year} {2015})}\BibitemShut {NoStop}%
\bibitem [{\citenamefont {Aivazian}\ \emph {et~al.}(2015)\citenamefont
  {Aivazian}, \citenamefont {Gong}, \citenamefont {Jones}, \citenamefont {Chu},
  \citenamefont {Yan}, \citenamefont {Mandrus}, \citenamefont {Zhang},
  \citenamefont {Cobden}, \citenamefont {Yao},\ and\ \citenamefont
  {Xu}}]{aivazian2015magnetic}%
  \BibitemOpen
  \bibfield  {author} {\bibinfo {author} {\bibfnamefont {G.}~\bibnamefont
  {Aivazian}}, \bibinfo {author} {\bibfnamefont {Z.}~\bibnamefont {Gong}},
  \bibinfo {author} {\bibfnamefont {A.~M.}\ \bibnamefont {Jones}}, \bibinfo
  {author} {\bibfnamefont {R.-L.}\ \bibnamefont {Chu}}, \bibinfo {author}
  {\bibfnamefont {J.}~\bibnamefont {Yan}}, \bibinfo {author} {\bibfnamefont
  {D.~G.}\ \bibnamefont {Mandrus}}, \bibinfo {author} {\bibfnamefont
  {C.}~\bibnamefont {Zhang}}, \bibinfo {author} {\bibfnamefont
  {D.}~\bibnamefont {Cobden}}, \bibinfo {author} {\bibfnamefont
  {W.}~\bibnamefont {Yao}},\ and\ \bibinfo {author} {\bibfnamefont
  {X.}~\bibnamefont {Xu}},\ }\bibfield  {title} {\bibinfo {title} {Magnetic
  control of valley pseudospin in monolayer {WSe$_2$}},\ }\href@noop {}
  {\bibfield  {journal} {\bibinfo  {journal} {Nature Physics}\ }\textbf
  {\bibinfo {volume} {11}},\ \bibinfo {pages} {148} (\bibinfo {year}
  {2015})}\BibitemShut {NoStop}%
\bibitem [{\citenamefont {Wojs}\ and\ \citenamefont
  {Hawrylak}(1995)}]{wojs1995negatively}%
  \BibitemOpen
  \bibfield  {author} {\bibinfo {author} {\bibfnamefont {A.}~\bibnamefont
  {Wojs}}\ and\ \bibinfo {author} {\bibfnamefont {P.}~\bibnamefont
  {Hawrylak}},\ }\bibfield  {title} {\bibinfo {title} {Negatively charged
  magnetoexcitons in quantum dots},\ }\href@noop {} {\bibfield  {journal}
  {\bibinfo  {journal} {Physical Review B}\ }\textbf {\bibinfo {volume} {51}},\
  \bibinfo {pages} {10880} (\bibinfo {year} {1995})}\BibitemShut {NoStop}%
\bibitem [{\citenamefont {Halonen}\ \emph {et~al.}(1992)\citenamefont
  {Halonen}, \citenamefont {Chakraborty},\ and\ \citenamefont
  {Pietil{\"a}inen}}]{halonen1992excitons}%
  \BibitemOpen
  \bibfield  {author} {\bibinfo {author} {\bibfnamefont {V.}~\bibnamefont
  {Halonen}}, \bibinfo {author} {\bibfnamefont {T.}~\bibnamefont
  {Chakraborty}},\ and\ \bibinfo {author} {\bibfnamefont {P.}~\bibnamefont
  {Pietil{\"a}inen}},\ }\bibfield  {title} {\bibinfo {title} {Excitons in a
  parabolic quantum dot in magnetic fields},\ }\href@noop {} {\bibfield
  {journal} {\bibinfo  {journal} {Physical Review B}\ }\textbf {\bibinfo
  {volume} {45}},\ \bibinfo {pages} {5980} (\bibinfo {year}
  {1992})}\BibitemShut {NoStop}%
\bibitem [{\citenamefont {Que}(1992)}]{que1992excitons}%
  \BibitemOpen
  \bibfield  {author} {\bibinfo {author} {\bibfnamefont {W.}~\bibnamefont
  {Que}},\ }\bibfield  {title} {\bibinfo {title} {Excitons in quantum dots with
  parabolic confinement},\ }\href@noop {} {\bibfield  {journal} {\bibinfo
  {journal} {Physical Review B}\ }\textbf {\bibinfo {volume} {45}},\ \bibinfo
  {pages} {11036} (\bibinfo {year} {1992})}\BibitemShut {NoStop}%
\bibitem [{\citenamefont {Maksym}\ and\ \citenamefont
  {Chakraborty}(1990)}]{maksym1990quantum}%
  \BibitemOpen
  \bibfield  {author} {\bibinfo {author} {\bibfnamefont {P.}~\bibnamefont
  {Maksym}}\ and\ \bibinfo {author} {\bibfnamefont {T.}~\bibnamefont
  {Chakraborty}},\ }\bibfield  {title} {\bibinfo {title} {Quantum dots in a
  magnetic field: Role of electron-electron interactions},\ }\href@noop {}
  {\bibfield  {journal} {\bibinfo  {journal} {Physical review letters}\
  }\textbf {\bibinfo {volume} {65}},\ \bibinfo {pages} {108} (\bibinfo {year}
  {1990})}\BibitemShut {NoStop}%
\bibitem [{\citenamefont {Yu}\ \emph {et~al.}(2014)\citenamefont {Yu},
  \citenamefont {Liu}, \citenamefont {Gong}, \citenamefont {Xu},\ and\
  \citenamefont {Yao}}]{yu2014dirac}%
  \BibitemOpen
  \bibfield  {author} {\bibinfo {author} {\bibfnamefont {H.}~\bibnamefont
  {Yu}}, \bibinfo {author} {\bibfnamefont {G.-B.}\ \bibnamefont {Liu}},
  \bibinfo {author} {\bibfnamefont {P.}~\bibnamefont {Gong}}, \bibinfo {author}
  {\bibfnamefont {X.}~\bibnamefont {Xu}},\ and\ \bibinfo {author}
  {\bibfnamefont {W.}~\bibnamefont {Yao}},\ }\bibfield  {title} {\bibinfo
  {title} {Dirac cones and {Dirac} saddle points of bright excitons in
  monolayer transition metal dichalcogenides},\ }\href@noop {} {\bibfield
  {journal} {\bibinfo  {journal} {Nature communications}\ }\textbf {\bibinfo
  {volume} {5}},\ \bibinfo {pages} {1} (\bibinfo {year} {2014})}\BibitemShut
  {NoStop}%
\bibitem [{\citenamefont {Glazov}\ \emph {et~al.}(2014)\citenamefont {Glazov},
  \citenamefont {Amand}, \citenamefont {Marie}, \citenamefont {Lagarde},
  \citenamefont {Bouet},\ and\ \citenamefont {Urbaszek}}]{glazov2014exciton}%
  \BibitemOpen
  \bibfield  {author} {\bibinfo {author} {\bibfnamefont {M.}~\bibnamefont
  {Glazov}}, \bibinfo {author} {\bibfnamefont {T.}~\bibnamefont {Amand}},
  \bibinfo {author} {\bibfnamefont {X.}~\bibnamefont {Marie}}, \bibinfo
  {author} {\bibfnamefont {D.}~\bibnamefont {Lagarde}}, \bibinfo {author}
  {\bibfnamefont {L.}~\bibnamefont {Bouet}},\ and\ \bibinfo {author}
  {\bibfnamefont {B.}~\bibnamefont {Urbaszek}},\ }\bibfield  {title} {\bibinfo
  {title} {Exciton fine structure and spin decoherence in monolayers of
  transition metal dichalcogenides},\ }\href@noop {} {\bibfield  {journal}
  {\bibinfo  {journal} {Physical Review B}\ }\textbf {\bibinfo {volume} {89}},\
  \bibinfo {pages} {201302} (\bibinfo {year} {2014})}\BibitemShut {NoStop}%
\bibitem [{\citenamefont {Yu}\ and\ \citenamefont {Wu}(2014)}]{yu2014valley}%
  \BibitemOpen
  \bibfield  {author} {\bibinfo {author} {\bibfnamefont {T.}~\bibnamefont
  {Yu}}\ and\ \bibinfo {author} {\bibfnamefont {M.}~\bibnamefont {Wu}},\
  }\bibfield  {title} {\bibinfo {title} {Valley depolarization due to
  intervalley and intravalley electron-hole exchange interactions in monolayer
  {MoS$_2$}},\ }\href@noop {} {\bibfield  {journal} {\bibinfo  {journal}
  {Physical Review B}\ }\textbf {\bibinfo {volume} {89}},\ \bibinfo {pages}
  {205303} (\bibinfo {year} {2014})}\BibitemShut {NoStop}%
\bibitem [{\citenamefont {Wu}\ \emph {et~al.}(2015)\citenamefont {Wu},
  \citenamefont {Qu},\ and\ \citenamefont {Macdonald}}]{wu2015exciton}%
  \BibitemOpen
  \bibfield  {author} {\bibinfo {author} {\bibfnamefont {F.}~\bibnamefont
  {Wu}}, \bibinfo {author} {\bibfnamefont {F.}~\bibnamefont {Qu}},\ and\
  \bibinfo {author} {\bibfnamefont {A.~H.}\ \bibnamefont {Macdonald}},\
  }\bibfield  {title} {\bibinfo {title} {Exciton band structure of monolayer
  {MoS$_2$}},\ }\href@noop {} {\bibfield  {journal} {\bibinfo  {journal}
  {Physical Review B}\ }\textbf {\bibinfo {volume} {91}},\ \bibinfo {pages}
  {075310} (\bibinfo {year} {2015})}\BibitemShut {NoStop}%
\bibitem [{\citenamefont {Korm{\'a}nyos}\ \emph {et~al.}(2015)\citenamefont
  {Korm{\'a}nyos}, \citenamefont {Burkard}, \citenamefont {Gmitra},
  \citenamefont {Fabian}, \citenamefont {Z{\'o}lyomi}, \citenamefont
  {Drummond},\ and\ \citenamefont {Fal’ko}}]{kormanyos2015k}%
  \BibitemOpen
  \bibfield  {author} {\bibinfo {author} {\bibfnamefont {A.}~\bibnamefont
  {Korm{\'a}nyos}}, \bibinfo {author} {\bibfnamefont {G.}~\bibnamefont
  {Burkard}}, \bibinfo {author} {\bibfnamefont {M.}~\bibnamefont {Gmitra}},
  \bibinfo {author} {\bibfnamefont {J.}~\bibnamefont {Fabian}}, \bibinfo
  {author} {\bibfnamefont {V.}~\bibnamefont {Z{\'o}lyomi}}, \bibinfo {author}
  {\bibfnamefont {N.~D.}\ \bibnamefont {Drummond}},\ and\ \bibinfo {author}
  {\bibfnamefont {V.}~\bibnamefont {Fal’ko}},\ }\bibfield  {title} {\bibinfo
  {title} {$k\cdot p$ theory for two-dimensional transition metal
  dichalcogenide semiconductors},\ }\href@noop {} {\bibfield  {journal}
  {\bibinfo  {journal} {2D Materials}\ }\textbf {\bibinfo {volume} {2}},\
  \bibinfo {pages} {022001} (\bibinfo {year} {2015})}\BibitemShut {NoStop}%
\bibitem [{\citenamefont {Mak}\ \emph {et~al.}(2013)\citenamefont {Mak},
  \citenamefont {He}, \citenamefont {Lee}, \citenamefont {Lee}, \citenamefont
  {Hone}, \citenamefont {Heinz},\ and\ \citenamefont {Shan}}]{mak2013tightly}%
  \BibitemOpen
  \bibfield  {author} {\bibinfo {author} {\bibfnamefont {K.~F.}\ \bibnamefont
  {Mak}}, \bibinfo {author} {\bibfnamefont {K.}~\bibnamefont {He}}, \bibinfo
  {author} {\bibfnamefont {C.}~\bibnamefont {Lee}}, \bibinfo {author}
  {\bibfnamefont {G.~H.}\ \bibnamefont {Lee}}, \bibinfo {author} {\bibfnamefont
  {J.}~\bibnamefont {Hone}}, \bibinfo {author} {\bibfnamefont {T.~F.}\
  \bibnamefont {Heinz}},\ and\ \bibinfo {author} {\bibfnamefont
  {J.}~\bibnamefont {Shan}},\ }\bibfield  {title} {\bibinfo {title} {Tightly
  bound trions in monolayer {MoS$_2$}},\ }\href@noop {} {\bibfield  {journal}
  {\bibinfo  {journal} {Nature materials}\ }\textbf {\bibinfo {volume} {12}},\
  \bibinfo {pages} {207} (\bibinfo {year} {2013})}\BibitemShut {NoStop}%
\bibitem [{\citenamefont {Ross}\ \emph {et~al.}(2013)\citenamefont {Ross},
  \citenamefont {Wu}, \citenamefont {Yu}, \citenamefont {Ghimire},
  \citenamefont {Jones}, \citenamefont {Aivazian}, \citenamefont {Yan},
  \citenamefont {Mandrus}, \citenamefont {Xiao}, \citenamefont {Yao} \emph
  {et~al.}}]{ross2013electrical}%
  \BibitemOpen
  \bibfield  {author} {\bibinfo {author} {\bibfnamefont {J.~S.}\ \bibnamefont
  {Ross}}, \bibinfo {author} {\bibfnamefont {S.}~\bibnamefont {Wu}}, \bibinfo
  {author} {\bibfnamefont {H.}~\bibnamefont {Yu}}, \bibinfo {author}
  {\bibfnamefont {N.~J.}\ \bibnamefont {Ghimire}}, \bibinfo {author}
  {\bibfnamefont {A.~M.}\ \bibnamefont {Jones}}, \bibinfo {author}
  {\bibfnamefont {G.}~\bibnamefont {Aivazian}}, \bibinfo {author}
  {\bibfnamefont {J.}~\bibnamefont {Yan}}, \bibinfo {author} {\bibfnamefont
  {D.~G.}\ \bibnamefont {Mandrus}}, \bibinfo {author} {\bibfnamefont
  {D.}~\bibnamefont {Xiao}}, \bibinfo {author} {\bibfnamefont {W.}~\bibnamefont
  {Yao}}, \emph {et~al.},\ }\bibfield  {title} {\bibinfo {title} {Electrical
  control of neutral and charged excitons in a monolayer semiconductor},\
  }\href@noop {} {\bibfield  {journal} {\bibinfo  {journal} {Nature
  communications}\ }\textbf {\bibinfo {volume} {4}},\ \bibinfo {pages} {1}
  (\bibinfo {year} {2013})}\BibitemShut {NoStop}%
\bibitem [{\citenamefont {Jones}\ \emph {et~al.}(2013)\citenamefont {Jones},
  \citenamefont {Yu}, \citenamefont {Ghimire}, \citenamefont {Wu},
  \citenamefont {Aivazian}, \citenamefont {Ross}, \citenamefont {Zhao},
  \citenamefont {Yan}, \citenamefont {Mandrus}, \citenamefont {Xiao} \emph
  {et~al.}}]{jones2013optical}%
  \BibitemOpen
  \bibfield  {author} {\bibinfo {author} {\bibfnamefont {A.~M.}\ \bibnamefont
  {Jones}}, \bibinfo {author} {\bibfnamefont {H.}~\bibnamefont {Yu}}, \bibinfo
  {author} {\bibfnamefont {N.~J.}\ \bibnamefont {Ghimire}}, \bibinfo {author}
  {\bibfnamefont {S.}~\bibnamefont {Wu}}, \bibinfo {author} {\bibfnamefont
  {G.}~\bibnamefont {Aivazian}}, \bibinfo {author} {\bibfnamefont {J.~S.}\
  \bibnamefont {Ross}}, \bibinfo {author} {\bibfnamefont {B.}~\bibnamefont
  {Zhao}}, \bibinfo {author} {\bibfnamefont {J.}~\bibnamefont {Yan}}, \bibinfo
  {author} {\bibfnamefont {D.~G.}\ \bibnamefont {Mandrus}}, \bibinfo {author}
  {\bibfnamefont {D.}~\bibnamefont {Xiao}}, \emph {et~al.},\ }\bibfield
  {title} {\bibinfo {title} {Optical generation of excitonic valley coherence
  in monolayer {WSe$_2$}},\ }\href@noop {} {\bibfield  {journal} {\bibinfo
  {journal} {Nature nanotechnology}\ }\textbf {\bibinfo {volume} {8}},\
  \bibinfo {pages} {634} (\bibinfo {year} {2013})}\BibitemShut {NoStop}%
\bibitem [{\citenamefont {Anisimovas}\ and\ \citenamefont
  {Peeters}(2003)}]{anisimovas2003excitonic}%
  \BibitemOpen
  \bibfield  {author} {\bibinfo {author} {\bibfnamefont {E.}~\bibnamefont
  {Anisimovas}}\ and\ \bibinfo {author} {\bibfnamefont {F.}~\bibnamefont
  {Peeters}},\ }\bibfield  {title} {\bibinfo {title} {Excitonic trions in
  vertically coupled quantum dots},\ }\href@noop {} {\bibfield  {journal}
  {\bibinfo  {journal} {Physical Review B}\ }\textbf {\bibinfo {volume} {68}},\
  \bibinfo {pages} {115310} (\bibinfo {year} {2003})}\BibitemShut {NoStop}%
\bibitem [{\citenamefont {Xie}\ and\ \citenamefont
  {Chen}(2000)}]{xie2000excitonic}%
  \BibitemOpen
  \bibfield  {author} {\bibinfo {author} {\bibfnamefont {W.}~\bibnamefont
  {Xie}}\ and\ \bibinfo {author} {\bibfnamefont {C.}~\bibnamefont {Chen}},\
  }\bibfield  {title} {\bibinfo {title} {Excitonic trion $x^-$ in gaas quantum
  dots},\ }\href@noop {} {\bibfield  {journal} {\bibinfo  {journal} {Physica E:
  Low-dimensional Systems and Nanostructures}\ }\textbf {\bibinfo {volume}
  {8}},\ \bibinfo {pages} {77} (\bibinfo {year} {2000})}\BibitemShut {NoStop}%
\bibitem [{\citenamefont {Bracker}\ \emph {et~al.}(2005)\citenamefont
  {Bracker}, \citenamefont {Stinaff}, \citenamefont {Gammon}, \citenamefont
  {Ware}, \citenamefont {Tischler}, \citenamefont {Park}, \citenamefont
  {Gershoni}, \citenamefont {Filinov}, \citenamefont {Bonitz}, \citenamefont
  {Peeters} \emph {et~al.}}]{bracker2005binding}%
  \BibitemOpen
  \bibfield  {author} {\bibinfo {author} {\bibfnamefont {A.}~\bibnamefont
  {Bracker}}, \bibinfo {author} {\bibfnamefont {E.}~\bibnamefont {Stinaff}},
  \bibinfo {author} {\bibfnamefont {D.}~\bibnamefont {Gammon}}, \bibinfo
  {author} {\bibfnamefont {M.}~\bibnamefont {Ware}}, \bibinfo {author}
  {\bibfnamefont {J.}~\bibnamefont {Tischler}}, \bibinfo {author}
  {\bibfnamefont {D.}~\bibnamefont {Park}}, \bibinfo {author} {\bibfnamefont
  {D.}~\bibnamefont {Gershoni}}, \bibinfo {author} {\bibfnamefont
  {A.}~\bibnamefont {Filinov}}, \bibinfo {author} {\bibfnamefont
  {M.}~\bibnamefont {Bonitz}}, \bibinfo {author} {\bibfnamefont
  {F.}~\bibnamefont {Peeters}}, \emph {et~al.},\ }\bibfield  {title} {\bibinfo
  {title} {Binding energies of positive and negative trions: From quantum wells
  to quantum dots},\ }\href@noop {} {\bibfield  {journal} {\bibinfo  {journal}
  {Physical Review B}\ }\textbf {\bibinfo {volume} {72}},\ \bibinfo {pages}
  {035332} (\bibinfo {year} {2005})}\BibitemShut {NoStop}%
\bibitem [{\citenamefont {Naik}\ \emph {et~al.}(2022)\citenamefont {Naik},
  \citenamefont {Regan}, \citenamefont {Zhang}, \citenamefont {Chan},
  \citenamefont {Li}, \citenamefont {Wang}, \citenamefont {Yoon}, \citenamefont
  {Ong}, \citenamefont {Zhao}, \citenamefont {Zhao} \emph
  {et~al.}}]{naik2022nature}%
  \BibitemOpen
  \bibfield  {author} {\bibinfo {author} {\bibfnamefont {M.~H.}\ \bibnamefont
  {Naik}}, \bibinfo {author} {\bibfnamefont {E.~C.}\ \bibnamefont {Regan}},
  \bibinfo {author} {\bibfnamefont {Z.}~\bibnamefont {Zhang}}, \bibinfo
  {author} {\bibfnamefont {Y.-h.}\ \bibnamefont {Chan}}, \bibinfo {author}
  {\bibfnamefont {Z.}~\bibnamefont {Li}}, \bibinfo {author} {\bibfnamefont
  {D.}~\bibnamefont {Wang}}, \bibinfo {author} {\bibfnamefont {Y.}~\bibnamefont
  {Yoon}}, \bibinfo {author} {\bibfnamefont {C.~S.}\ \bibnamefont {Ong}},
  \bibinfo {author} {\bibfnamefont {W.}~\bibnamefont {Zhao}}, \bibinfo {author}
  {\bibfnamefont {S.}~\bibnamefont {Zhao}}, \emph {et~al.},\ }\bibfield
  {title} {\bibinfo {title} {Nature of novel moir\'e exciton states in
  {WSe$_2$/WS$_2$} heterobilayers},\ }\href@noop {} {\bibfield  {journal}
  {\bibinfo  {journal} {arXiv preprint arXiv:2201.02562}\ } (\bibinfo {year}
  {2022})}\BibitemShut {NoStop}%
\bibitem [{\citenamefont {Gu}\ \emph {et~al.}(2021)\citenamefont {Gu},
  \citenamefont {Ma}, \citenamefont {Liu}, \citenamefont {Watanabe},
  \citenamefont {Taniguchi}, \citenamefont {Hone}, \citenamefont {Shan},\ and\
  \citenamefont {Mak}}]{gu2021dipolar}%
  \BibitemOpen
  \bibfield  {author} {\bibinfo {author} {\bibfnamefont {J.}~\bibnamefont
  {Gu}}, \bibinfo {author} {\bibfnamefont {L.}~\bibnamefont {Ma}}, \bibinfo
  {author} {\bibfnamefont {S.}~\bibnamefont {Liu}}, \bibinfo {author}
  {\bibfnamefont {K.}~\bibnamefont {Watanabe}}, \bibinfo {author}
  {\bibfnamefont {T.}~\bibnamefont {Taniguchi}}, \bibinfo {author}
  {\bibfnamefont {J.~C.}\ \bibnamefont {Hone}}, \bibinfo {author}
  {\bibfnamefont {J.}~\bibnamefont {Shan}},\ and\ \bibinfo {author}
  {\bibfnamefont {K.~F.}\ \bibnamefont {Mak}},\ }\bibfield  {title} {\bibinfo
  {title} {Dipolar excitonic insulator in a moir\'e lattice},\ }\href@noop {}
  {\bibfield  {journal} {\bibinfo  {journal} {arXiv preprint arXiv:2108.06588}\
  } (\bibinfo {year} {2021})}\BibitemShut {NoStop}%
\end{thebibliography}%
\end{document}